\documentclass[aps, prfluids, onecolumn, amssymb, eprint, preprintnumbers, 11pt, showkeys, tightenlines]{revtex4-2}

\usepackage{amsmath,amssymb}
\usepackage[ruled,vlined,english]{algorithm2e}
\usepackage[babel=true]{csquotes}
\usepackage{xcolor}
\usepackage{enumitem}
\usepackage{datetime}
\usepackage{caption}
\usepackage{subcaption}
\usepackage{setspace}
\usepackage{lineno,hyperref}
\modulolinenumbers[5]
\usepackage{mathrsfs}
\usepackage{amsbsy}
\usepackage{svg}
\usepackage{amsmath,amsfonts,amsthm,bm} 
\usepackage{multirow}
\usepackage{tabularx}
\usepackage{makecell}
\newcolumntype{C}{>{\centering\arraybackslash}X}   
\usepackage{booktabs}
\usepackage{graphicx}
\usepackage{float}

\begin{document}
\title{Synchronization and optimization of Large Eddy Simulation using an online Ensemble Kalman Filter}

\author{L. Villanueva}
\email[Corresponding author : ]{lucas.villanueva@ensma.fr}
\affiliation{Institut Pprime, CNRS -
ISAE-ENSMA - Universit\'{e} de Poitiers, 11 Bd. Marie et Pierre Curie,
Site du Futuroscope, TSA 41123, 86073 Poitiers Cedex 9, France}

\author{K. Truffin}
\affiliation{Institut Carnot IFPEN Transports Energie, IFP Energies nouvelles, 1-4 avenue de Bois-Préau, 92852 Rueil-Malmaison, France}

\author{M. Meldi}
\affiliation{Univ. Lille, CNRS, ONERA, Arts et Métiers ParisTech, Centrale Lille, UMR 9014- LMFL- Laboratoire de Mécanique des fluides de Lille - Kampé de Feriet, F-59000 Lille, France}
\date{October 2023}%

\preprint{\textit{Preprint submitted to Physical Review Fluids}}
\begin{abstract}
    An online Data Assimilation strategy based on the Ensemble Kalman Filter (EnKF) is used to improve the predictive capabilities of Large Eddy Simulation (LES) for the analysis of the turbulent flow in a plane channel, $Re_\tau \approx 550$. The algorithm sequentially combines the LES prediction with high-fidelity, sparse instantaneous data obtained from a Direct Numerical Simulation (DNS). It is shown that the procedure provides an augmented state which exhibits higher accuracy than the LES model and it synchronizes with the time evolution of the high-fidelity DNS data if the hyperparameters governing the EnKF are properly chosen. In addition, the data-driven algorithm is able to improve the accuracy of the subgrid-scale model included in the LES, the Smagorinsky model, via the optimization of a free coefficient. However, while the online EnKF strategy is able to reduce the global error of the LES prediction, a discrepancy with the reference DNS data is still observed because of structural flaws of the subgrid-scale model used.    
\end{abstract}


\maketitle

\section{Introduction}
Among the state-of-the-art tools in Computational Fluid Dynamics (CFD) for the analysis of complex flow configurations, the Large Eddy Simulation (LES) \cite{Pope2000_cambridge,Sagaut2006_springer} is arguably the most investigated strategy in the last decades. LES relies on the application of statistical hypothesis related to turbulence theory to filter out the smallest physical scales of motion, so that the number of degrees of freedom to be simulated is drastically reduced when compared with Direct Numerical Simulation. The effects of such filtered eddies and their interactions with the resolved flow are taken into account by a specific SubGrid-Scale (SGS) closure. One of the most interesting features of LES is that it can naturally represent the unstationary, three-dimensional features of the flow. This key property, which is not obtained by most of the closures used to simulate turbulent flows, is essential for example for the prediction of extreme events. These rare occurrences must be fully taken into account in industrial applications and they are observed in a large spectrum of applications, such as internal flows for the study of combustion cyclic variability~\cite{ding2023use,truffin2015using}, non-cyclic phenomena~\cite{robert2019large,robert2015prediction} or direct spray injection and aerodynamics in transient combustion engines~\cite{poubeau2017large} and external flows for wind / urban engineering \cite{Solari2020_fbe,SOLARI2020_jweia}. 

The representation of instantaneous features of the flow also exhibits a great potential for LES applications in the framework of Industry 4.0 \cite{Colombo2014_icb,Bai2020_ijpe}. Within this digital revolution, envisioned applications predict and control real configurations, usually referred to as physical twin, using a numerical counterpart, the digital twin \cite{Rasheed2020_IEEE,Semeraro2021_ci}. Most studies in the literature for fluid mechanics couple the physical system with reduced-order models or low-rank CFD \cite{Renganathan2020_AIAA,Thomas2021_aaps,Chetan2021_we,Molinaro2021_cf,Ahmed2021_cf,Li2022_e} and thus the communication and control is limited to statistical macro-features of the flow. Applications of LES in this context are potentially groundbreaking because the real-time coupling of a real flow with LES is consistent in terms of physical representation. Successful implementation of a LES-based digital twin could potentially anticipate extreme events via numerical simulation and prevent catastrophic occurrences for the physical twin. However, three barriers must be lifted to see the fruition of this futuristic application. First, computational resources required to perform LES are orders of magnitude larger than the real time of physical applications of industrial interest. While this barrier seems unbeatable, new technologies such as quantum computing \cite{givi2021machine,sammak2015quantum} may provide a needed breakthrough in terms of power needed for extended digital twin applications. Second, low-rank CFD is affected by a bias associated with the turbulence / SGS closures which often interact with the discretization error as well as explicit/implicit filtering for LES. These non-linear interactions between error sources may severely impact the accuracy of the results as they are often very sensitive to the test case of investigation. Therefore, general guidelines for applications are elusive. Third, CFD and in particular LES is extremely sensitive to perturbations and uncertainty in the initial and boundary conditions. Such perturbations, which also interact with the discretization error and the SGS modeling, may produce significant instantaneous decorrelation of initially identical fields in very short times. 

The second and third barriers listed, namely the accuracy of turbulence closures and the possibility for scale-resolved CFD to follow with good correlation a physical flow, have been recently investigated using data-driven methods. Uncertainty Quantification techniques have been extensively used to improve the predictive features of LES \cite{Meldi2011_pof,Meldi2012_pof,Khalil2015_pci,Safta2017_ijnmf} and, more recently, works in Data Assimilation \cite{Daley1991_cambridge,Asch2016_SIAM} optimized the behavior of SGS modeling in different numerical solvers \cite{Chandramouli2020_jcp,Mons2021_prf,Moldovan2022_jfm}. In particular, Mons et al. \cite{Mons2021_prf} have performed an advanced optimization of the Smagorinsky model \cite{Smagorinsky1963_mwr}, one of the most used SGS closures in the literature, for the test case of the plane channel flow. In their work, the DA procedure relies on statistical features of the flow for optimization. While the results obtained significantly increase the global accuracy of the LES solver, this procedure is not fit for on-the-fly optimization in the framework of a digital twin. A number of DA works have also targeted numerical synchronization and reconstruction of turbulent instantaneous flows from limited data. Using DA formalism, this procedure can be referred to as \textit{state augmentation}. Such studies have been relying on DNS \cite{Wang2022_jfm} as well as LES \cite{Meldi2017_jcp,Labahn2019_pci,Chandramouli2020_jcp}. The main conclusions that can be drawn by these studies is that the efficiency in the synchronization of the flow depends on the number and positioning of sensors, as well as on the DA technique used. Among the proposals in the literature, the Ensemble Kalman Filter \cite{Evensen2009_IEEE,Asch2016_SIAM}, which relies on an ensemble of numerical realizations to perform optimization and state reconstruction, appears to be a perfect candidate for this task. Thanks to its sequential features which allow to perform an instantaneous, on-the-fly update of the physical field, this tool shows potential for future integration in digital twins.

The present work proposes an extensive analysis of an EnKF-based tool application to LES in terms of i) optimization of the SGS model and ii) state augmentation. The test case of investigation is the turbulent channel flow, which has already been analyzed using DA techniques \cite{Mons2021_prf,Wang2022_jfm}. The novel point here is that both the optimization and the state augmentation are performed on-the-fly, progressively informing the LES ensemble members with time-resolved DNS data which are sampled at a limited amount of sensors near the wall. The objective here is to assess the robustness of the procedure, both in terms of optimization as well as flow reconstruction, when spatial-temporal sparse data are used. The on-the-fly coupling of LES simulation and DNS data is performed via CONES \cite{Villanueva2023_arxiv}, a library developed by the team to perform online coupling between different solvers.

The article is structured as follows. In section \ref{Sec:numsTools}, the numerical tools used for the analysis are going to be presented and discussed. This includes the numerical LES solver, the EnKF methodology, and the platform CONES. In section \ref{Sec:setup}, the test case and the set-up of the DA runs are going to be introduced. In section \ref{Sec:results}, the results of the optimization of the SGS model are discussed. In section \ref{Sec::synchro}, the global impact of the DA methodology over the instantaneous flow predicted and the correlation with the DNS data available is investigated. Finally, in section \ref{Sec::conclusions} concluding remarks are drawn and future perspectives are investigated.

\section{Numerical tools}
\label{Sec:numsTools}

All the numerical ingredients used to perform the present analysis are presented in this section. These tools include a description of the dynamic equations and the numerical solver used, details about the EnKF, and information about the platform CONES used to perform online DA.

\subsection{Dynamic equations and numerical solver}
\label{Sec:nums}

The Navier--Stokes equations for incompressible flows and Newtonian fluid can be formulated as:

\begin{eqnarray}
\frac{\partial u_j}{\partial x_j} &=& 0 \label{eq:NSMass1} \\
    \frac{\partial \, {u}_i}{\partial t} + \frac{\partial \, {u}_i \, {u}_j}{\partial x_j} &=& - \frac{1}{\rho} \frac{\partial \, {p}}{\partial x_i} + \nu \frac{\partial^2 \, {u}_i}{\partial x_j \partial x_j} + f_i \label{eq:NSMomentum}
\end{eqnarray}

where $\mathbf{u}=[u_1, \, u_2, \, u_3] = [u_x, \, u_y, \, u_z]$ is the velocity field, $\rho$ is the density, $p$ is the pressure, $\nu$ is the kinematic viscosity and $\mathbf{f}=[f_1, \, f_2, \, f_3]$ is a volume forcing. Repetition over the index $j$ is employed for the sake of conciseness. In the LES formalism, equations \ref{eq:NSMass1} and \ref{eq:NSMomentum} are filtered to obtain a global reduction of the degrees of freedom of the physical system:

\begin{eqnarray}
\frac{\partial \widetilde{u}_j}{\partial x_j} &=& 0 \label{eq:NSMassLES} \\
    \frac{\partial \, \tilde{u}_i}{\partial t} + \frac{\partial \, \widetilde{u}_i \, \widetilde{u}_j}{\partial x_j} &=& - \frac{1}{\rho} \frac{\partial \, \widetilde{p}}{\partial x_i} + \nu \frac{\partial^2 \, \widetilde{u}_i}{\partial x_j \partial x_j} - \frac{\partial \tau_{ij}}{\partial x_j} + \widetilde{f}_i\label{eq:NSMomentumLES}
\end{eqnarray}

The tilde symbol stands for filtered variables and $\tau_{ij}=\widetilde{u_i u_j} - \widetilde{u}_i \widetilde{u}_j$ is the subgrid scale stress tensor. In the Smagorinsky model \cite{Smagorinsky1963_mwr}, the deviatoric part of $\tau_{ij}$ is modelled as an eddy viscosity effect:

\begin{equation}
    \tau_{ij} - \frac{1}{3} \tau_{kk} \delta_{ij} = -2 \nu_{sgs} \widetilde{S}_{ij} \, , \quad \nu_{sgs} = (C_S \Delta)^2 \sqrt{2 \widetilde{S}_{ij} \widetilde{S}_{ij}} \label{eq:tauSGS_LES}
\end{equation}

where $\widetilde{S}_{ij} = \frac{1}{2} \left( \frac{\partial \widetilde{u}_i}{\partial x_j} + \frac{\partial \widetilde{u}_j}{\partial x_i} \right)$ is the rate-of-strain tensor of the resolved velocity field, $\Delta$ is the filter width and $C_S$ is a model coefficient that can be selected by the user. Classical values found in the literature are $C_S \in [0.1, \, 0.2]$. This formulation, which is derived from the asymptotic turbulence theory by Kolmogorov, fails to provide an accurate prediction of the interactions between the resolved and filtered physical variables. The reason is that the SGS stress tensor in equation \ref{eq:tauSGS_LES} is inherently dissipative and affects all the simulated scales of the flow \cite{Sagaut2006_springer}. Despite these negative features, the direct and simple implementation of such a model made it a popular choice for most solvers.

The numerical simulation of equations \ref{eq:NSMassLES} - \ref{eq:NSMomentumLES} is performed using the open-source code \textit{OpenFOAM} \cite{OpenFOAM}. This C++ library provides a finite volume \cite{Ferziger1996_springer} discretization of the dynamic equations and modules for turbulence / SGS closure are already implemented. The equations are resolved using a PISO loop \cite{Ferziger1996_springer} which employs a Poisson equation to iteratively obtain a solenoidal condition for the velocity field, starting from the prediction obtained by the resolution of the momentum equation \ref{eq:NSMomentumLES}. Second-order centered schemes have been used for the discretization of spatial derivatives. A second-order backward scheme has been used for the time advancement of the solution. The LES equations are closed using the classical Smagorinsky model previously introduced. The implementation in OpenFOAM relies on two model constants, the parameter $C_k$ and the normalized dissipation parameter $C_{\varepsilon}$. The latter usually exhibit high sensitivity to turbulence production effects and lack of homogeneity of the flow \cite{Vassilicos2015_arfm}. In the case of turbulent equilibrium, such as in Kolmogorov theory, $C_{\varepsilon}=const$, and its value can be set by the user. OpenFOAM suggests a default value of $C_{\varepsilon}=1.048$, which is in the range of experimental and numerical findings. Within this framework, the connection between $C_S$ and $C_k$ is:

\begin{equation}
    C_S^2 = C_k \sqrt{\frac{C_k}{C_{\varepsilon}}}
\end{equation}

The LES filtering is performed implicitly using the grid resolution. The filter width $\Delta$ is thus locally proportional to the volume of each cell $V_c$ (\textit{cube-root volume} filter option in OpenFOAM) and more precisely $\Delta =\sqrt[3]{V_c}$.

\subsection{Data Assimilation}

Data Assimilation \cite{Daley1991_cambridge,Asch2016_SIAM} includes a large spectrum of data-driven techniques whose main goal is to obtain an \textit{augmented} prediction of a random process investigated, combining different sources of information. The tools are usually grouped in two main categories. The \textit{variational} approaches perform the DA strategy via an optimization problem. The \textit{sequential} approaches usually rely on probabilistic approaches which are based on Bayes' theorem. This work will be performed using the Ensemble Kalman Filter \cite{Evensen2009_IEEE,Asch2016_SIAM}. This tool, which has been extensively used in meteorological applications in the last decades, has seen numerous recent applications for problems in fluid mechanics \cite{Rochoux2014_nhess,Labahn2019_pci,Zhang2020_cf,Mons2021_jfm,Moldovan2022_jfm,Zhao2022_be}. The most interesting feature of the present work is that the EnKF operates \textit{sequentially} i.e. it can combine data in-streaming obtained from different sources. This key feature will be exploited for on-the-fly coupling of high-precision, localized DNS data with running LES calculations. 

\subsubsection{Ensemble Kalman Filter (EnKF)}
\label{Sec:EnKF}

The Kalman Filter (KF) is a well-known DA tool first introduced in 1960 by R.E. Kalman \cite{Kalman1960_jbe} to estimate an augmented system state from sparse external data, or observations. Both sources of information are affected by uncertainties, which are approximated to be Gaussian random variables. The augmented state is obtained by combining a set of observations and a state vector obtained via a model. In the present work, the physical quantity updated is the velocity field $\mathbf{u}$, which is obtained via LES (the model). Observation is sampled at specific locations from a high-resolution simulation (DNS) and indicated as $\alpha$. Corresponding sampled quantities at the same locations for the state vector are indicated as $\mathbf{s} = \mathbf{H} \mathbf{u}$. $\mathbf{H}$ is a projection matrix that maps the values of the model state to the observation space. Let us consider the time advancement of the model from the time step $k$ to $k+1$ in the case observation is available for the latter time. The augmented state is obtained as:

\begin{equation}
    \mathbf{u}_{k+1}^a = \mathbf{u}_{k+1}^f + \mathbf{K}_{k+1}(\mathbf{\alpha}_{k+1}-\mathbf{s}_{k+1})
    \label{eq:KFstate}
\end{equation}

The superscript \textit{f} (\textit{forecast}) represents the time advancement of the physical quantities by the model from time $k$ to $k+1$. The superscript  \textit{a} (\textit{analysis}) represents the final augmented state of the algorithm. The Kalman gain $\mathbf{K}$ is obtained from manipulation of the error covariance matrix $\mathbf{P}=\mathbb{E}((\mathbf{u}-\mathbb{E}(\mathbf{u}))(\mathbf{u}-\mathbb{E}(\mathbf{u}))^T)$, which measures the correlations between the state vector and the observations. It takes into account the level of confidence in the model and in the observation, respectively, which is measured by the variance of the uncertainties affecting the two sources of information. More precisely, the model and observation uncertainties can be described by an unbiased Gaussian distribution with variances $\mathbf{Q_k}$ and $\mathbf{R_k}$, respectively. The main drawback of the classical KF resides in the costly manipulations of the matrix $\mathbf{P}$ and also the necessity to use linear models. 

The Ensemble Kalman Filter (EnKF) \cite{Evensen2009_IEEE}, which is an advanced DA tool based on the KF, is extensively used in weather sciences \cite{Asch2016_SIAM}. It overcomes the aforementioned drawbacks by using the Monte Carlo method to estimate the error covariance matrix $\mathbf{P}$ through the use of an ensemble of pseudo-random realizations. 
An ensemble of $N_e$ physical states $\mathbf{u}$, each of them described by $N$ degrees of freedom, is advanced in time using a model $\mathcal{M}$, which can in this case be non-linear. A state matrix $\mathbf{U}$ of size $[N, \, N_e]$ is assembled at each analysis phase. Each column $i = 1, \cdots, N_e$ of the state matrix represents a physical state $\mathbf{u}_i$ obtained by the $i^{th}$ ensemble member. 
Considering the time advancement of the solution from the instant $k$ to $k+1$ such as in equation \ref{eq:KFstate} for the KF, the EnKF provides an ensemble estimation of the error covariance matrix $\mathbf{P}$ using the hypothesis of statistical independence of the members :  

 \begin{equation}
    \mathbf{P} = \mathbf{\Gamma}(\mathbf{\Gamma})^T
\end{equation}
where $\mathbf{\Gamma}$ is the anomaly matrix, which is derived from the state matrix $\mathbf{U}$ of the ensemble members. It quantifies the deviation of the state vectors from 
their ensemble means:
\begin{equation}
    \mathbf{\Gamma}_{k+1} = \frac{\mathbf{u}_{i,k+1}^f-\langle\mathbf{u}\rangle_{k+1}^f}{\sqrt{N_e-1}} \; , \qquad \langle\mathbf{u}\rangle_{k+1}^f = \frac{1}{N_e}\sum_{i = 1}^{N_e}\mathbf{u}_{i,k+1}^f
\end{equation}

In order to obtain a well-posed mathematical problem, the array of $N_o$ available observations is artificially perturbed to obtain $N_e$ sets of values. To do so, a Gaussian noise based on the covariance matrix of the measurement error $\mathbf{R}_{k+1}$ is added to the observation vector:
\begin{equation}
    \label{eqn:EnKF_obs}
    \mathbf{\alpha}_{i,k+1} = \mathbf{\alpha}_{k+1} + \mathbf{e}_{i,k+1},\; \text{with} \; \mathbf{e}_{i,k+1} \thicksim \mathcal{N}(0, \mathbf{R}_{k+1})
\end{equation}
The model realizations and the observations are combined over the observation space using the projection matrix $\mathbf{H}$:
\begin{equation}
    \label{eqn:EnKF_Projobs}
    \mathbf{s}_{i,k+1} = \mathbf{H} \mathbf{u}_{i,k+1}^f
\end{equation}

These elements provide a closed form for the Kalman gain:
\begin{equation}
    \label{eqn:EnKF_gain_R}
    \mathbf{K}_{k+1} = \mathbf{\Gamma}_{k+1}(\mathbf{S}_{k+1})^T \left[\mathbf{S}_{k+1}(\mathbf{S}_{k+1})^T + \mathbf{R}_{k+1}\right]^{-1}\\
\end{equation}
with
\begin{equation}
\mathbf{S}_{k+1}= \frac{\mathbf{s}_{i,k+1}-\langle\mathbf{s}\rangle_{k+1}}{\sqrt{N_e-1}} \; , \qquad \langle\mathbf{s}\rangle_{k+1} = \frac{1}{N_e}\sum_{i = 1}^{N_e}\mathbf{s}_{i,k+1} \\
\end{equation}

In a limited ensemble size, $\mathbf{R}_{k+1}$ is preferred to the anomaly matrices product of the errors $\mathbf{E}_{k+1}(\mathbf{E}_{k+1})^T$ in equation \ref{eqn:EnKF_gain_R}. It provides a simplified algorithm and reduced computational cost \cite{carrassi2018_wcc, Hoteit2015_mwr}. Finally, the physical state predicted by each ensemble member is updated using the Kalman Gain: 
\begin{equation}
    \mathbf{u}_{i,k+1}^a = \mathbf{u}_{i,k+1}^f + \mathbf{K}_{k+1}(\mathbf{\alpha}_{i,k+1}-\mathbf{s}_{i,k+1})
\end{equation}

The approaches based on the EnKF can also simultaneously optimize the free parameters of the model to minimize the discrepancy between the model and observation during the analysis phase. These parameters are usually assembled in an array referred to as $\theta$. A straightforward strategy to perform such optimization is the so-called \textit{extended state} \cite{Asch2016_SIAM}. Here the EnKF problem is resolved for a state vector $\mathbf{{u}_{ext}}$ defined as: 

\begin{equation}
\mathbf{{u}_{ext}} = \begin{bmatrix} 
	\mathbf{u} \\
	\mathbf{\theta}
	\end{bmatrix}
\end{equation}

The size of the extended state is now equal to $N_{ext}=N + N_\theta$, where $N_\theta$ is the number of parameters to be optimized. This modification brings a negligible increase in computational costs if $N_\theta << N$ and it simultaneously provides an updated state estimation and optimized parametric description for the model at the end of the analysis phase.

\subsubsection{Inflation}
\label{Sec:inflation}
One of the major drawbacks of the Ensemble Kalman Filter is the fast collapse of the state matrix variability. The consequence of the unwanted reduction of the variability is the convergence of the state matrix towards a localized optimum, which is strongly tied with the prior state provided. If the latter is not accurate, then the precision of the optimization via EnKF can be severely impacted. One can increase the global variability of the system and decrease the sampling errors using a higher number of members in the ensemble, gaining accuracy in the prediction of the EnKF. However, this strategy is not conceivable for fluid dynamics applications where computational costs preclude the usage of large ensembles. In fact, the number of members generally used for three-dimensional runs is around $N_e \in [40,100]$ \cite{Mons2021_prf,Moldovan2022_jfm}, which is pretty far from classical Monte-Carlo convergence. 

This problem is usually mitigated by inflating the variance of the ensemble \textit{after} the analysis phase. This can be easily obtained by increasing the discrepancy between each state vector ${u}_{i,k+1}^a$ and the ensemble mean $\langle\mathbf{u}^a\rangle$ by algebraic operations driven via a coefficient $\lambda$. This procedure is referred to as \textit{multiplicative inflation}. The way this procedure is performed can be \textit{deterministic} or \textit{stochastic}: 

\begin{equation}
    deterministic \qquad \mathbf{u}_{i}^a \longrightarrow \langle\mathbf{u}^a\rangle + \lambda_i(\mathbf{u}_{i}^a-\langle\mathbf{u}^a\rangle) \qquad with \, \lambda_i > 1
\end{equation}

\begin{equation}
    stochastic \qquad \mathbf{u}_{i}^a \longrightarrow (1+\lambda_i) \mathbf{u}_{i}^a \qquad with \, \lambda_i \thicksim \mathcal{N}(0,\sigma)
\end{equation}

The deterministic implementation can be very efficient during the initial analysis phases of the calculation. Considering it is applied to the discrepancy from the mean values of the ensemble, the process is quite stable, and higher values of $\lambda$ can be used. Nonetheless, it is less efficient when the ensemble exhibits a strong collapse of the physical solution ($\mathbf{u}_{i}^a-\langle\mathbf{u}^a\rangle \approx 0$). On the other hand, stochastic inflation is very useful to mitigate a fast collapse of the state matrix, allowing it to target a global optimum solution. The Gaussian distribution used to determine $\lambda_i$ is usually truncated to avoid the generation of outliers which could lead to the divergence of the EnKF.

\subsubsection{Localization}
\label{Sec:localization}
The coefficients of the state matrix correspond to values of the flow variables (namely the velocity field) in specific points of the physical domain, usually the center of the mesh elements. As discussed in Sec. \ref{Sec:EnKF} and shown in eq. \ref{eqn:EnKF_gain_R}, the Kalman gain establishes a correlation between those values and the values of the state matrix projected in the observation space i.e. sensors where high-fidelity data is available. Considering that the physical correlation naturally decays with distance in continuous systems, the approximations used to determine an ensemble Kalman gain can lead to spurious effects on the analyzed state matrix for large domains. These effects can be responsible for critical problems such as unphysical solutions, which can lead to the divergence of the calculations. Again, these problems can be reduced by increasing the number of ensemble members, which is not a cost-efficient solution for applications involving CFD. Therefore, different strategies need to be employed to mitigate the effects of spurious correlations. The most used strategy to reduce them is to operate on the coefficients correlating variables in the EnKF which are calculated in points far from each other. In this case, one would expect that the physical phenomena are completely decorrelated. Two possible strategies may be adopted to obtain this result \cite{Asch2016_SIAM}. The \textit{Covariance localization} directly operates on the coefficients of the error covariance matrix $\mathbf{P}_{k+1}^f$, pre-multiplying them with a term that tends to zero as the physical distance between observations sensors and elements of the state increases. This process is mathematically performed using a coefficient-wise multiplication between the covariance matrix and a correction matrix referred to as $\mathbf{L}$. This expression can be directly added in the algorithm without any structural modification. The localized Kalman gain becomes:
\begin{equation}
    [\mathbf{P}^f_{k+1}]_{i,j}[\mathbf{L}]_{i,j} \longrightarrow {\mathbf{K}_{k+1}^{loc} = [\mathbf{L}]_{i,j}[\mathbf{K}_{k+1}]_{i,j}}
\end{equation}
The structure of the matrix $\mathbf{L}$ must be set by the user. In fluid systems, and in particular for turbulence, the correlation decreases fast in space. Therefore, a generally used structure for the localization matrix is an exponential decay form: 
\begin{equation}
    \label{eq:loc_matrix}
    \mathbf{L}(i,j) = e^{-\Delta^2_{i,j}/l}
\end{equation}
where $\Delta_{i,j}$ is the distance between the given observation sensor and the point of evaluation of the model (center of the mesh element in CFD). $l$ is a correlation length scale that can be tuned accordingly to the local characteristics of the test case. 

Another way to localize the Kalman gain is to use \textit{physical localization}. The principle is quite straightforward. Instead of performing the EnKF on the entire physical domain, one can proceed to do the calculation on a \textit{clipped} domain. The reduced space must contain the observation sensors. This strategy also has the advantage of reducing the number of degrees of freedom operating in the DA procedure, which can produce a significant gain in terms of computational resources required. \textit{Covariance localization} is commonly used together with physical localization to avoid discontinuities of the updated physical state, in particular at the interface of the clipped domain. This strategy prevents potential divergence of the model runs. This method is very efficient in speeding up the calculation and simultaneously improving the stability of the calculation and the accuracy of the prediction for a reduced ensemble size such as the ones currently usable for CFD-based studies \cite{Villanueva2023_arxiv}.

\begin{figure}
    \centering
    \includegraphics[width=0.9\textwidth]{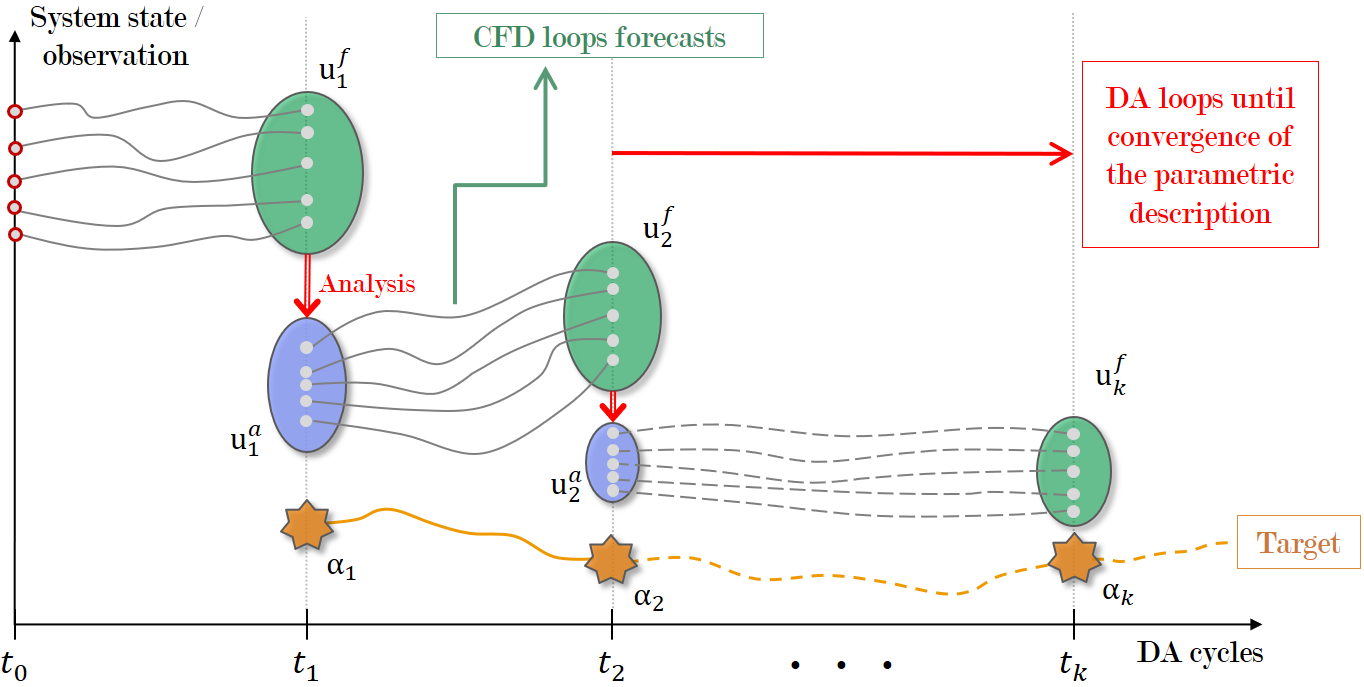}
    \caption{Scheme representing the ensemble Kalman filter}
    \label{Fig:EnKF_plot}
\end{figure}

The DA procedure used in this study is qualitatively shown in Fig. \ref{Fig:EnKF_plot} and a detailed algorithm of the EnKF (including state-of-the-art modifications) is provided in Alg. \ref{Alg:EnKF}.

\begin{algorithm}
    \caption{Algorithm for the Ensemble Kalman Filter}
    \label{Alg:EnKF}
    \textbf{Input:} $\mathcal{M}$, $\mathcal{H}$, $\mathbf{R}_{k+1}$, and some priors for the state system $\mathbf{u}_{i,0}^a$, where usually $\mathbf{u}_{i,0}^a \sim \mathcal{N}(\mu_N, \sigma_N^2)$ \\
    \For{$k = 0$ to $K-1$}{
        \For{$i = 1$ to $N_e$}{
    \nl Advancement in time of the state vectors:\\
    \qquad $\mathbf{u}_{i,k+1}^f = \mathcal{M}\mathbf{u}_{i,k}^a$ \\
    \nl Creation of an observation matrix from the observation data by introducing errors:\\
    \qquad$\mathbf{\alpha}_{i,k+1} = \mathbf{\alpha}_{k+1} + \mathbf{e}_{i, k+1}$, with $\mathbf{e}_{i, k+1} \thicksim \mathcal{N}(0,\mathbf{R}_{k+1})$\\
    \nl Calculation of the predicted observation:\\
    \qquad$\mathbf{s}_{i,k+1} = \mathcal{H}\mathbf{u}_{i,k+1}^f$\\
    \nl Calculation of the ensemble means:\\
    \qquad$\langle\mathbf{u}\rangle_{k+1}^f = \frac{1}{N_e}\sum_{i = 1}^{N_e}\mathbf{u}_{i,k+1}^f$,\,
    $\langle\mathbf{s}\rangle_{k+1} = \frac{1}{N_e}\sum_{i = 1}^{N_e}\mathbf{s}_{i,k+1}$,\\
    \nl Calculation of the anomaly matrices:\\
    \qquad$\mathbf{\Gamma}_{k+1} = \frac{\mathbf{u}_{i,k+1}^f-\langle\mathbf{u}\rangle_{k+1}^f}{\sqrt{N_e-1}}$,\,
    $\mathbf{S}_{k+1} = \frac{\mathbf{s}_{i,k+1}-\langle\mathbf{s}\rangle_{k+1}}{\sqrt{N_e-1}}$,\\
    \nl Calculation of the Kalman gain:\\
    \qquad$\mathbf{K}_{k+1} = \mathbf{\Gamma}_{k+1}(\mathbf{S}_{k+1})^T \left[\mathbf{S}_{k+1}(\mathbf{S}_{k+1})^T + \mathbf{R}_{k+1}\right]^{-1}$\\
    \nl Localization of the Kalman gain:\\
    \qquad$\mathbf{K}_{k+1}^{loc} = [\mathbf{L}]_{i,j}[\mathbf{K}_{k+1}]_{i,j}$\\
    \nl Update of the state matrix:\\
    \qquad$\mathbf{u}_{i,k+1}^a = \mathbf{u}_{i,k+1}^f + \mathbf{K}_{k+1}^{loc}(\mathbf{\alpha}_{i,k+1}- \mathbf{s}_{i,k+1})$\\
    \nl Inflation of the state matrix:\\
    \qquad$\mathbf{u}_{i,k+1}^a = (1+\lambda_i)\mathbf{u}_{i,k+1}^a$
    }
    }
\end{algorithm}

\newpage
\subsubsection{CONES}
\textit{Coupling OpenFOAM with Numerical EnvironmentS} (CONES) is a C++ library add-on to the open-source CFD software OpenFOAM. CONES allows OpenFOAM to exchange field data through MPI communications \cite{Villanueva2023_arxiv}. The coupling of OpenFOAM with other numerical environments is operated by CWIPI (Coupling With Interpolation Parallel Interface) developed by CERFACS and ONERA \cite{Reflox2011_aerlab}. CONES has been developed by the team in order to perform on-the-fly DA with OpenFOAM, which has been coupled with a tailored EnKF code for this purpose. 
The main advantages CONES provides to perform DA with OpenFOAM are: 
\begin{itemize}
    \item Data Assimilation is performed \textit{online} without stopping the CFD runs, which represent the ensemble members. The computational resources required to restart the simulations after an analysis phase are large, usually more than the total computational cost for the DA run if several analysis steps have to be performed.
    \item Communication of large physical fields (arrays of millions of elements such as the velocity field) is performed rapidly and efficiently.
    \item Compilation of additional functions is performed via \textit{wmake} routine in the user-dedicated library of OpenFOAM. 
    \item Coupling between codes is performed preserving the original structure of the existing CFD solvers. Every CONES-related function is contained in a Pstream (Part of OpenFOAM) modified library, hence, data exchange is done at the end of the solver loop by calling specific functions, and the calculation loop remains unmodified.
    \item Direct HPC communications are established between multiple processors, which handle partitions of the numerical simulations and the DA process.
\end{itemize}

\begin{figure}
    \centering
    \includegraphics[width=1\textwidth]{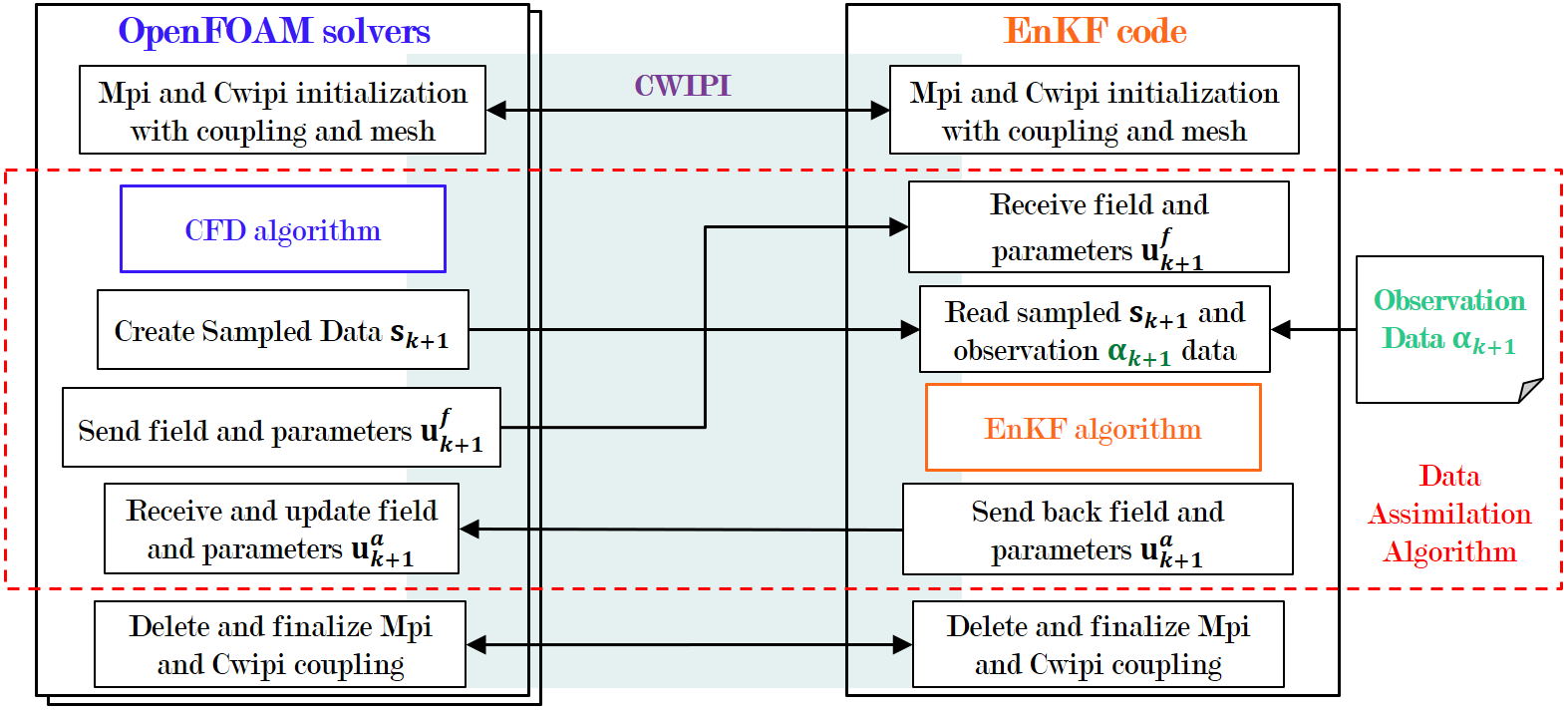}
    \caption{Scheme of the library CONES}
    \label{Fig:CONES}
\end{figure}

Data flow and exchanges between codes are summarized in Fig.~\ref{Fig:CONES}. As CWIPI is based on the MPI library, both MPI and CWIPI environments have to be initialized when launching the calculation. Similarly, they have to be finalized at the end. Once the forecast(s) step(s) of the EnKF algorithm is performed, sampled data $\boldsymbol{s}_{k+1}$ is interpolated for each member and transferred to the EnKF code for the analysis step. The entire velocity field and the studied parameters are also sent in order to perform the EnKF algorithm. CWIPI exchanges data through coincident meshes in CONES. However, in case the mesh is not coincident, the field data is interpolated automatically. This is an important feature for the potential use of \textit{multigrid}-based DA algorithms in the future \cite{Moldovan2022_jfm}. The observation is uploaded just before the analysis step. After the state vectors $\mathbf{u}_{i,k+1}^a$ have been updated, the information is sent back to each member to resume the forecast steps with the updated physical states and/or values of the model constants. The state matrix contains the velocity fields of all the members and the constant $C_k$ of the turbulence model optimized in this study. Details about the optimization of this parameter will be provided in Sec. \ref{Sec:setup}. The observation, containing velocities of the reference data for all times available, is stored in a single .txt file that is read at each analysis phase. The related computational cost is negligible compared to the calculation of the Kalman gain when performing the EnKF algorithm as shown in appendix \ref{Appendix::cost}.

\newpage
\section{Test case and set-up of the DA analysis}
\label{Sec:setup}

\subsection{Turbulent plane channel flow, $Re_\tau\approx550$}
\label{Sec:testcasePrior}

The test case chosen to perform the DA analysis is the turbulent plane channel flow for $Re_\tau = u_\tau h / \nu = 546$. Here $u_\tau = \sqrt{\tau_w / \rho}$ is the friction velocity and $\tau_w$ is the shear stress at the wall. $h$ is the half-height of the channel and $\nu$ is the kinematic viscosity. This academic test case, which is driven by shear mechanisms at the wall and naturally excludes complex aspects associated with favorable/adverse mean pressure gradients \cite{Pope2000_cambridge}, is nonetheless problematic for LES \cite{Meyers2007_pf}. Complex non-linear interaction occurs between two main error sources, namely those associated with the numerical discretization and the SGS closure. These mechanisms are responsible for very high sensitivity to relatively small variations in the grid discretization and the SGS closure selected. Therefore, this test case is an excellent candidate to study the objectives presented in the introduction. Results obtained from the large-eddy simulations performed in this work will be compared with DNS data on the same test case previously performed by the research team \cite{Martinez2023_arxiv}.

\begin{figure}
    \centering
    \includegraphics[width=0.7\textwidth]{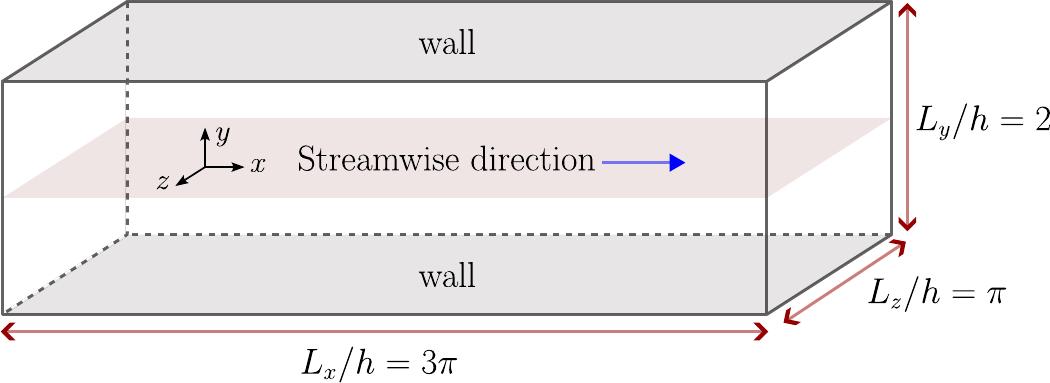}
    \caption{Size of the physical domain investigated.}
    \label{Fig:channel_scheme}
\end{figure}

The geometric features are shown in Fig. \ref{Fig:channel_scheme}. The size of the domain investigated is $3 \pi h \times 2h \times \pi h$. $x$ is the streamwise direction, $y$ the normal direction and $z$ the spanwise direction. 
The top and bottom boundaries are no-slip walls. A periodic boundary condition is applied on the four lateral sides. A source term, already integrated within the solver of \textit{OpenFOAM}, is included in the dynamic equations to preserve the global mass flow rate in time. More precisely, the source term targets the conservation of the bulk streamwise velocity $u_b = \iiint_{V_D} u_x \, dV^{\prime} / V_D$, where $V_D$ is the volume of the physical domain investigated. The targeted criterion used for all simulations is $u_b = 0.899 u_c$, where $u_c$ is the mean streamwise velocity at the center of the channel obtained by the DNS. The kinematic viscosity $\nu$ is the same for the DNS and LES calculations. The bulk Reynolds number obtained by the DNS is equal to $Re = 2hu_b/\nu = 20124$.

A baseline LES is performed using the well-known Smagorinsky subgrid-scale model \cite{Smagorinsky1963_mwr} (see Sec. \ref{Sec:nums}). This simulation is run by the pimpleFoam solver of the OpenFOAM CFD library. It is a solver tailored for the simulation of incompressible turbulent flows using the PIMPLE algorithm. The grid is composed of $350\,000$ cells, whose details are reported in Tab. \ref{Tab:grid_ resolutions} along with the reference DNS. The size of the grid elements is adimensionalized with respect to the viscous wall unit $\delta_{\nu} = \nu / u_{\tau}$. Superscript $\star$ is used when normalizations are performed using the $u_\tau$ calculated by the $DNS$. On the other hand, the superscript $+$ is used when $u_{\tau}$ is obtained by each LES simulation. $\Delta x^\star$ and $\Delta z^\star$ are obtained using a uniform distribution. A geometric expansion is used to control the size of the elements $\Delta y$ in the normal direction to grant higher resolution at the wall. The size of the smallest element $\Delta y^\star_1$ at the wall and the largest element $\Delta y^\star_c$ at the centerline are reported. The size of the mesh elements used for the calculation of the baseline simulation is larger than typical values observed in LES for this case, which are $\Delta x^\star \approx 50$, $\Delta y^\star_1 \approx 1$ and $\Delta z^\star \approx 20$ \cite{Mons2021_prf}. This choice was made in order to i) assess the capabilities of the DA method to provide an accurate state estimation and parametric inference even in under-resolved conditions and ii) to obtain faster runs of the DA algorithm using a sufficiently large ensemble of simulations. The initial conditions for the baseline LES case were set using an interpolated field from a DNS solution. The simulation was carried out for a duration of $50$ advective times, calculated as $t_A = h / u_c$, in order to dissipate the initial field. Then, average quantities have been calculated over a time window of $900 t_A$. The time step for the advancement of the solution is constant and equal to $\Delta t = 0.02 t_A$.

\begin{table}
\centering
\begin{tabular}{@{}cccccccccccc@{}}
\toprule
Type    & $L_x$ & $L_y$ & $L_z$ & $N_x$ & $N_y$ & $N_z$ & $\Delta x^\star$ & $\Delta y^\star_1$ & $\Delta y^\star_c$ & $\Delta z^\star$ & Cells \\ \midrule 
LES & $3\pi$        & $h$        & $\pi$       & $70$       & $100$      & $50$       & $73$        & $2.6$  & $27.8$      & $34$       & $3.5 \times 10^5$     \\
DNS  & $6\pi$       & $h$       & $2\pi$       & $1024$      & $256$      & $512$      & $9.9$        & $0.96$ & $11.2$       & $6.6$       & $13.4 \times 10^7$         \\ \bottomrule
\end{tabular}%
\caption{Details of the grids used for LES calculations as well as for the DNS}
\label{Tab:grid_ resolutions}
\end{table}

\begin{figure}
    \centering
    \begin{subfigure}{.45\textwidth}
        \centering 
        \includegraphics[width=1\textwidth]{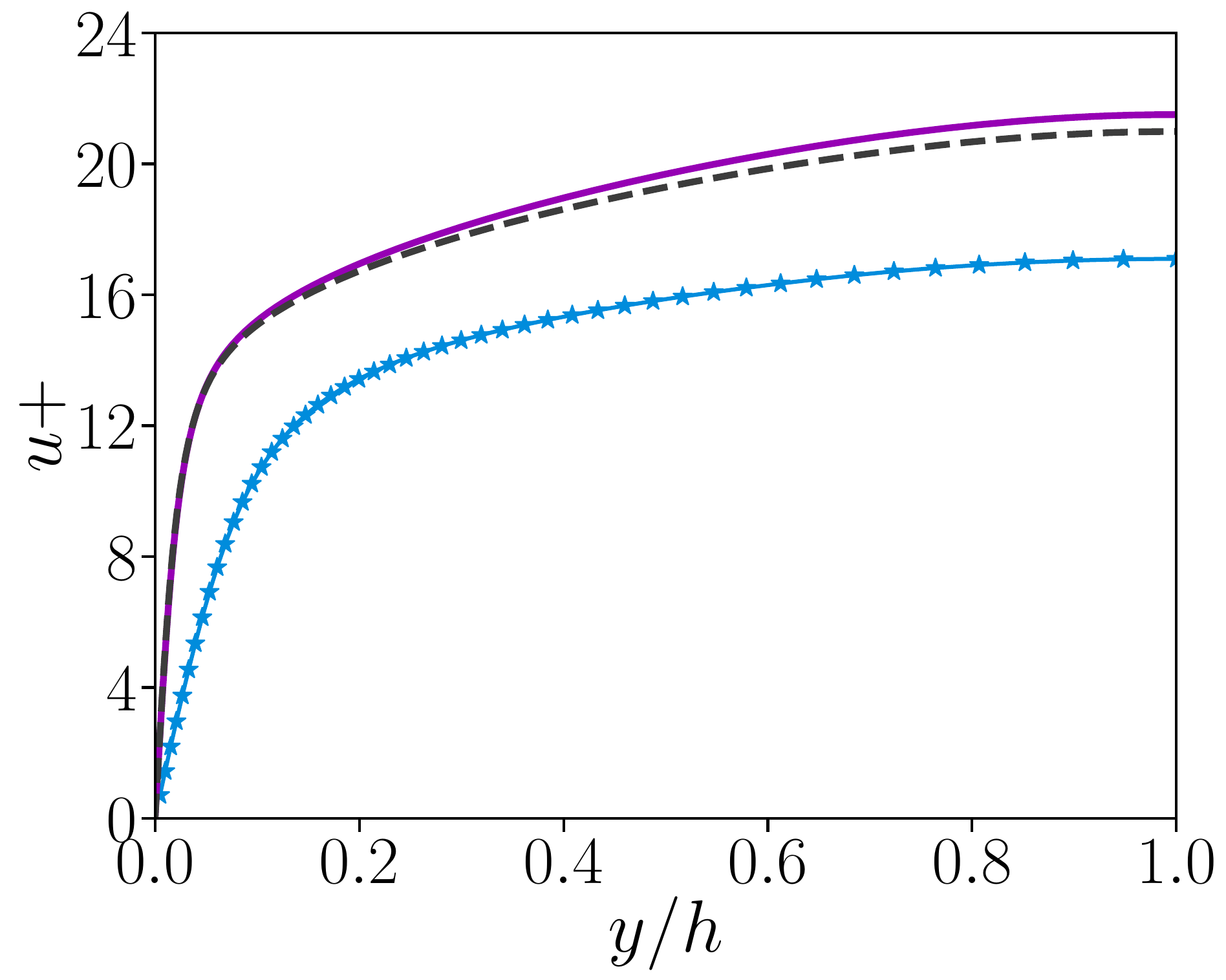}
        \caption{}
        \label{Fig:UPlusOverYh_Ref}
    \end{subfigure}
    \begin{subfigure}{.44\textwidth}
        \centering 
        \includegraphics[width=1\textwidth]{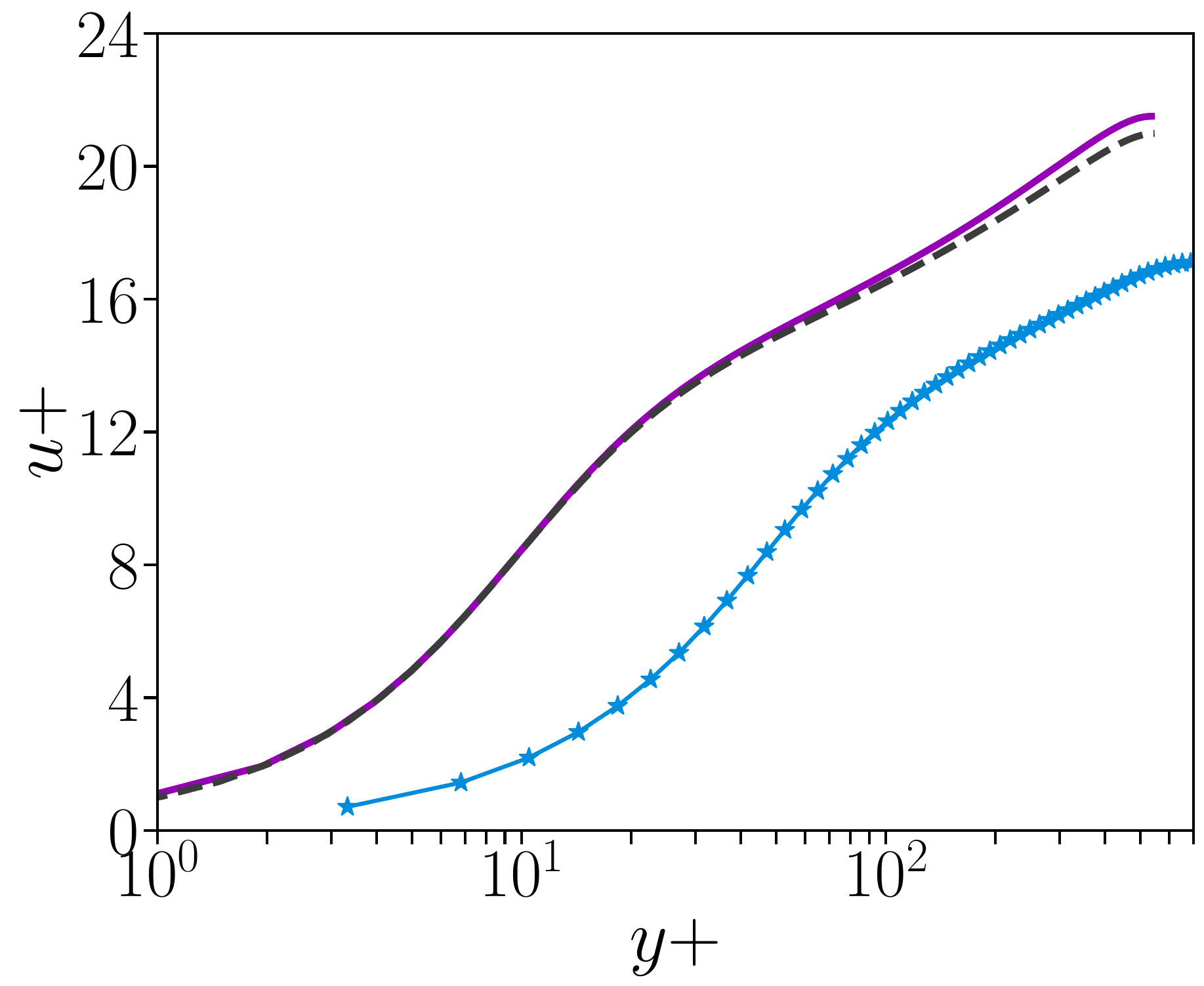}
        \caption{}
         \label{Fig:UPlusOverYPlus_Ref}
    \end{subfigure}
    
    \caption{Averaged streamwise velocity profiles $u^+$ for the baseline Smagorinsky LES (\textcolor{cyan}{$\star$}) and the DNS (\textcolor{violet}{$-$}) compared to reference data ($--$)}
    \label{Fig:reference_profiles}
\end{figure}

\begin{figure}
    \centering
    \includegraphics[width=0.45\textwidth]{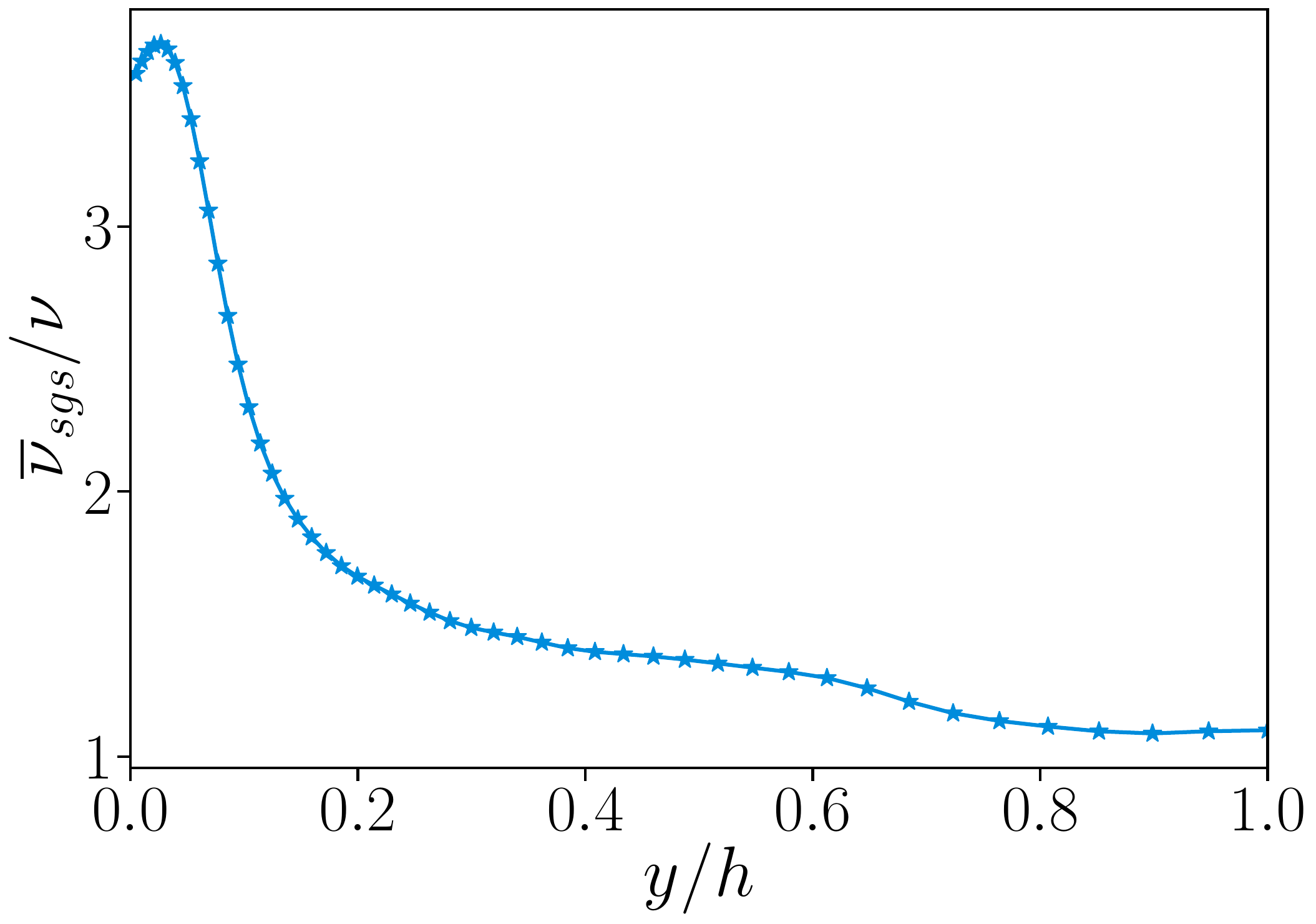}
    \caption{Distribution in the wall-normal direction of $\overline{\nu}_{sgs}/\nu$ for the baseline LES}
    \label{Fig:nut_mean}
\end{figure}

\begin{figure}
    \centering
    \begin{subfigure}{.40\textwidth}
        \centering 
        
        \includegraphics[width=1\textwidth]{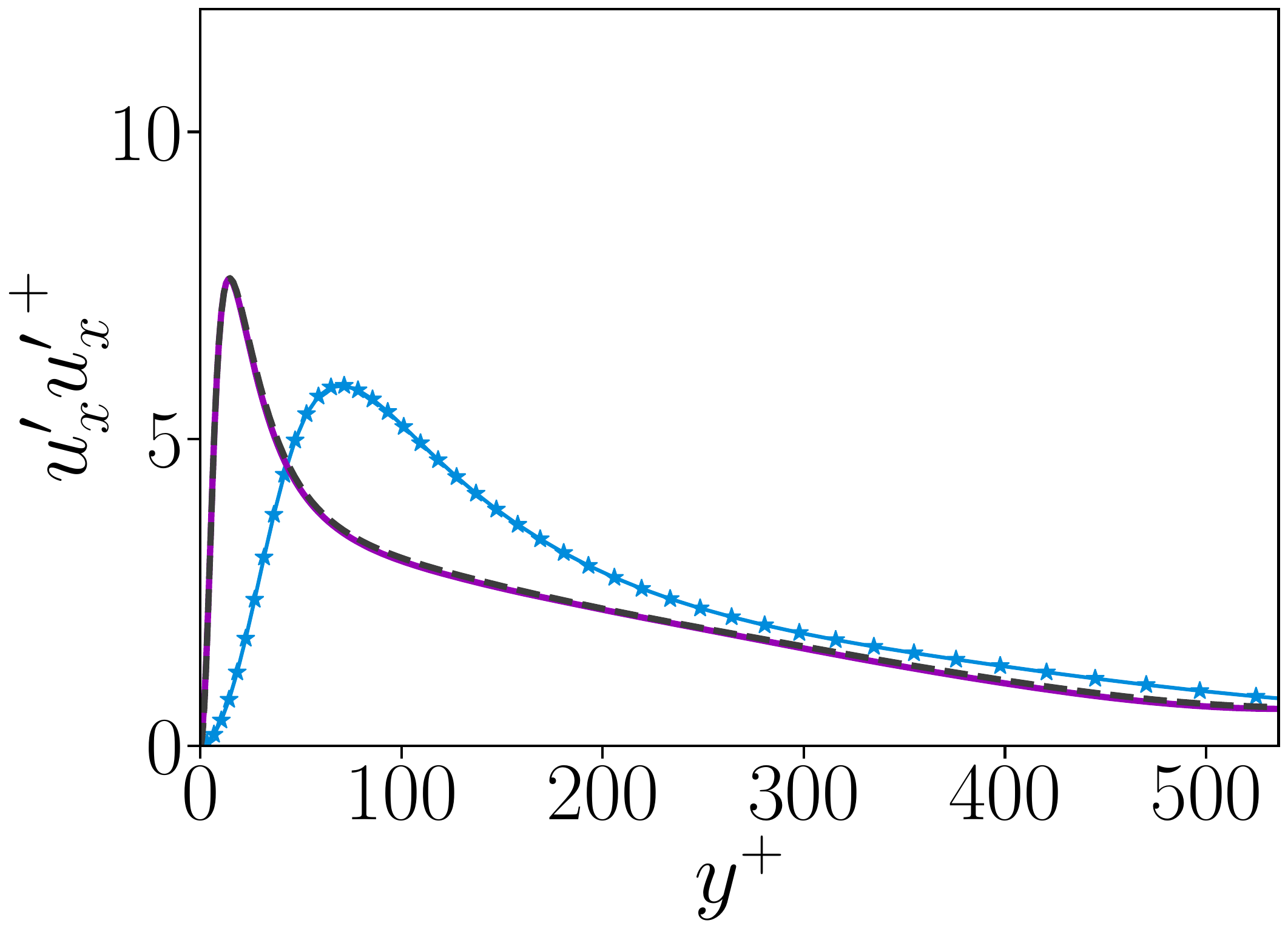}
        \caption{}
        \label{Fig:uuReference}
    \end{subfigure}
    \begin{subfigure}{.40\textwidth}
        \centering 
        
        \includegraphics[width=1\textwidth]{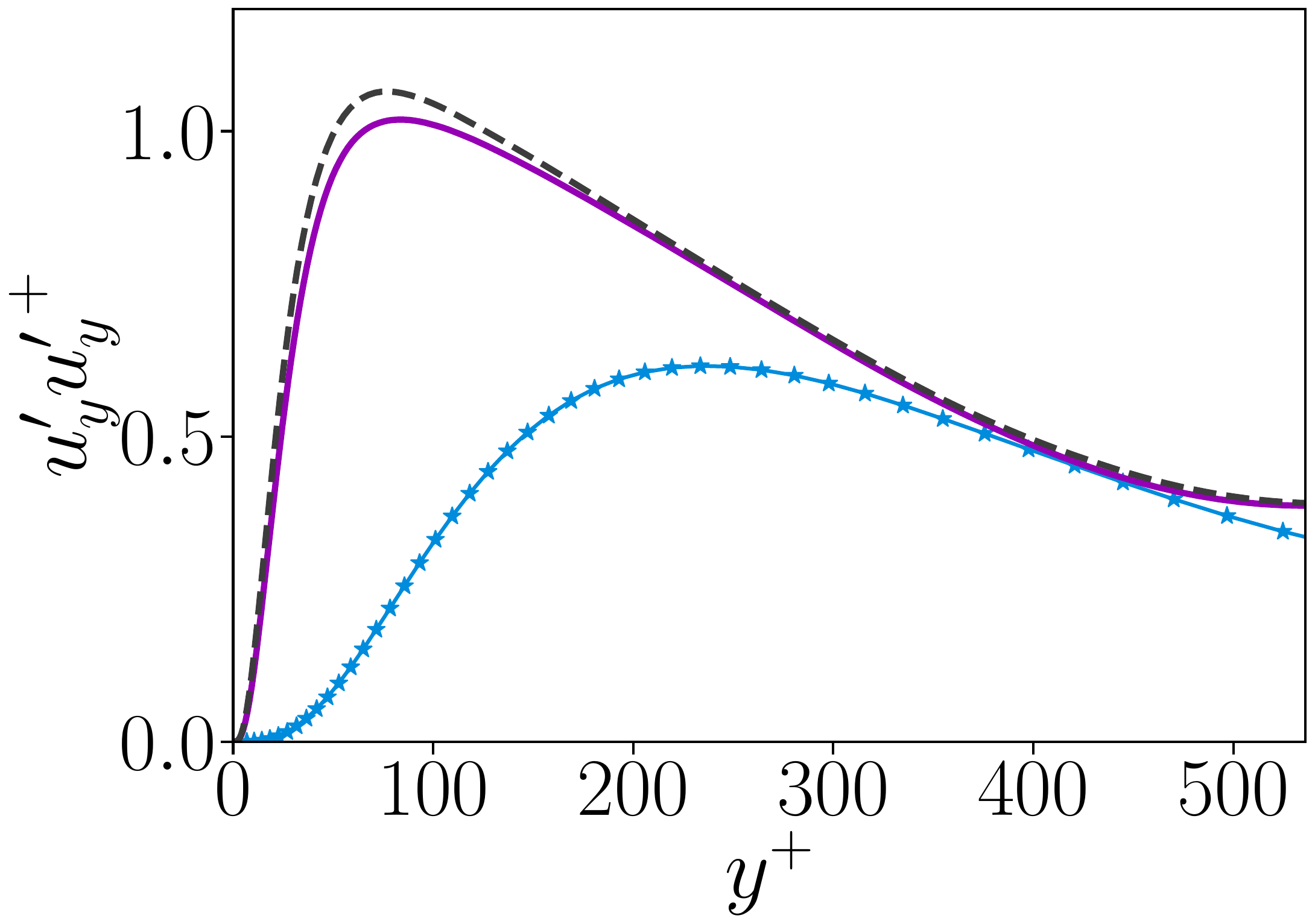}
        \caption{}
        \label{Fig:vvReference}
    \end{subfigure}
    \begin{subfigure}{.40\textwidth}
        \centering 
        
        \includegraphics[width=1\textwidth]{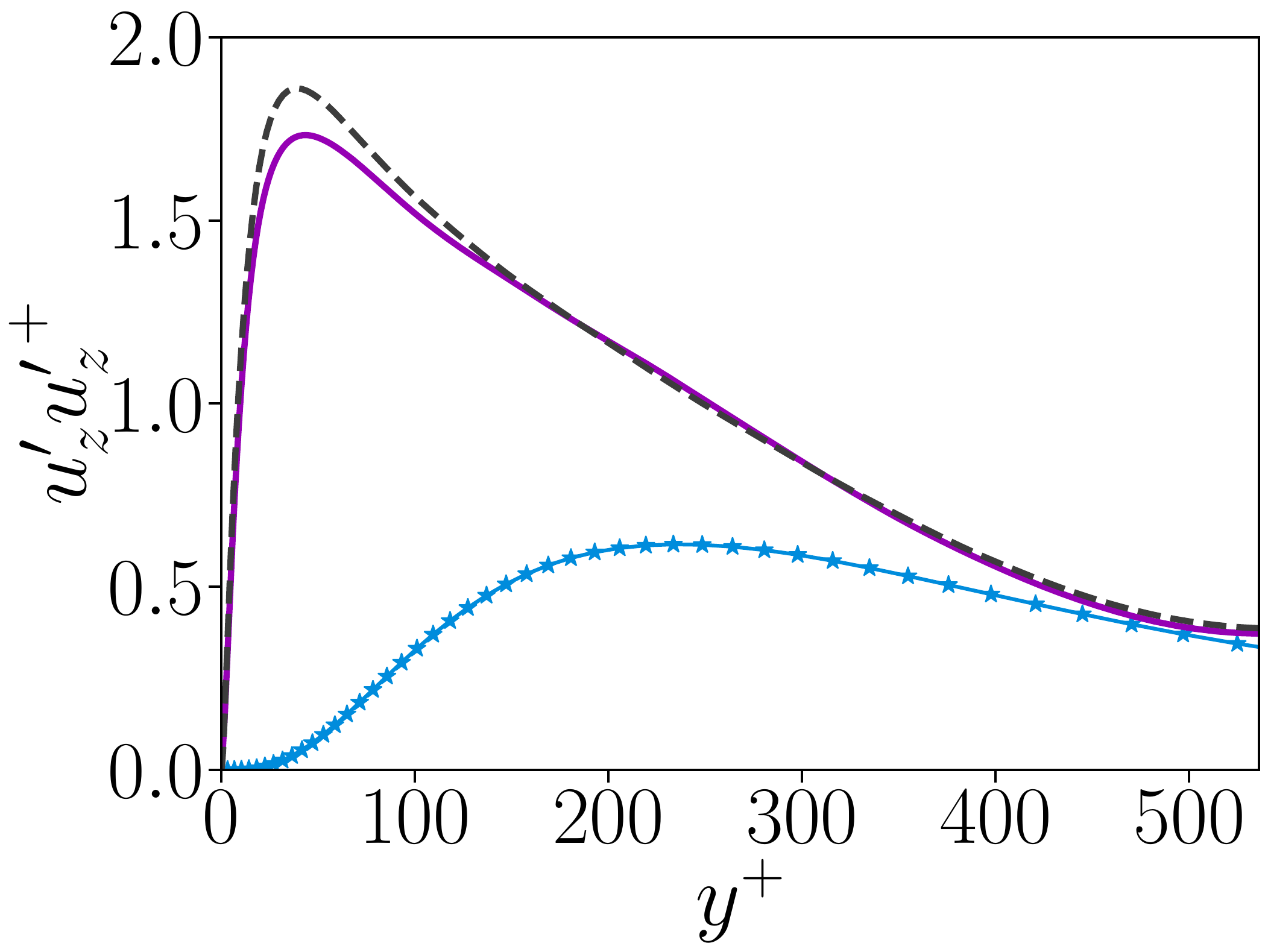}
        \caption{}
        \label{Fig:wwReference}
    \end{subfigure}
    \begin{subfigure}{.42\textwidth}
        \centering 
        
        \includegraphics[width=1\textwidth]{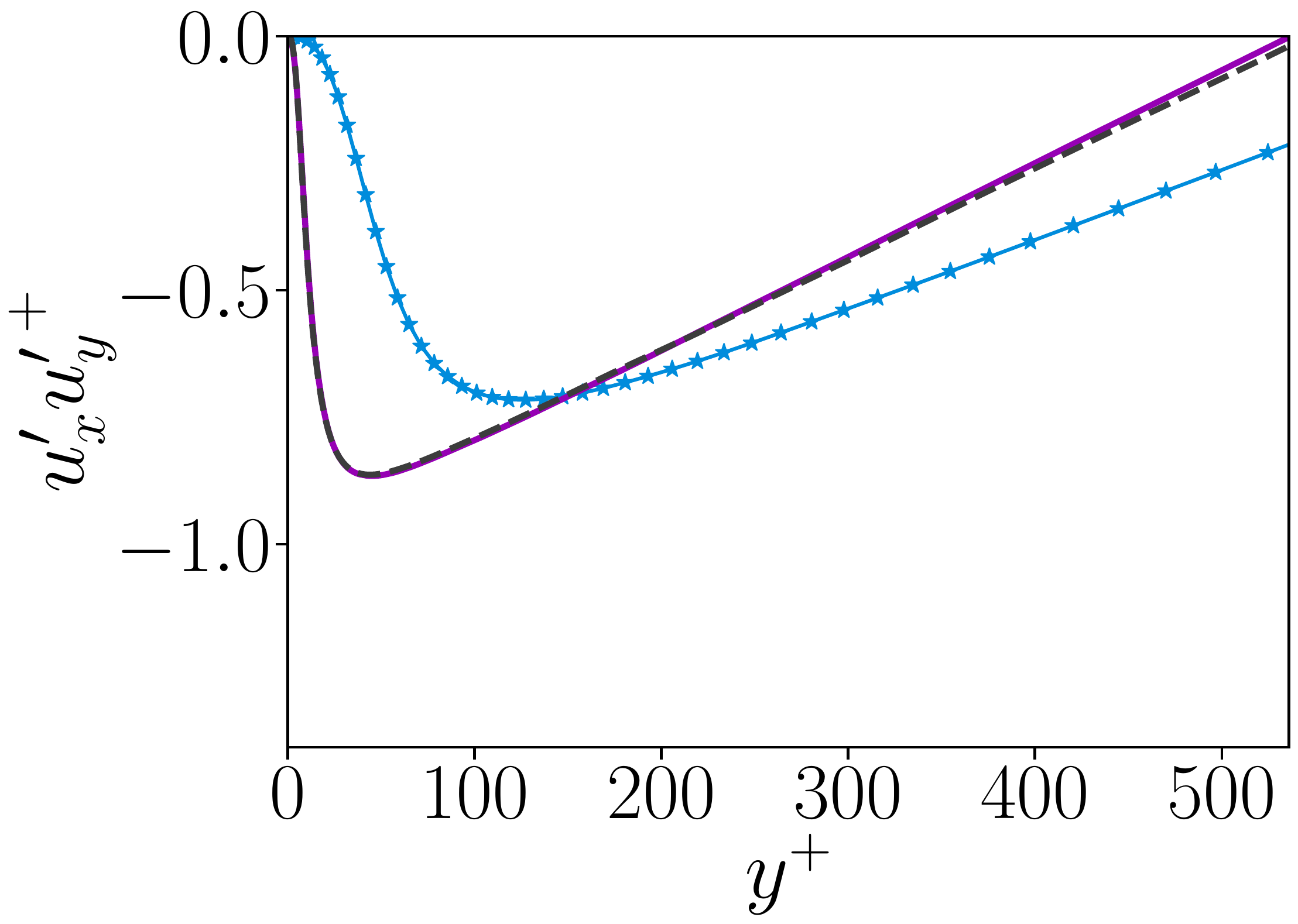}
        \caption{}
         \label{Fig:uvReference}
    \end{subfigure}
    \caption{Components of the Reynolds stress tensor for the baseline Smagorinsky LES (\textcolor{cyan}{$\star$}) and the DNS (\textcolor{violet}{$-$}) compared to reference data ($--$)}
    \label{Fig:RST_reference_profiles}
\end{figure}

Results from the baseline LES are now compared with the DNS and additional reference DNS results freely available online for a very similar $Re_\tau$ \cite{DelAlamo2003_pf}. Fig. \ref{Fig:reference_profiles} show the normalized mean streamwise velocity profile $u^+ = \overline{u_x} / u_\tau$. Averages (indicated with the overline) are performed in time as well as in the streamwise and spanwise direction, in order to obtain improved statistical convergence. One can see that the discrepancy between the LES prediction and the DNS results is significant. One of the key elements affecting this lack of accuracy is the erroneous prediction of the shear stress at the wall $\tau_w$ and thus of the friction velocity $u_\tau$. For this parameter, a discrepancy of $28 \%$ with DNS results is observed. The main source of error for the prediction of this quantity is related to the SGS closure used, for which $\nu_{sgs}$ does not correctly scale to zero approaching the wall, as shown in Fig. \ref{Fig:nut_mean}. A large discrepancy is also observed for the accuracy in the prediction of the components of the resolved Reynolds stress tensor $\overline{u_i^\prime u_j^\prime}$. The quantities  ${u_x^\prime u_x^\prime}^+ = \overline{u_x^\prime u_x^\prime} / u_\tau^2$, ${u_y^\prime u_y^\prime}^+ = \overline{u_y^\prime u_y^\prime} / u_\tau^2$, ${u_z^\prime u_z^\prime}^+ = \overline{u_z^\prime u_z^\prime} / u_\tau^2$ and ${u_x^{\prime} u_y^\prime}^+ = \overline{u_x^\prime u_y^\prime} / u_\tau^2$ are shown in Fig. \ref{Fig:RST_reference_profiles}. One can see that both the magnitude and the position of the peak are not accurately predicted. 

The results obtained via the baseline LES indicate that, using the numerical set-up described combined with the Smagorinsky SGS closure, an accurate prediction of the statistical moments of the flow field is not obtained. In subsection \ref{Sec:DAsetup}, the DA procedure used to improve the flow prediction using this LES setup is detailed.

\subsection{Data Assimilation strategy}
\label{Sec:DAsetup}

The DA simulations performed in this work aim to provide instantaneous augmented states of the test case investigated. This objective will be achieved by coupling on-the-fly the numerical prediction of the LES solver with localized information sampled from the DNS reference. This strategy relies on three main ingredients:

\begin{itemize}
\item \textit{The model}, which provides a quasi-continuous description of the physical phenomenon investigated. In this analysis, the \textit{model} is the LES setup presented in section \ref{Sec:testcasePrior}.  
\item \textit{The observation}. Time-resolved samples of the instantaneous velocity field from the reference DNS are used for this purpose. The samples are collected over $10800$ sensors in the physical domain for $0.48 \leq y^+ \leq 56.4$ i.e. in the viscous sublayer, in the buffer region, and in the inertial range. Sampling in time is performed at a constant rate of $\Delta t_{DA} = 0.04 t_A$.  
\item \textit{The coupler}. CONES is used to couple the incompressible OpenFOAM solver pimpleFoam with an EnKF algorithm as presented in Sec. \ref{Sec:nums}. 
\end{itemize}

\begin{figure}
    \centering
    \includegraphics[width=0.7\textwidth]{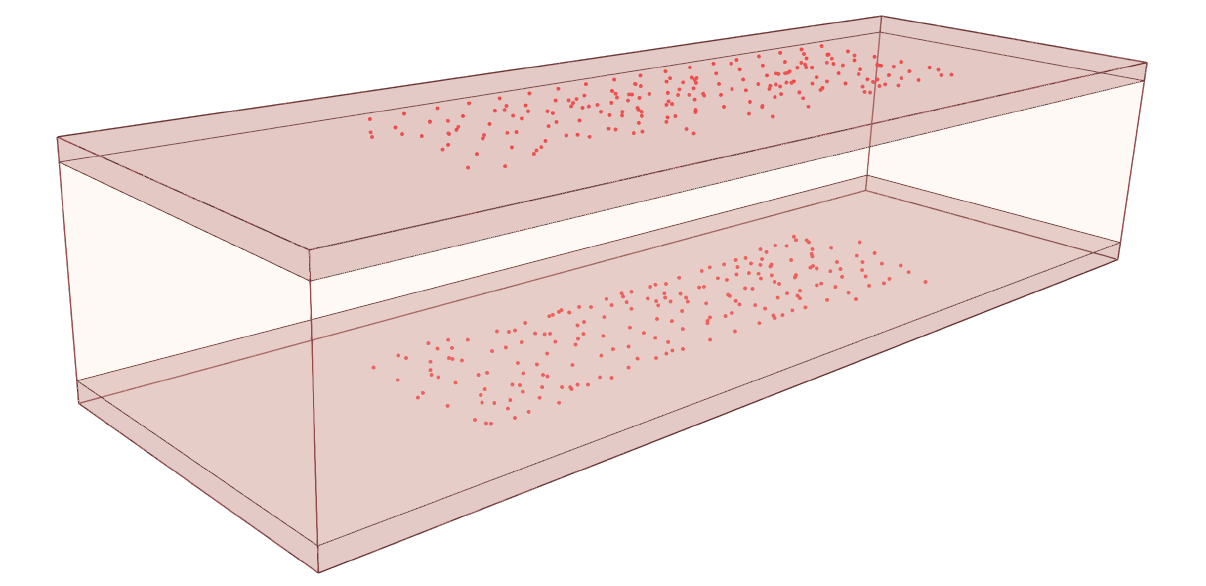}
    \caption{Location of the sensors used to obtain observation for data assimilation. The region in red corresponds to the physical clipping for the EnKF i.e. the region in space where the state augmentation is performed.}
    \label{Fig:observations}
\end{figure}

The setup of the EnKF procedure is now detailed. The size $[N_{ext}\, , \, N_e]$ of the state matrix $\mathbf{U}$ is given by $N_e=40$ (number of ensemble members) and $N_{ext} = N + N_{\theta}$. Here $N_{\theta}$ is the number of parameters optimized by the EnKF and it is different for the two DA runs which will be presented in the following. $N = 3 \, n_{cells}$ is equal to three times the number of grid elements that are used in the DA procedure. This is because the number of degrees of freedom considered in the DA procedure is the three components of the velocity field for each of the $n_{cells}$ mesh elements. The value of $n_{cells}$ is strictly connected with the physical localization performed by clipping the numerical domain analyzed. This procedure, which is illustrated in Fig. \ref{Fig:observations}, consists in excluding from the DA calculation the grid elements for $0.18 < y/h < 1.82$. These elements are relatively far from the sensors and therefore the risk of spurious correlation affecting the stability of the DA algorithm is high. In addition, the excluded domain represents around $56 \%$ of the total number of cells used by the LES model. The computational gain for the calculation of the Kalman gain is also approximately $56\%$ as shown in Tab. \ref{Tab:gain_comp}. Covariance localization is applied as well to the calculation so that discontinuities associated with the physical clipping are smoothed out. The structure of the matrix $L$ used for covariance localization is the one presented in equation \ref{eq:loc_matrix}, where the parameter $l=0.175$ in streamwise and spanwise directions and $l=0.000985$ in wall-normal direction.

Observation is obtained from $408$ sensors which have been selected among the $10800$ available. The constraint $x \in [0.6\pi, \, 2.4 \pi], \; z \in [0.25\pi, \, 0.75 \pi] $ has been applied in the selection to take into account the different domain sizes for the LES and the DNS and to exclude potential problems emerging with the periodic boundary conditions. The location of the probes, which are indicated as red dots, is shown in Fig. \ref{Fig:observations}. As previously discussed, the three components of the instantaneous velocity field are sampled. However, in the following configurations, the observation array is composed of 408 samples of the streamwise velocity only. The confidence in the DNS data is driven by the matrix $R$ presented in Sec. \ref{Sec:EnKF}. The matrix is diagonal and expressed as $R = \sigma_m^2I$, where $\sigma_m$ quantifies the uncertainty of the measurements. An accuracy of $20\%$ is applied as a percentage to the values for each observation. This implies that the variance of the velocity field oscillates between $0.003$ and $0.188$ depending on the distance from the wall of the sensor considered. These last remarks also imply that the weight given to each observation is the same. A specific DA run (DA-LESA), presented in appendix \ref{Appendix::DA-LESA}, has been performed taking as observation the three components of the velocity field for each sensor.

The general algorithm for the DA run is now presented. The ensemble members are initialized with a \textit{prior} state in terms of initial physical field and parametric description of the SGS model. The former is the field of the converged Smagorinsky simulation shown Sec. \ref{Sec:testcasePrior} and is the same for every member of the ensemble. 
The initial conditions for the parametric description of the SGS model are different for the two DA simulations performed and they will be described in sections \ref{Sec::run1} and \ref{Sec::run2}. Once the initial state is provided, the DA procedure advances in time the LES ensemble members for a total of $300 t_A$ times, performing an analysis phase each $0.12 t_A$. This choice, which implies that only one out of every three observation samples is integrated within the DA scheme, results in a total of $2500$ analysis phases. If one considers that the time step for the LES simulations is $\Delta t = 0.02 t_A$, this indicates that one analysis is performed every six forecast steps. No \textit{state} inflation is used in the DA runs. However, a time-varying \textit{parametric} stochastic inflation is included to improve the efficiency of the DA optimization. No inflation was used from $t_A = 0$ to $t_A = 12$. Then, a relatively strong inflation was included for $t_A = [12; \, 24]$ with $\lambda = 10\%$, followed by $\lambda = 5\%$ for $t_A = [24;\, 36]$. Finally, $\lambda = 1\%$ was used to carry out the averaging for the calculation of the statistical moments. Statistical averages are calculated in the range $t \in [50, \, 300]$, in order to safely dissipate high levels of variance previously used for the convergence of the parametric description.

The two main DA simulations are now presented in detail, highlighting the differences among the procedures.

\subsubsection{DA run $1$ (DA-LES1): optimization of the coefficient $C_k$}
\label{Sec::run1}

In this first DA run (referred to as DA-LES1) the vector of the parameters to be optimized consists of one element, which is the model constant $C_k$ of Smagorinsky's SGS closure. This is equivalent to optimizing the well-known coefficient $C_S$, which has been studied in the literature in particular in the framework of UQ analyses \cite{Meldi2011_pof,Meldi2012_pof}. As previously stated, the value of this global constant is updated at each analysis phase. Initial values of the $N_e = 40$ ensemble simulations are determined using a bounded Gaussian distribution $\mathcal{N}(\mu_u,\sigma_u^2)$. Considering data in the literature \cite{Meldi2011_pof}, $\mu_u = 0.094$ and $\sigma_u = 0.03$ were chosen to investigate a suitably large parametric space. The Gaussian distribution is constrained to values in the range $\mu_u \pm 2\sigma_u$, in order to avoid initial nonphysical parametrization which could lead to the divergence of the algorithm. 

\subsubsection{DA run 2 (DA-LES2): model spatial expansion for $C_k$}
\label{Sec::run2}
 
Following the results of DA-LES1, a more complex optimization is targeted to improve the predictive capabilities of the LES solver. Exploiting the homogeneity features of the test case in the streamwise direction $x$ and the spanwise direction $z$, the optimization of this second run (referred to as DA-LES2) targets the behavior of a functional expression for $C_k(y)$. More precisely, the free coefficients in a Gaussian expansion of $C_k$ are considered as variables to be optimized: 
\begin{equation}
    C_k = \sum_{i=1}^n \exp{\left(a_i-\frac{(y-y_i)^2}{\sigma_i^2} \right)} 
\end{equation}

For each of the $n$ Gaussian functions used in the decomposition, the free parameters to be determined are $a_i$ (intensity of the peak), $\sigma_i$ (width of the function), and $y_i$ (position of the peak). The functions are considered to be symmetric with respect to the half channel height, owing to the statistical symmetry in the wall-normal direction $y$. The decomposition is performed using $n=5$ Gaussian functions. This adds up to $15$ parameters in the control vector $\theta$ to be optimized via the EnKF. The average prior distribution for these functions is shown in Fig. \ref{Fig:gaussian_prior}. This initial distribution is chosen so that the peak of three functions is closer to the wall, in order to provide a suitable representation of $\nu_{sgs}$ in this region. For each ensemble member, the value of the $15$ free coefficients is determined using a Gaussian truncated ($\pm 2\sigma$) perturbation so that $a_i = \mathcal{N}(-4.5,0.3^2)$, $\sigma_i = \mathcal{N}(0.18,0.04^2)$ and $y_i = \mathcal{N}(0.15,0.05^2)$.

\section{Prediction of the statistical features using on-the-fly DA}
\label{Sec:results}
The previous discussion stressed how the DA tools provide an update of the physical state as well as an optimization of the model. In this section, particular attention is focused on the latter aspect. Results from DA-LES1 and DA-LES2 are investigated to observe how the DA procedure dynamically affects the value of the parameter $C_k$ as well as to assess the effects of the parametric optimization over the flow statistical behavior. One important point that must be stressed is that such statistical moments are not directly observed by the DA algorithms. In fact, unlikely recent analyses in the literature \cite{Mons2021_prf}, the DA procedure relies on instantaneous flow fields obtained from the model and sampled as observation.  

First, the optimized behavior of the parameter $C_k$ is investigated. For classical simulations using the prescribed values of the numerical code, one has $C_k=0.094$, $C_\varepsilon=1.048$ which corresponds to $C_S \approx 0.17$. 
Once the convergence of the parametric description is obtained, the coefficients exhibit a very weak time evolution. Results obtained from the run DA-LES1, which targets a global $C_k$ optimization, show that the time-averaged optimized value for $C_k \approx 0.014$. This result, which corresponds to $C_S \approx 0.04$, is $7$ times smaller than the default $C_k$ value provided by the code. The uncertainty associated with the limited amount of ensemble members has been assessed repeating the initial DA phases using different random distributions for $C_k$. These results indicated that optimized values fall in the range $C_k \in [0.012, \, 0.017]$. Within these ranges, variations in the predicted physical quantities are very small and they fall within the confidence threshold (i.e. values of the matrix $R$) provided for this study.

\begin{figure}
    \centering
    \begin{subfigure}{.45\textwidth}
        \centering
        \includegraphics[width=1\textwidth]{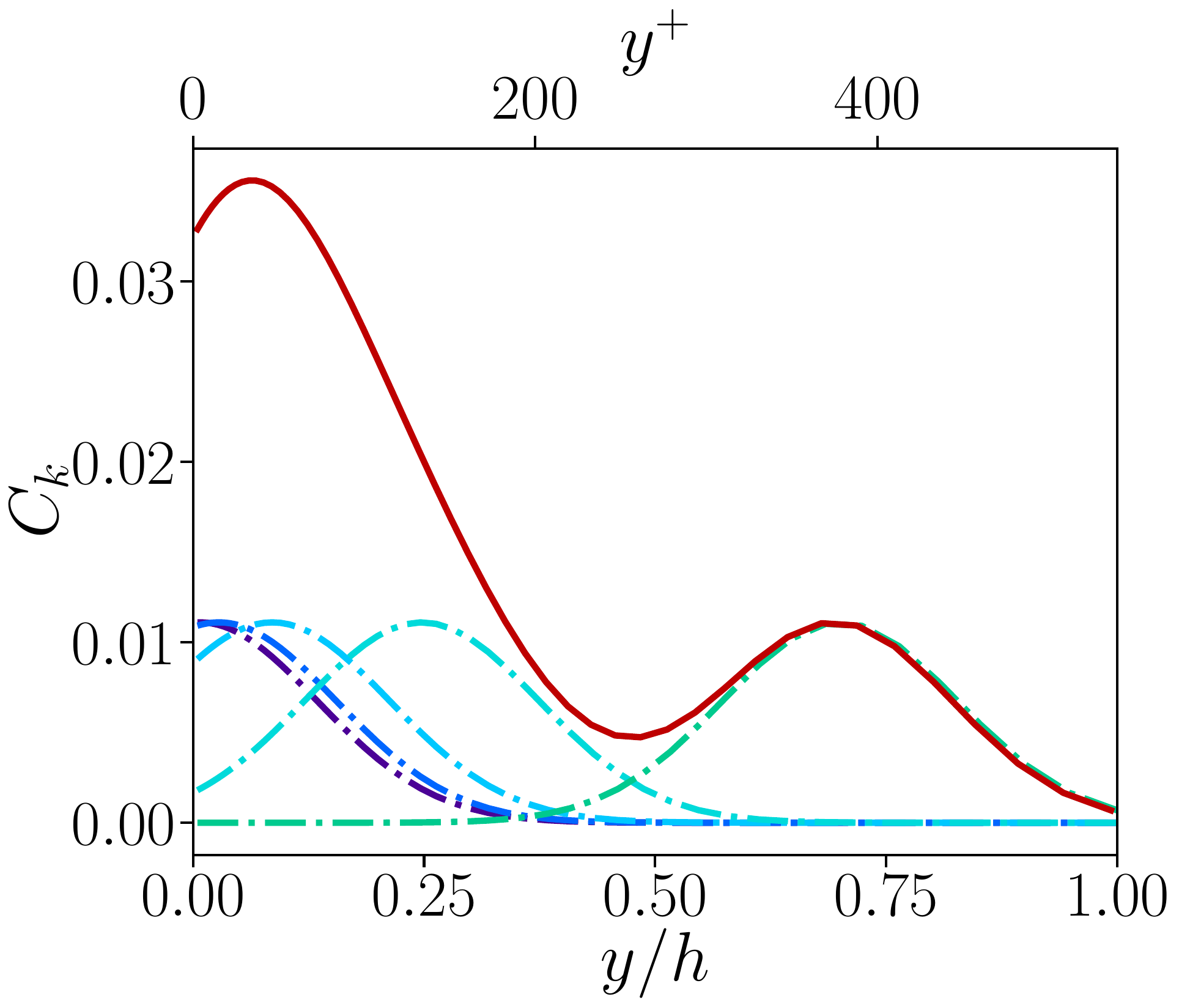}
        \caption{}
        \label{Fig:gaussian_prior}
    \end{subfigure}
    \begin{subfigure}{.45\textwidth}
        \centering
        \includegraphics[width=1\textwidth]{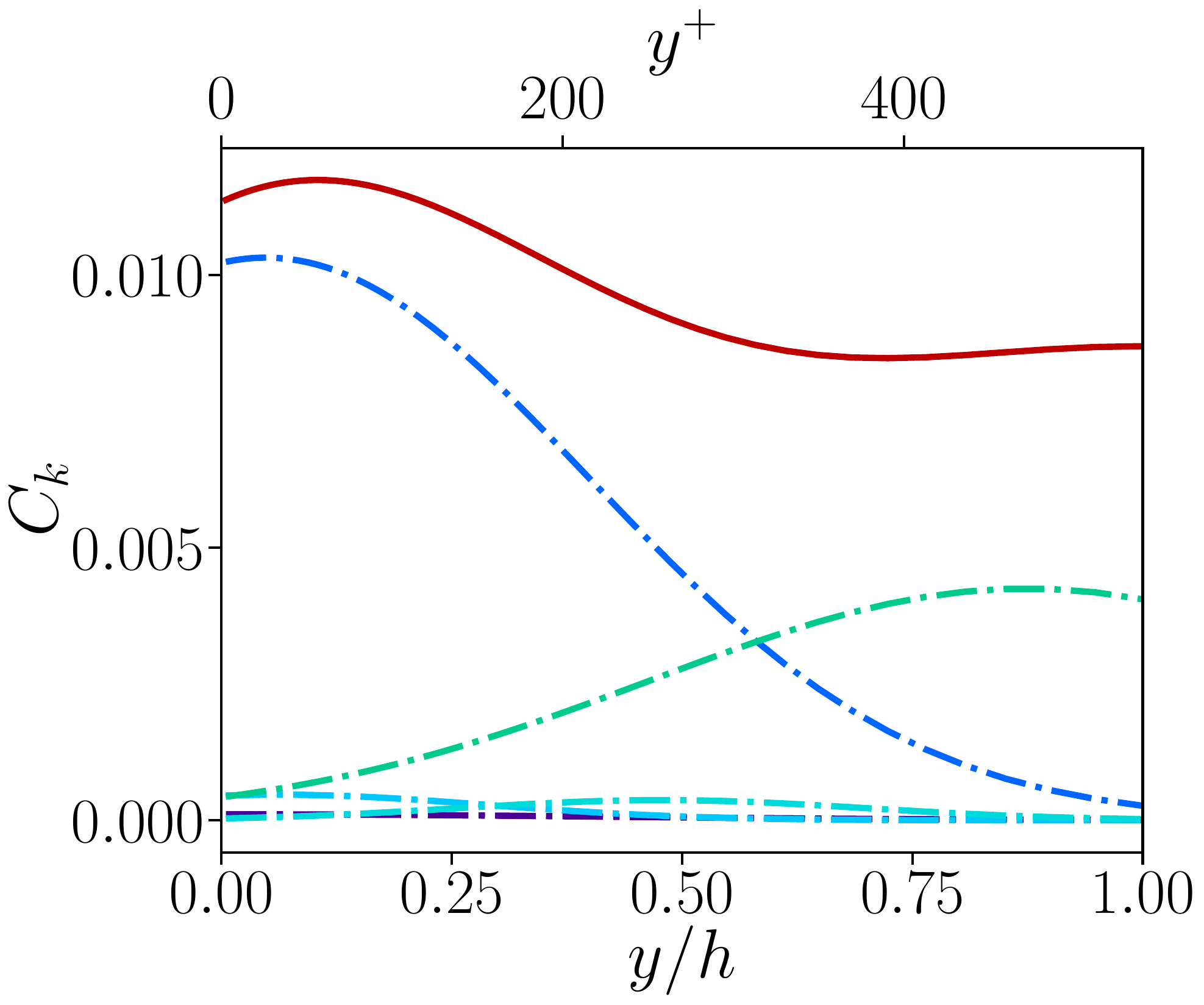}
        \caption{}
        \label{Fig:final_Ck_profiles}
    \end{subfigure}
    \caption{Profiles of the five Gaussian functions used to represent $C_k$ in blue. Distribution of $C_k$ in red. Left: prior distribution. Right: optimized distribution}
\label{Fig:gaussian_distrib}
\end{figure}

The results for the run DA-LES2 are shown in Fig. \ref{Fig:final_Ck_profiles}, where the profile of $C_k$ and each function of the Gaussian spatial distribution are shown. The final $C_k$ profile in red is again significantly lower than the distribution used as \textit{prior}. The values range from $0.008$ in the core flow region to a maximum value around $0.012$ reached close to the wall around $y^+ = 60$. In addition, the contributions of the modes of the Gaussian Expansion to describe the augmented $C_k$ profile appear to be very different. Two main modes govern the shape of $C_k$. The first mode exhibits a slow, quasi-linear decrease moving from the centerline towards the wall. On the other hand, the second one exhibits a maximum in the near-wall region ( $y^\star \approx 30$). The magnitude of the other three modes is significantly lower and they mainly smooth out the profile for $C_k$. Despite the higher complexity of this strategy, one can see that the distribution in the $y$ direction of $C_k$ is quasi-constant and $C_k \approx 0.01$ i.e. very similar to the global value obtained in the DA-LES1 run. 

\begin{figure}
    \centering
    \begin{subfigure}{.45\textwidth}
        \centering 
        \includegraphics[width=1\textwidth]{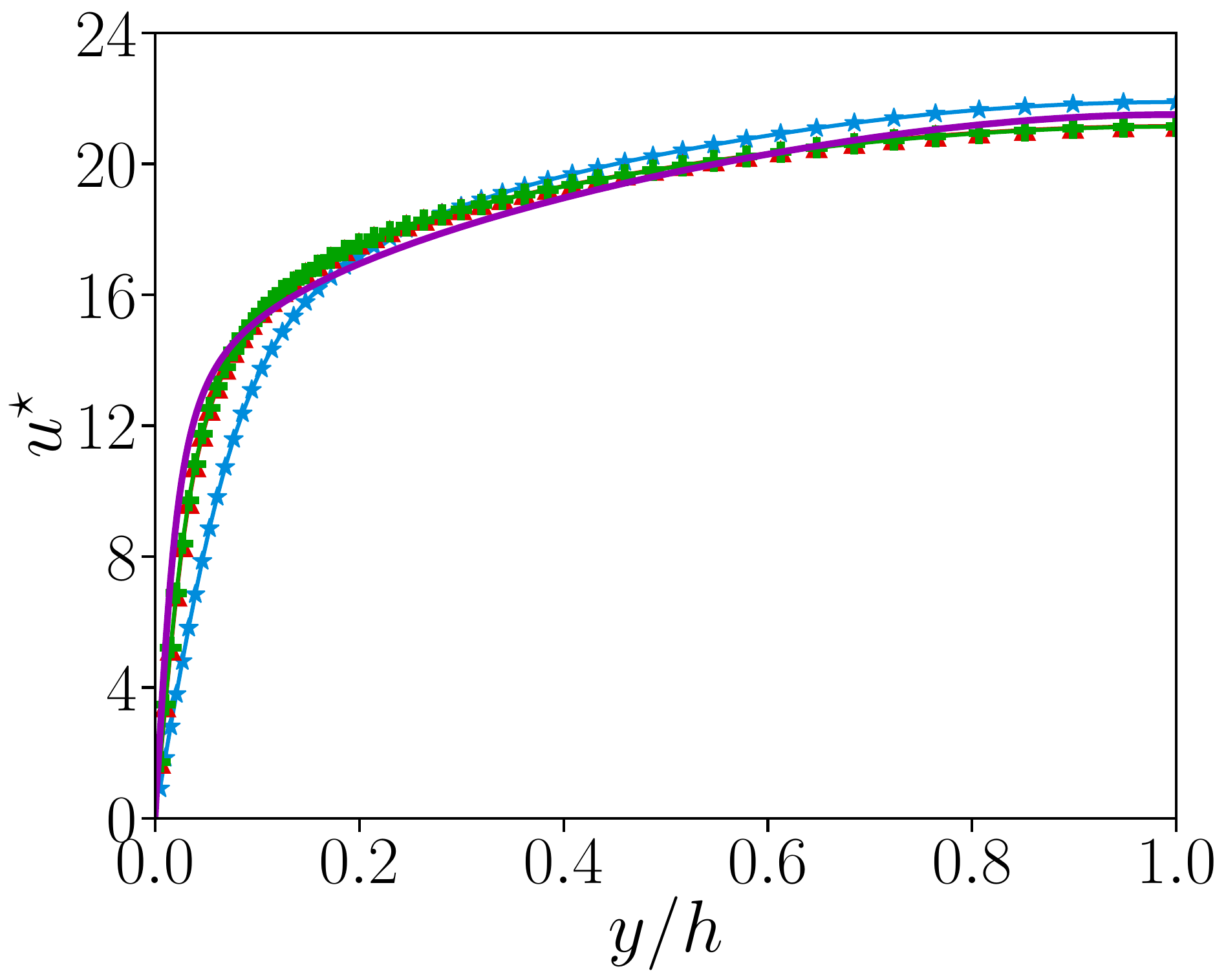}
        \caption{}
        \label{Fig:UStarOverYh}
    \end{subfigure}
    \begin{subfigure}{.44\textwidth}
        \centering 
        \includegraphics[width=1\textwidth]{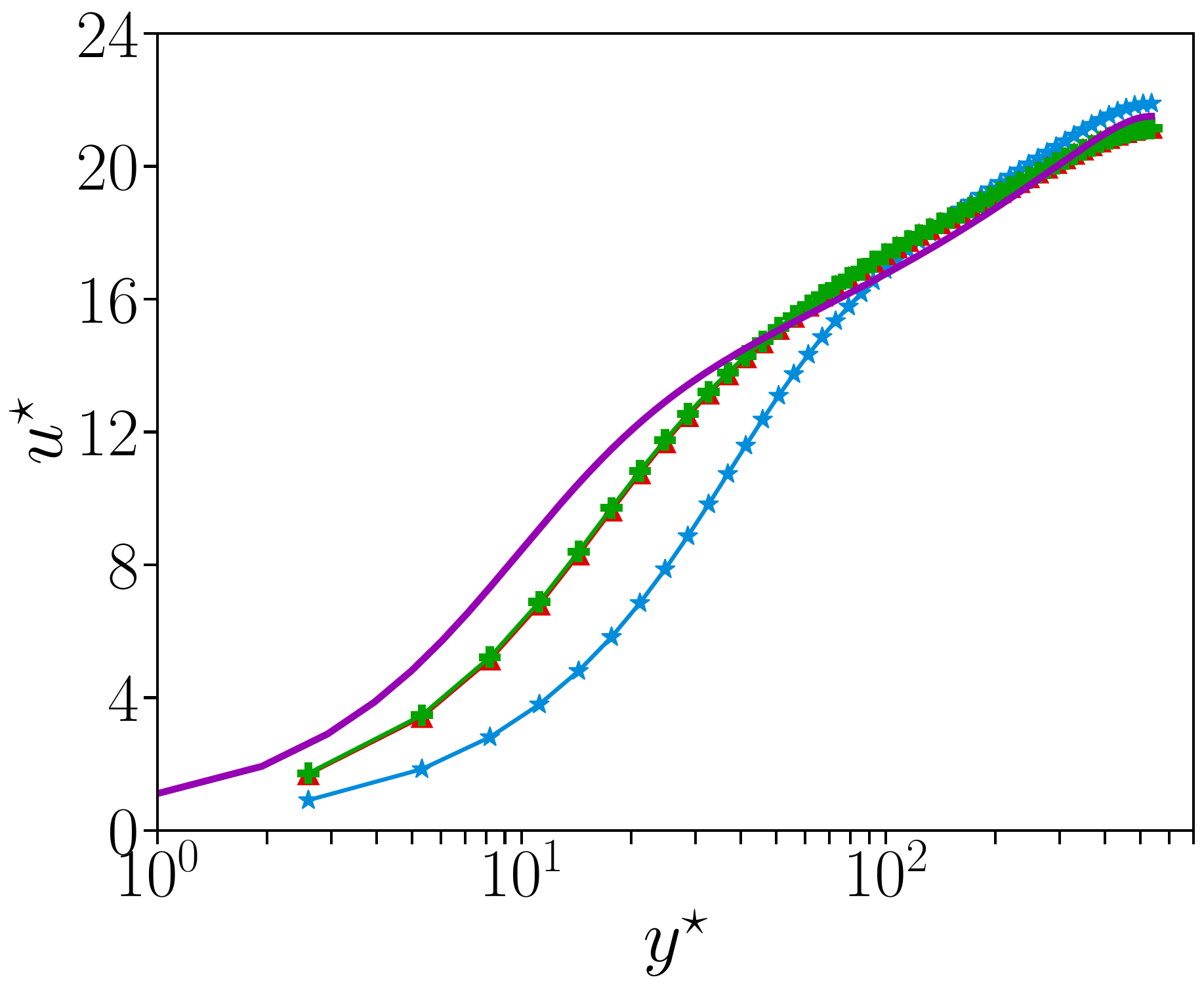}
        \caption{}
         \label{Fig:UStarOverYStar}
    \end{subfigure}
\caption{Half channel velocity profiles for DA-LES1 (\textcolor{red}{$\blacktriangleright$}), DA-LES2 (\textcolor{green}{$+$}), baseline LES (\textcolor{cyan}{$\star$}) and the DNS (\textcolor{violet}{$-$}). Adimensionalization is performed using the friction velocity $u_\tau$ obtained in the DNS run.}
\label{Fig:mean_profiles_star}
\end{figure}

The enhancement of the predictive capabilities via DA optimization is now assessed by comparing the statistical moments of the velocity field with the available DNS results as well as with the baseline LES. First of all, the prediction of the friction velocity $u_\tau$, which is one of the key features of this test case, is significantly improved for all DA runs. In fact, the targeted DNS friction velocity is  $u_\tau = 0.048$ and baseline LES simulation predicts a $u_\tau = 0.0614$, which represents an over-prediction of $28\%$. DA-LES1 and DA-LES2 predict almost the same friction velocity, with $u_\tau^{DL1} = 0.04617$ and $u_\tau^{DL2} = 0.04619$. In this case, the friction velocity is under-predicted when compared with the DNS, but the discrepancy is only $4\%$. This increase in accuracy comes with a significant reduction of the subgrid-scale viscosity in the near-wall region, which does not scale correctly for the Smagorinsky LES. Considering that the values obtained for $C_k$ in the near wall region with the two DA procedures are almost identical, it is not surprising to observe minimal variations in the prediction of $u_\tau$. Similar conclusions can be drawn by the analysis of Fig.~\ref{Fig:mean_profiles_star}, where the normalized mean velocity profile $u^\star$ against $y^\star$ are shown. Averages for the DA procedures are performed so that $u^\star = \langle \overline{u_x} \rangle/ u_\tau$ where $\langle . \rangle$ is the ensemble average operator and $\overline{.}$ is the time-average operator. The results obtained via the two DA procedures show a global improvement in the prediction of the velocity field, reducing on average the discrepancy with the DNS data. This observation is a direct consequence of the improved prediction for $u_\tau$. The apparently more accurate behavior of the baseline LES close to the center of the channel is actually a compensation of errors between the local numerical error and the erroneous prediction of $u_\tau$, which can be observed in Fig.~\ref{Fig:UStarOverYh}. In fact, with a more accurate prediction of $u_\tau$, the baseline LES would almost exactly collapse on the results obtained by DA. Minor discrepancies can be observed between the runs DA-LES1 and DA-LES2, which are arguably associated with the rate of convergence of the EnKF using $40$ ensemble members.

\begin{figure}
    \centering
    \begin{subfigure}{.40\textwidth}
        \centering 
        
        \includegraphics[width=1\textwidth]{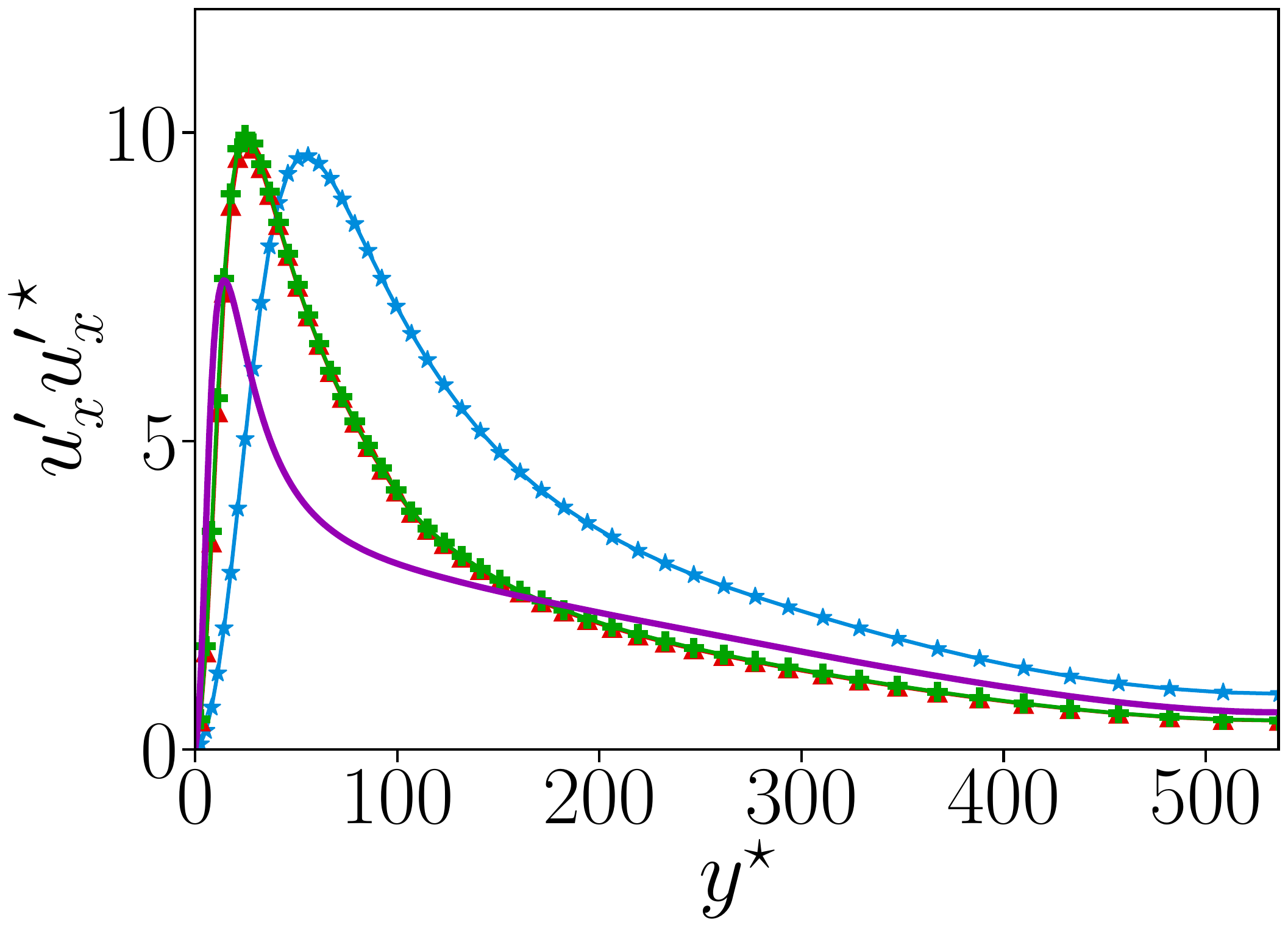}
        \caption{}
        \label{Fig:uuStarOverYStar}
    \end{subfigure}
    \begin{subfigure}{.41\textwidth}
        \centering 
        
        \includegraphics[width=1\textwidth]{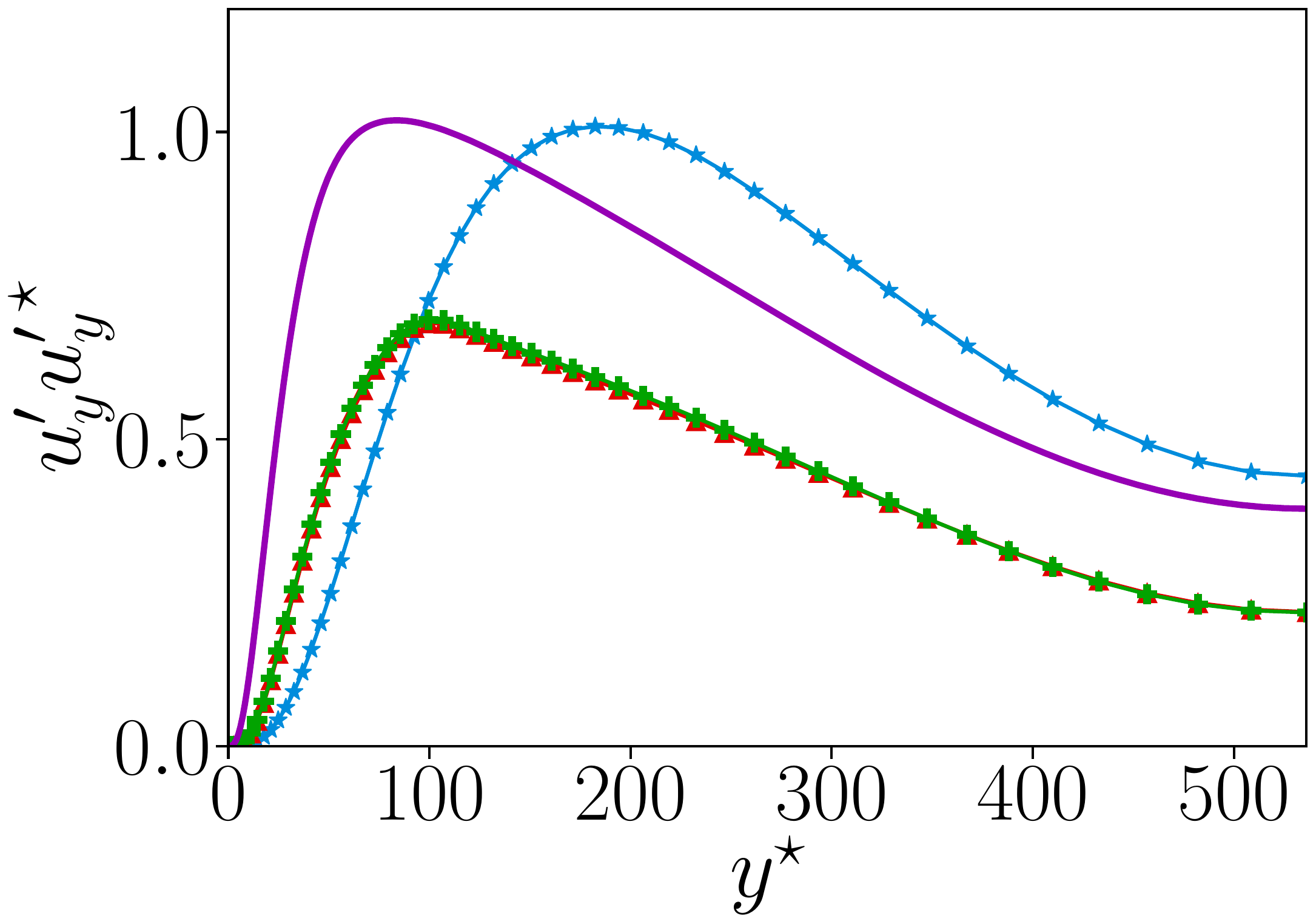}
        \caption{}
        \label{Fig:vvStarOverYStar}
    \end{subfigure}
        \begin{subfigure}{.40\textwidth}
        \centering 
        
        \includegraphics[width=1\textwidth]{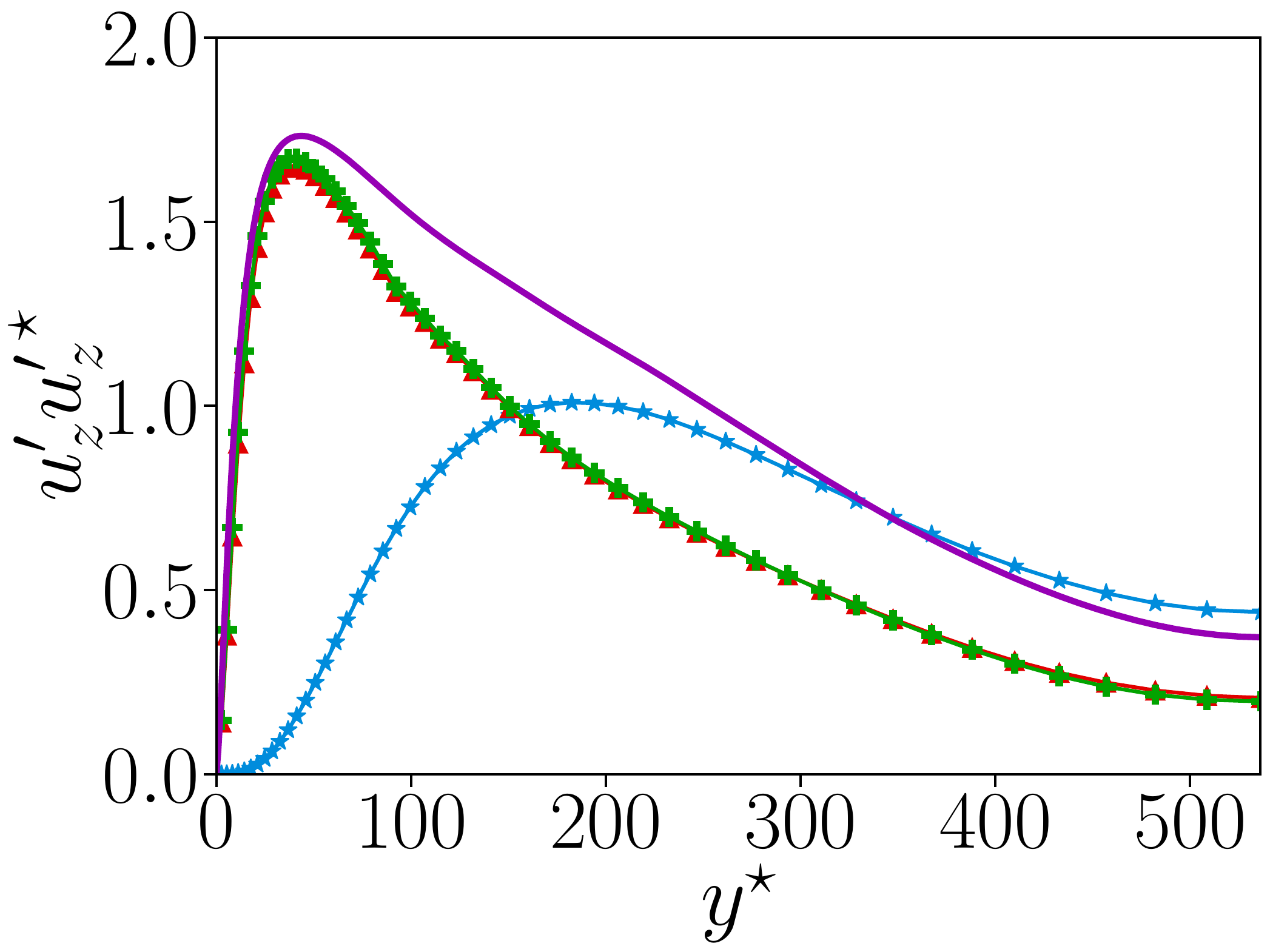}
        \caption{}
        \label{Fig:wwStarOverYStar}
    \end{subfigure}
        \begin{subfigure}{.42\textwidth}
        \centering 
       
        \includegraphics[width=1\textwidth]{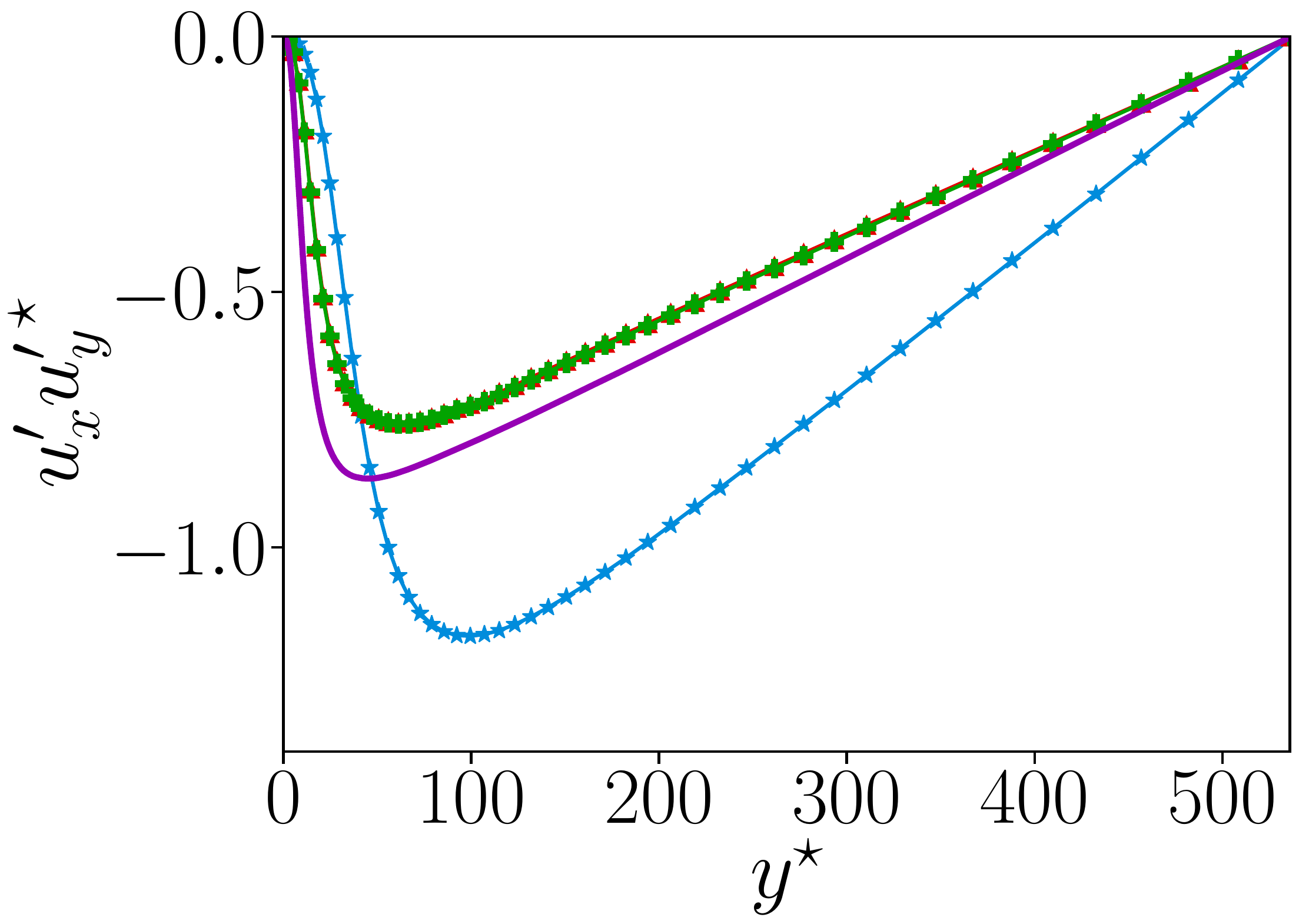}
        \caption{}
         \label{Fig:uvStarOverYStar}
    \end{subfigure}
    \caption{Components of the Reynolds stress tensor for DA-LES1 (\textcolor{red}{$\blacktriangleright$}), DA-LES2 (\textcolor{green}{$+$}), baseline LES (\textcolor{cyan}{$\star$}) and the DNS (\textcolor{violet}{$-$}). Adimensionalization is performed using the friction velocity $u_\tau$ obtained in the DNS run.}
    \label{Fig:RST_profiles_star} 
\end{figure}

The normalized components of the resolved Reynolds stress tensor are shown Fig. \ref{Fig:RST_profiles_star}. Again, for the DA runs, ${u_i^\prime u_j^\prime}^\star = \langle \overline{u_i^\prime u_j^\prime} \rangle / u_{\tau}^2$. A global improvement in the accuracy of the prediction of such quantities is observed. For all the components, the location of the peak is accurately predicted. The magnitude of the components also exhibits a general improvement, which is however dependent on the component considered. In fact, while a very good agreement with DNS data is observed for ${u_z^\prime u_z^\prime}^\star$, a slight decrease in accuracy is instead obtained for ${u_y^\prime u_y^\prime}^\star$. The almost identical results obtained with the two runs DA-LES1 and DA-LES2 suggest that the variations of $C_k$ in the $y$ direction for the latter do not affect the flow prediction. One could argue that the present optimization reached the best performances obtainable with Smagorinsky LES, whose subgrid-scale representation is affected by strong, intrinsic limitations \cite{Mons2021_prf}. Another possibility is the combination of prior state and inflation employed in the present analysis for the model coefficients was not sufficient to perform a complete exploration of the parametric space, and the final solution for DA-LES2 was drawn to the same local optimized state obtained for DA-LES1. 

\begin{figure}
    \centering
    \begin{subfigure}{.45\textwidth}
        \centering 
        
        \includegraphics[width=1\textwidth]{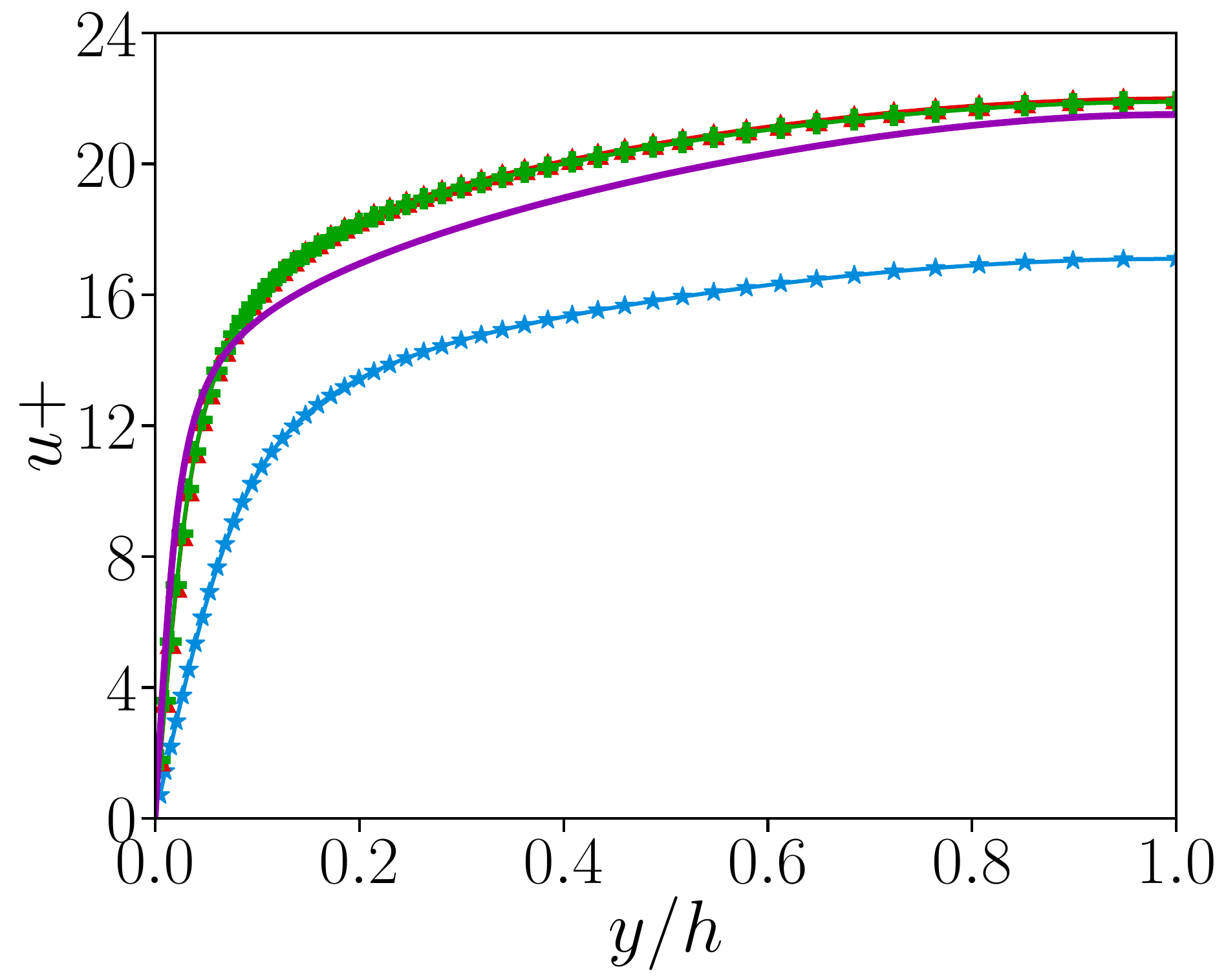}
        \caption{}
        \label{Fig:UPlusOverYh}
    \end{subfigure}
    \begin{subfigure}{.44\textwidth}
        \centering 
        
        \includegraphics[width=1\textwidth]{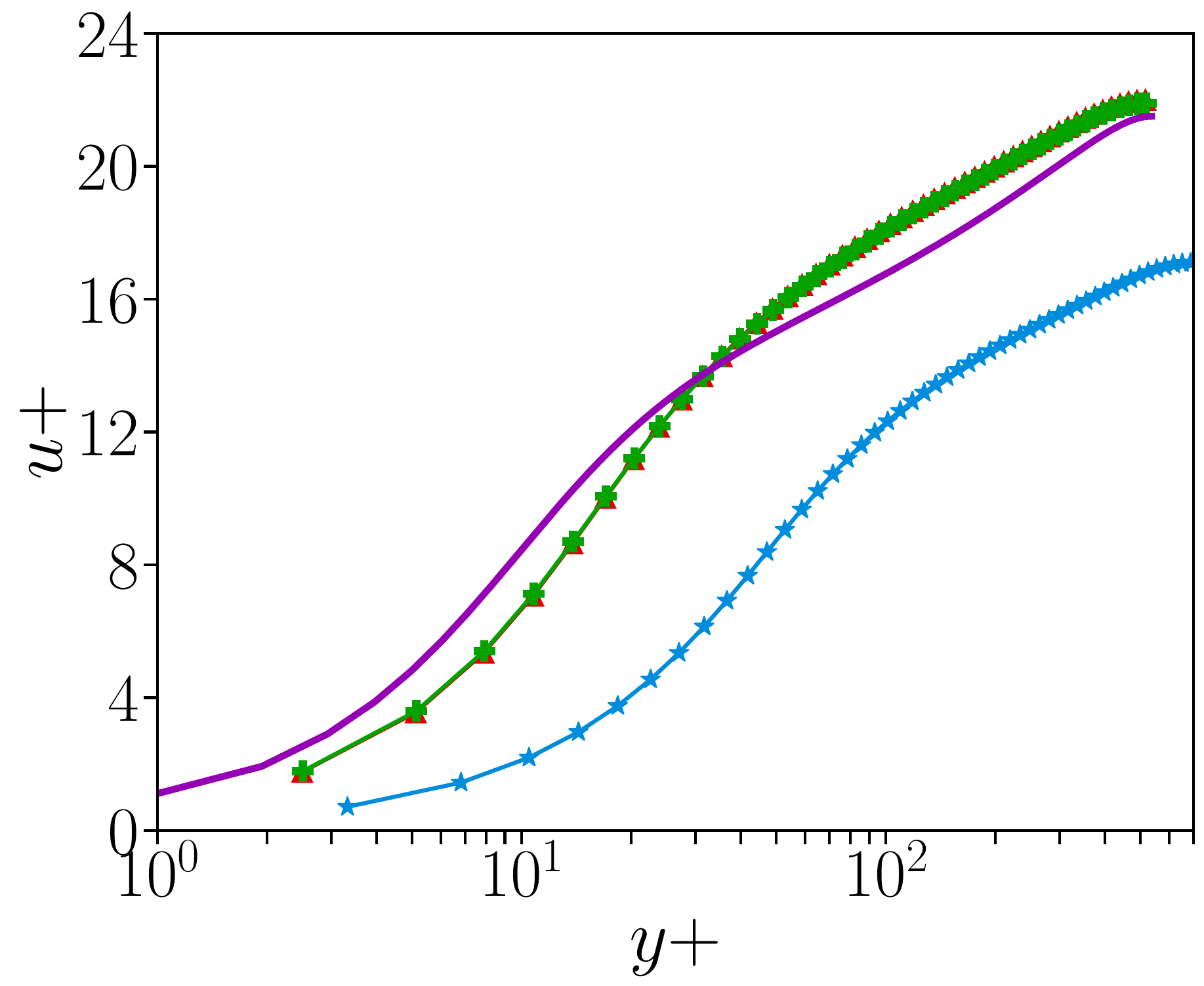}
        \caption{}
        \label{Fig:UPlusOverYPlus}
    \end{subfigure}
\caption{Half channel velocity profiles for DA-LES1 (\textcolor{red}{$\blacktriangleright$}), DA-LES2 (\textcolor{green}{$+$}), baseline LES (\textcolor{cyan}{$\star$}) and the DNS (\textcolor{violet}{$-$}).}    
\label{Fig:mean_profiles_plus}
\end{figure}

\begin{figure}
    \centering
    \begin{subfigure}{.40\textwidth}
        \centering 
        
        \includegraphics[width=1\textwidth]{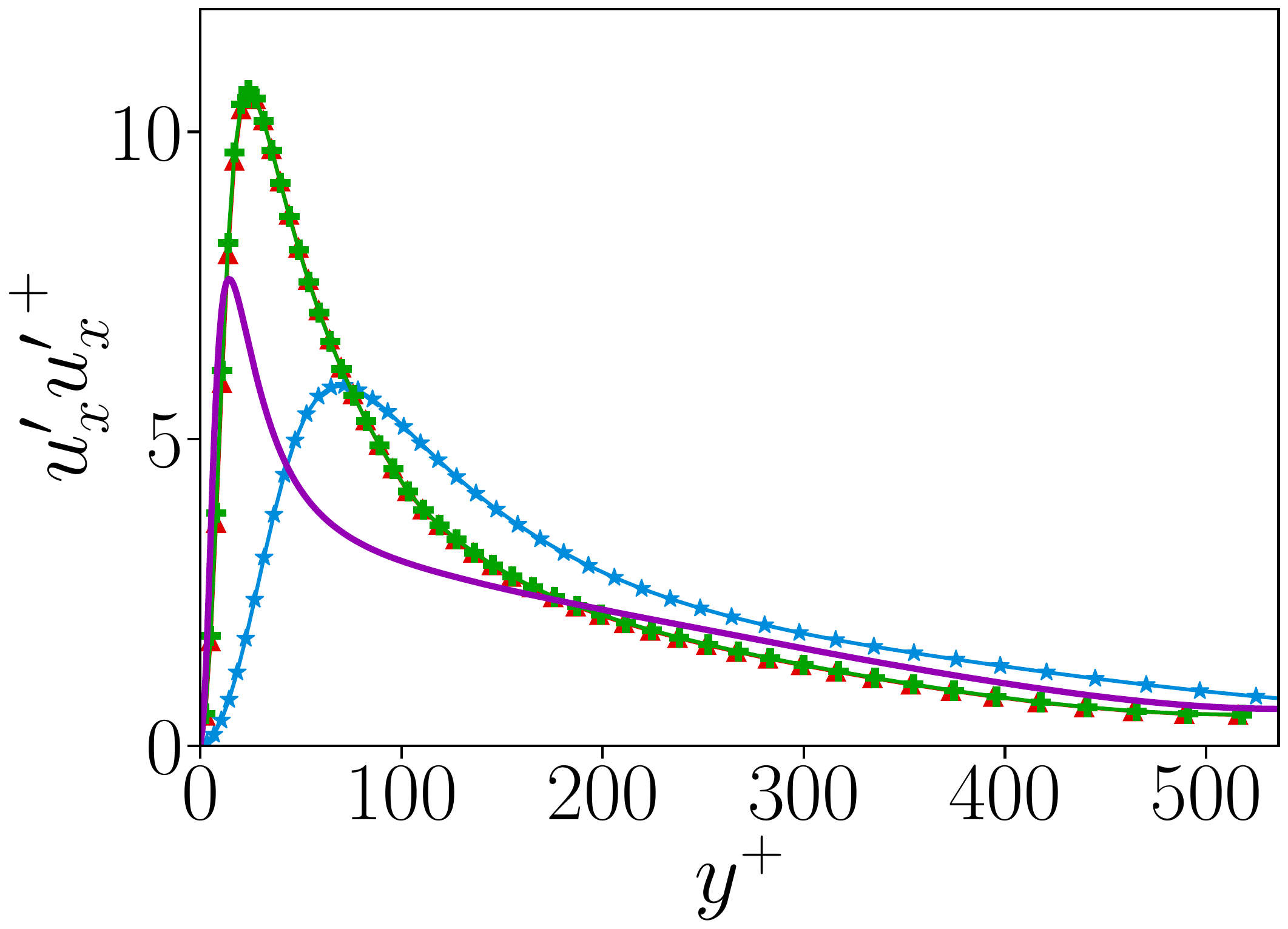}
        \caption{}
        \label{Fig:uuPlusOverYPlus}
    \end{subfigure}
    \begin{subfigure}{.41\textwidth}
        \centering 
        
        \includegraphics[width=1\textwidth]{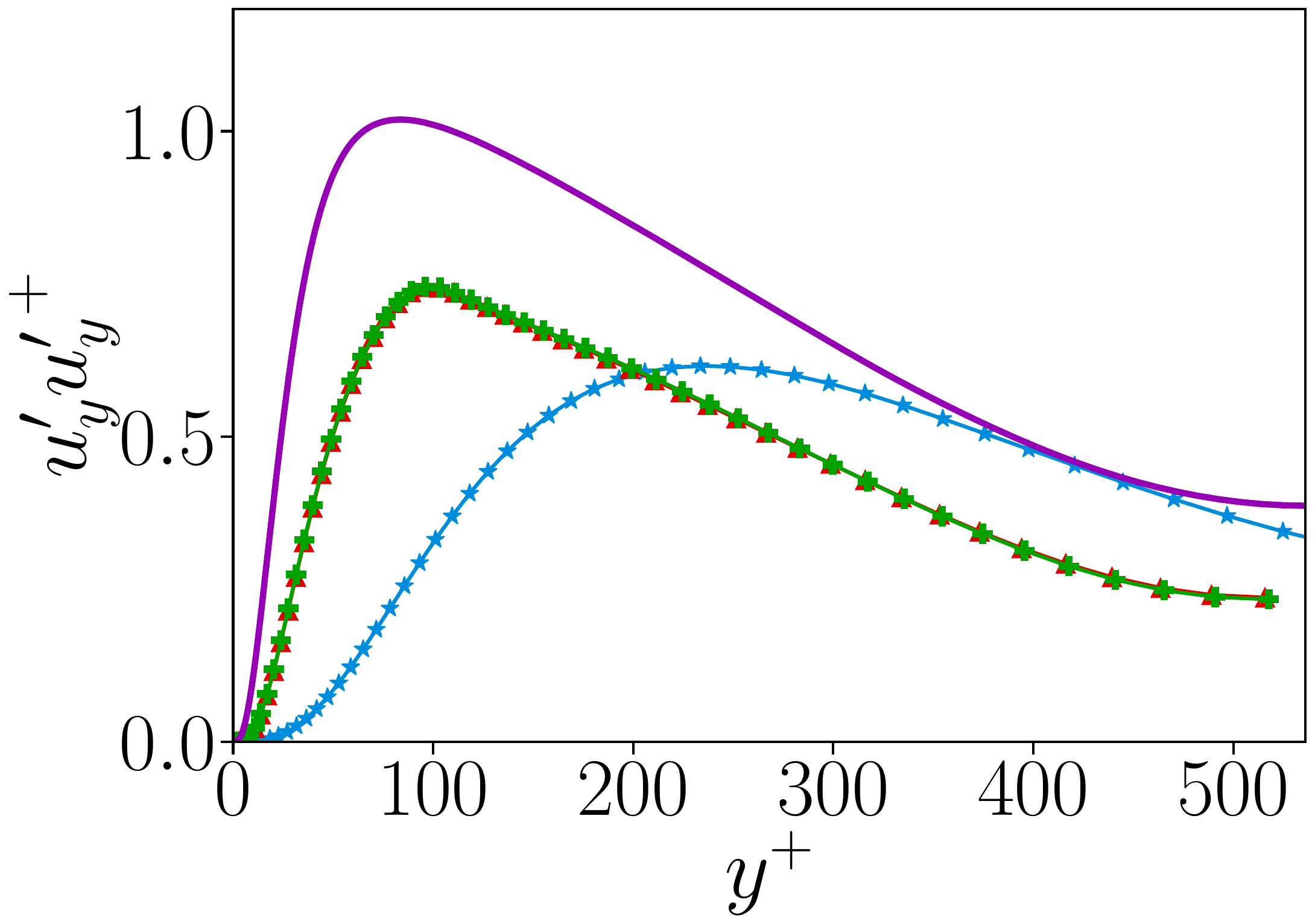}
        \caption{}
        \label{Fig:vvPlusOverYPlus}
    \end{subfigure}
        \begin{subfigure}{.40\textwidth}
        \centering 
       
        \includegraphics[width=1\textwidth]{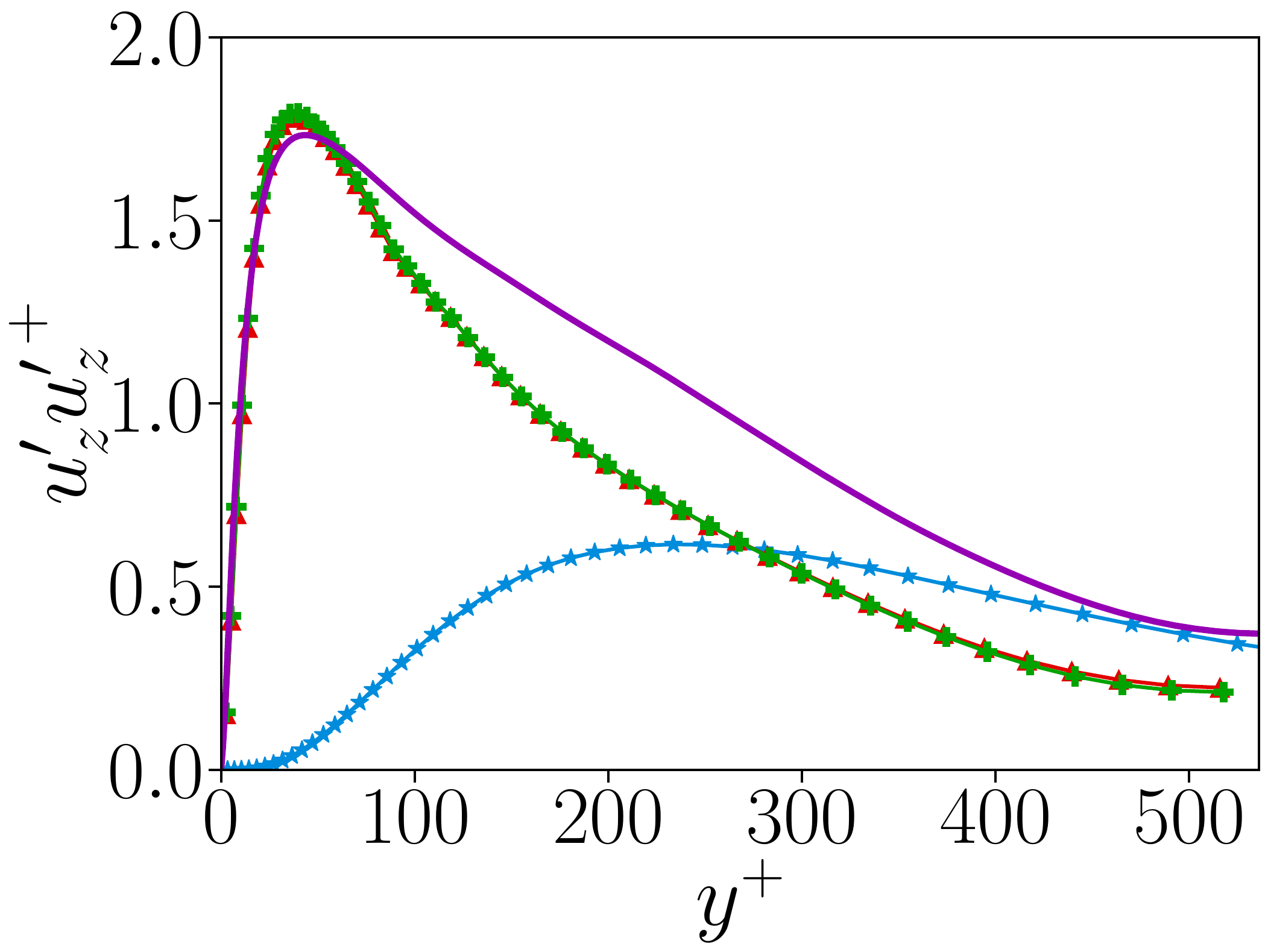}
        \caption{}
         \label{Fig:wwPlusOverYPlus}
    \end{subfigure}
        \begin{subfigure}{.42\textwidth}
        \centering 
       
        \includegraphics[width=1\textwidth]{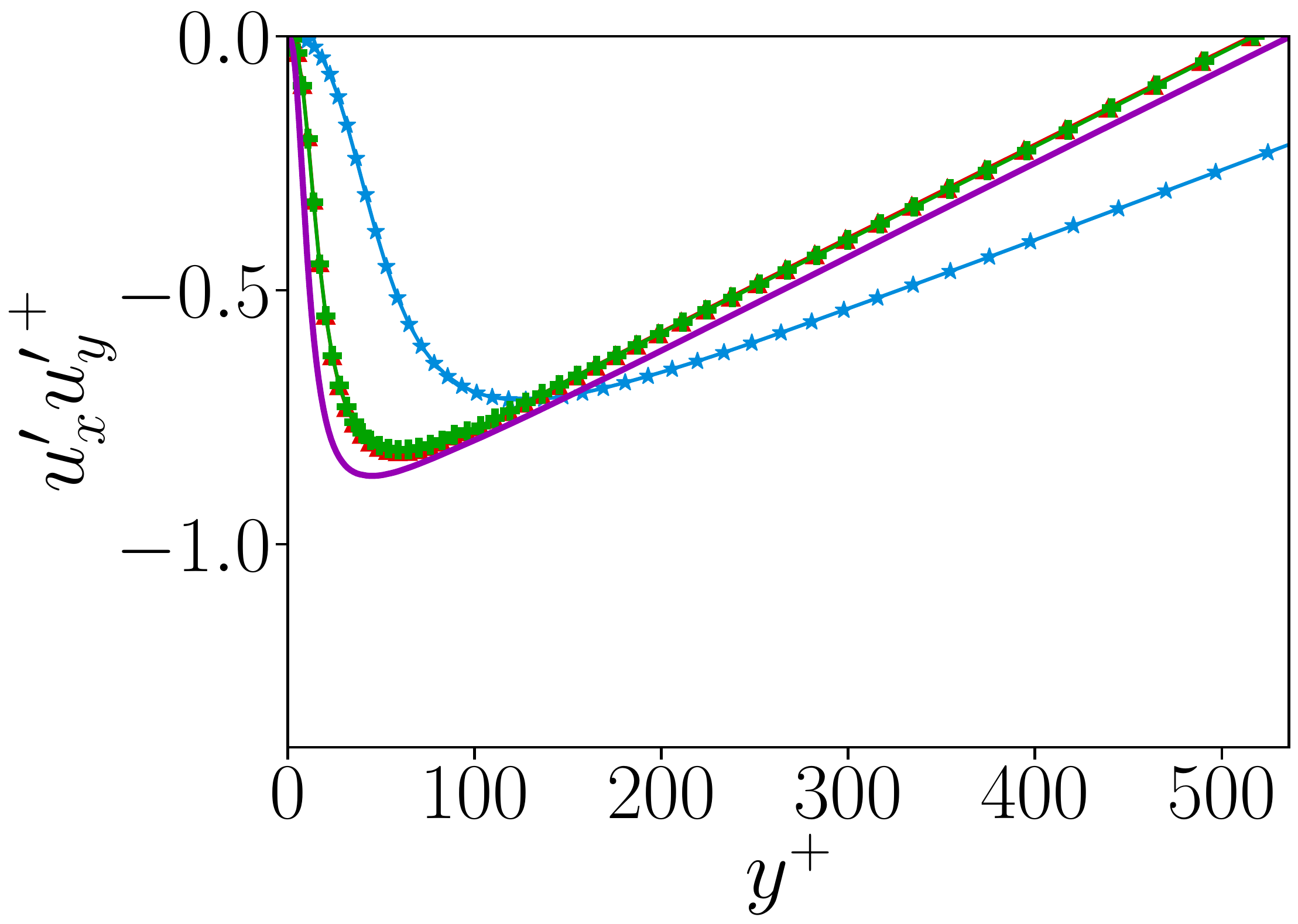}
        \caption{}
         \label{Fig:uvPlusOverYPlus}
    \end{subfigure}
    \caption{Components of the Reynolds stress tensor for DA-LES1 (\textcolor{red}{$\blacktriangleright$}), DA-LES2 (\textcolor{green}{$+$}), baseline LES (\textcolor{cyan}{$\star$}) and the DNS (\textcolor{violet}{$-$}).}
    \label{Fig:RST_profiles_plus}
\end{figure}

The analysis of the physical quantities normalized over the $u_\tau$ calculated by each simulation (suffix $+$) leads to similar conclusions. The mean streamwise velocity profiles, which are shown in Fig. \ref{Fig:mean_profiles_plus}, confirm the global lack of accuracy of the baseline simulation, which is now even more magnified by the significant error in the prediction of $u_\tau$. The components of the resolved Reynolds stress tensor, which are reported in Fig.~\ref{Fig:RST_profiles_plus}, also provide very similar indications.

\begin{figure}
        \centering
    \begin{subfigure}{.32\textwidth}
        \centering 
       
        \includegraphics[width=1\textwidth]{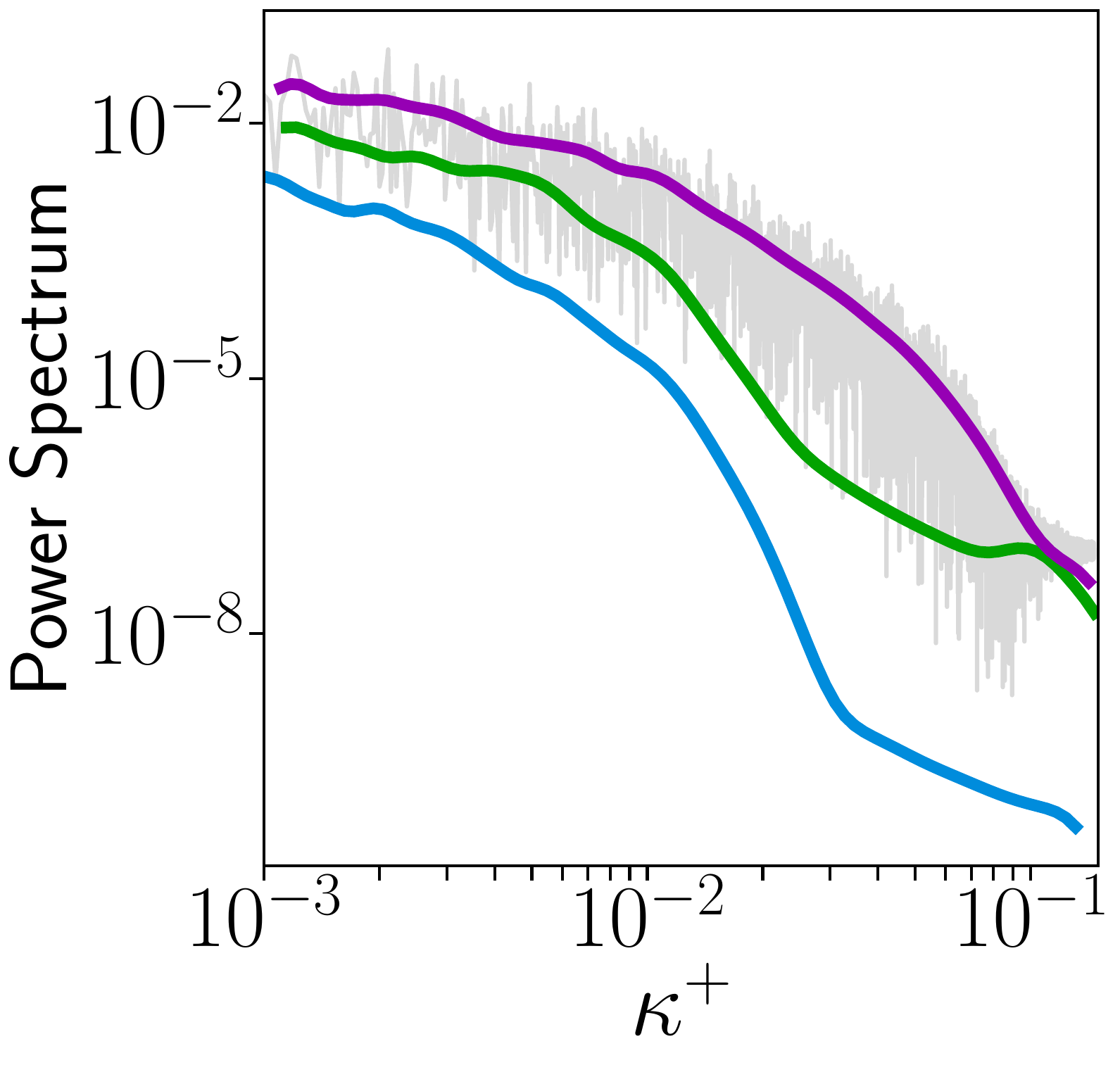}
        \caption{U-DA, $u_x'$, $y^\star=1.45$}
         \label{Fig:probe1_X}
    \end{subfigure}
    \begin{subfigure}{.3\textwidth}
        \centering 
        
        \includegraphics[width=1\textwidth]{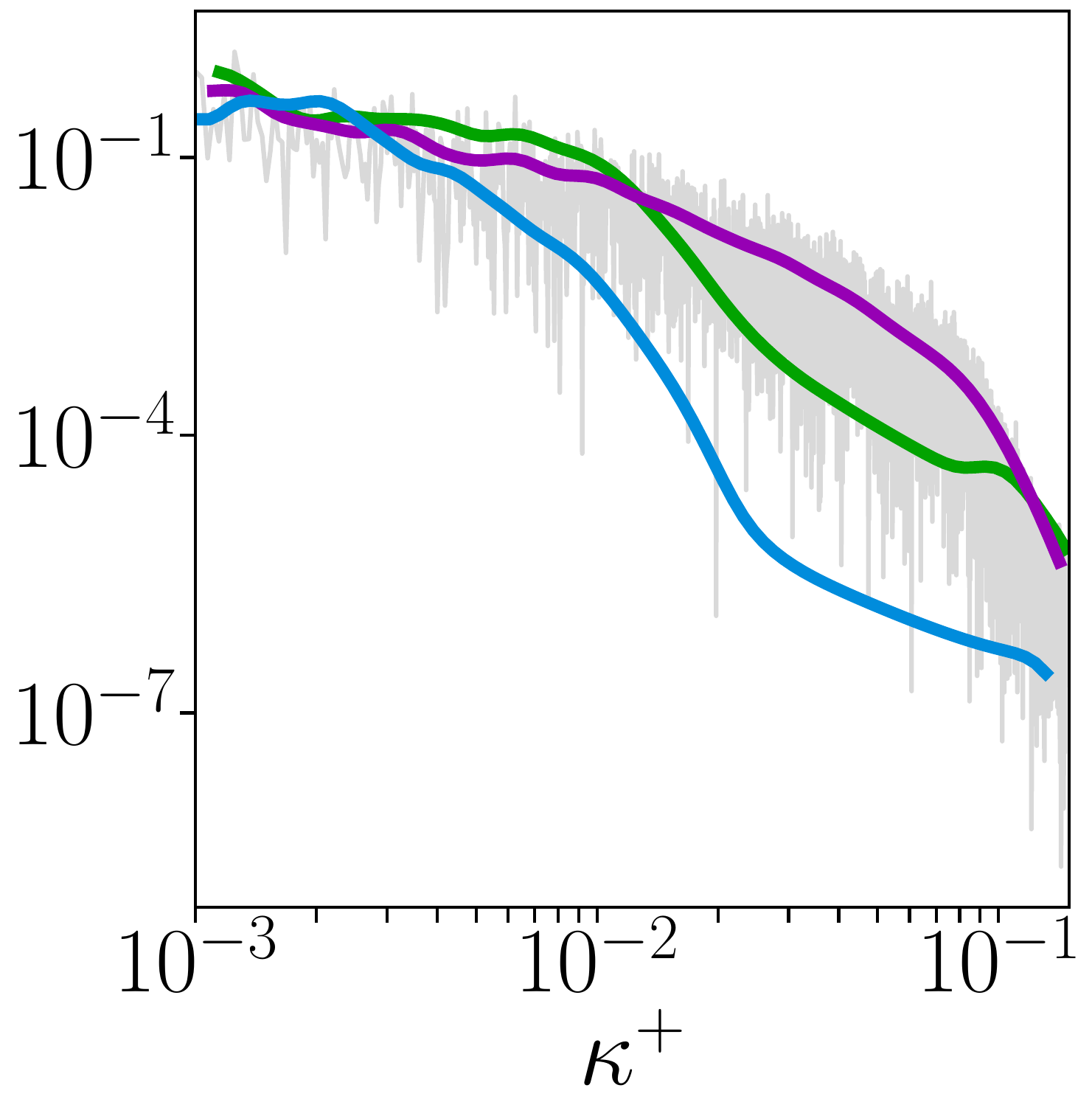}
        \caption{U-DA, $u_x'$, $y^\star=26.85$}
        \label{Fig:probe3_X}
    \end{subfigure}
    \begin{subfigure}{.3\textwidth}
        \centering 
        
        \includegraphics[width=1\textwidth]{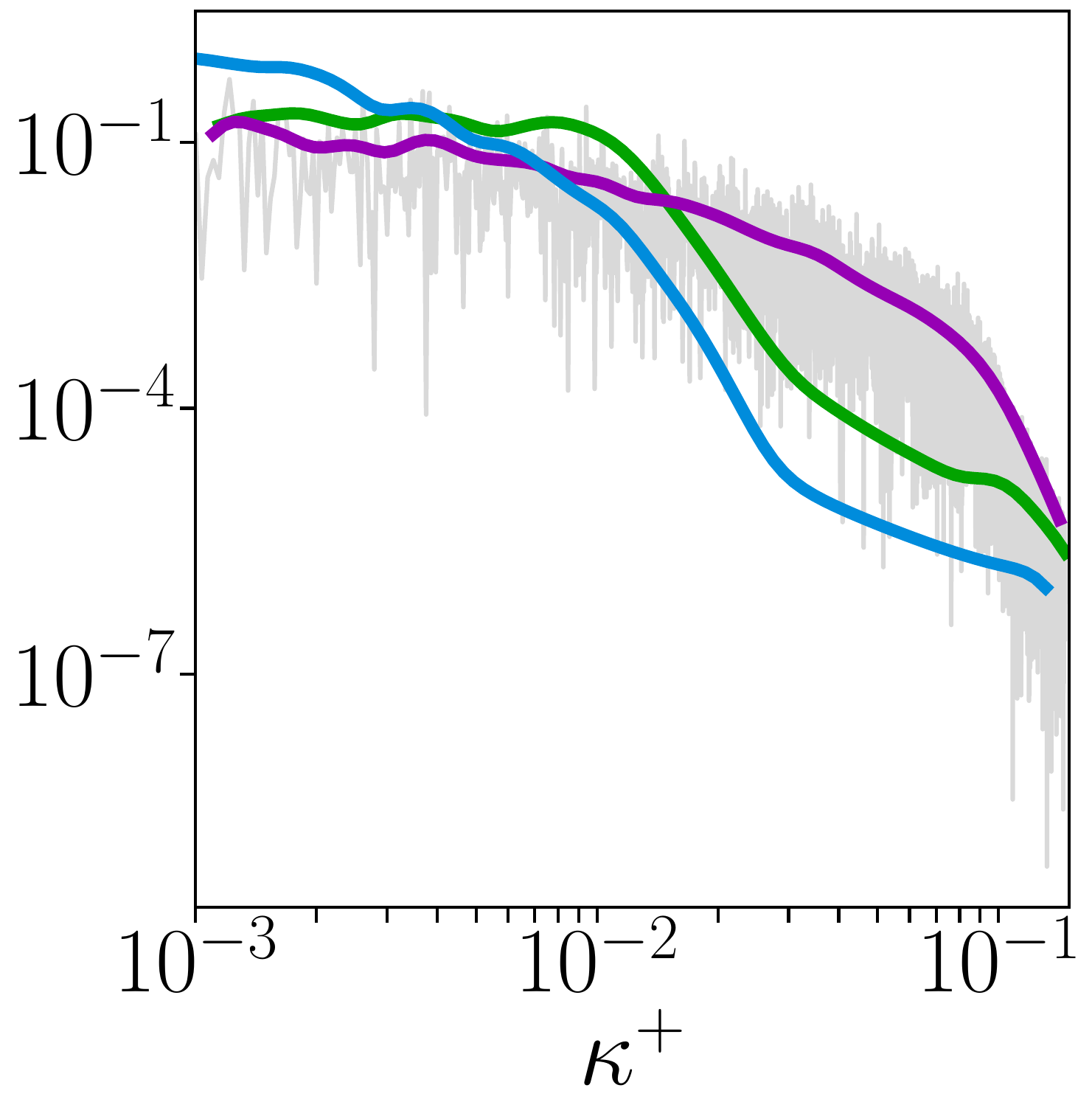}
        \caption{U-DA, $u_x'$, $y^\star=46.74$}
        \label{Fig:probe5_X}
    \end{subfigure}
    \begin{subfigure}{.32\textwidth}
        \centering 
       
        \includegraphics[width=1\textwidth]{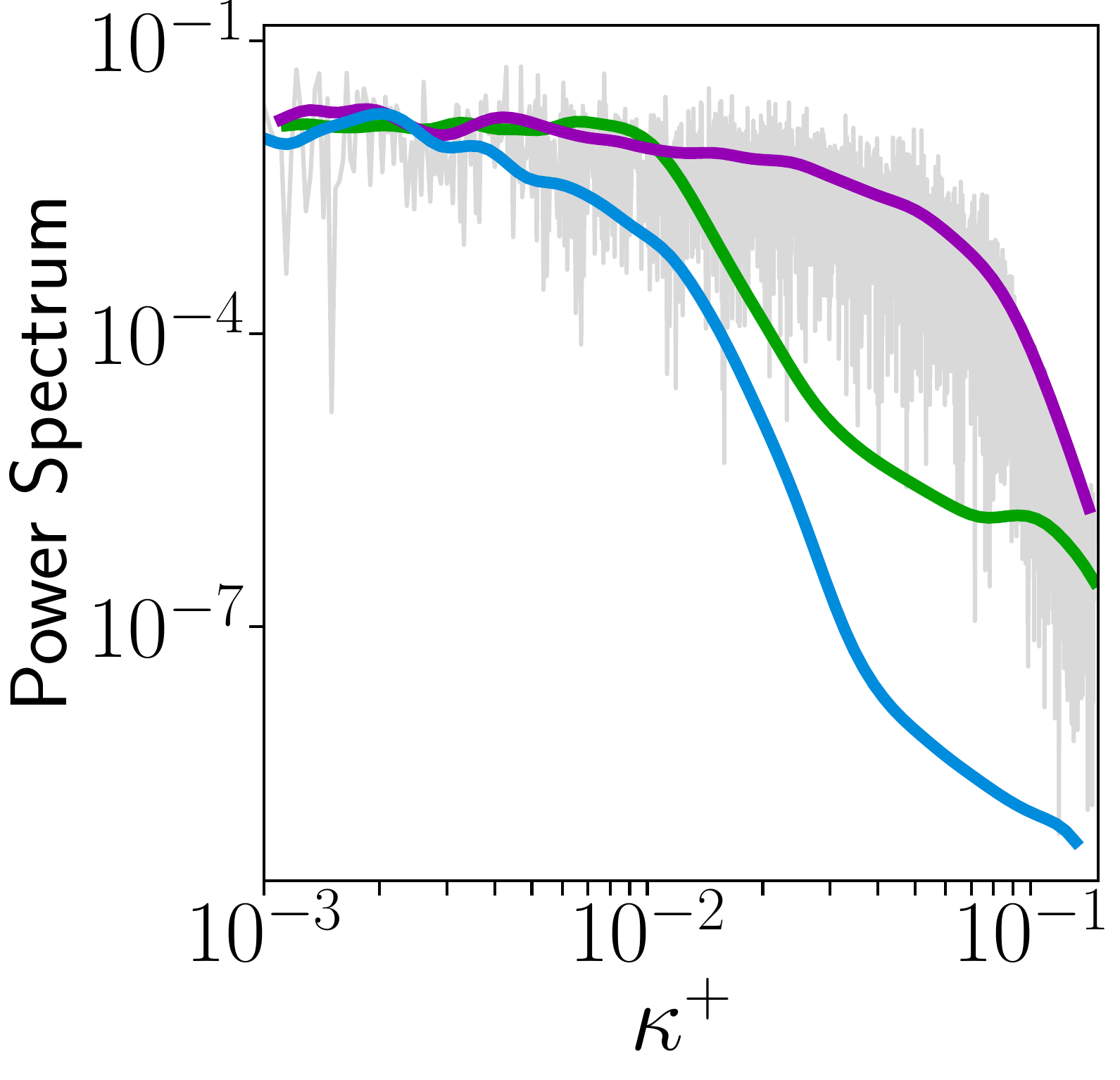}
        \caption{U-DA, $u_y'$, $y^\star=46.74$}
         \label{Fig:probe5_Y}
    \end{subfigure}
    \begin{subfigure}{.3\textwidth}
        \centering 
        \includegraphics[width=1\textwidth]{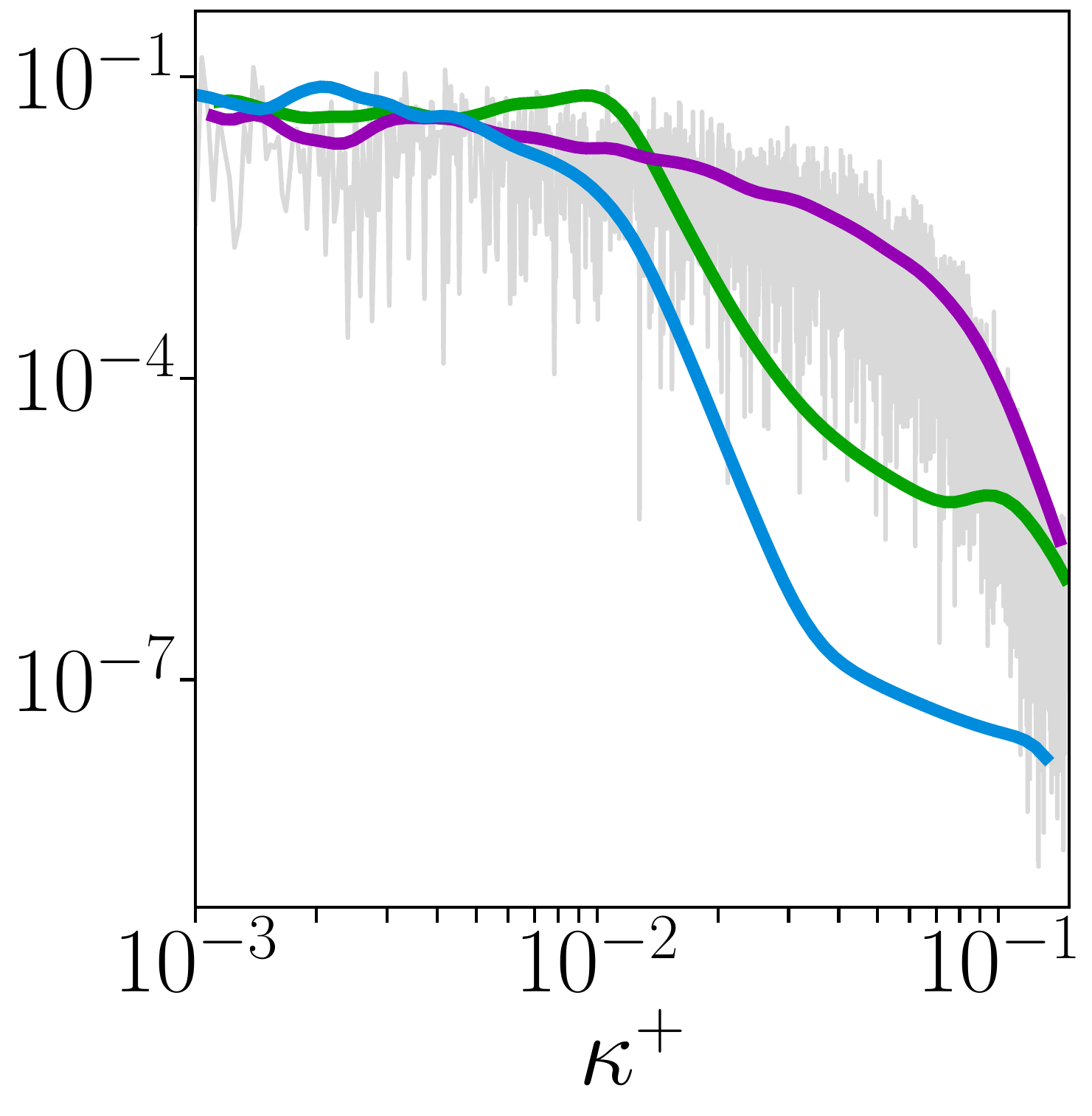}
        \caption{U-DA, $u_z'$, $y^\star=46.74$}
        \label{Fig:probe5_Z}
    \end{subfigure}
    \begin{subfigure}{.3\textwidth}
        \centering 
        
        \includegraphics[width=1\textwidth]{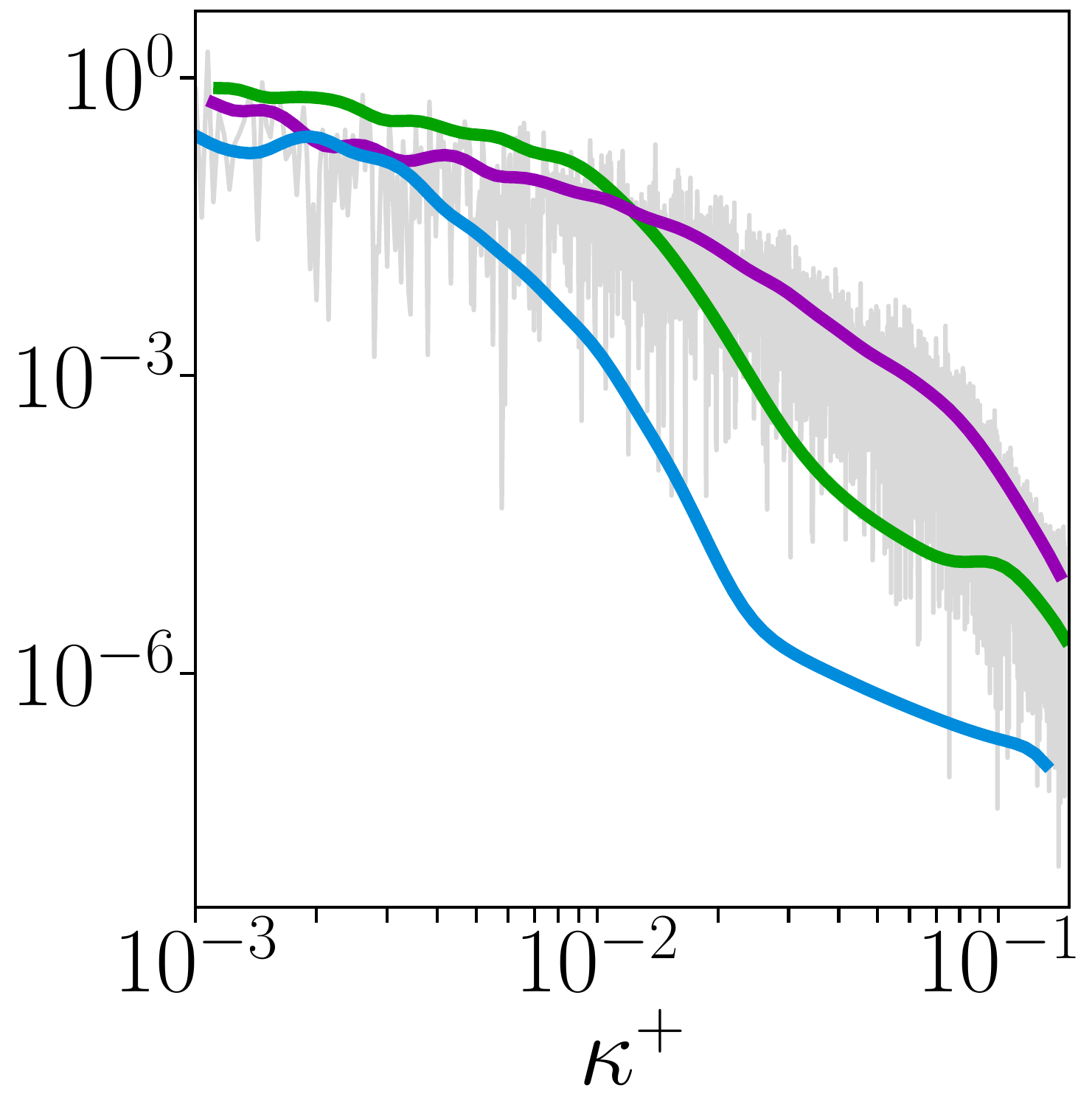}
        \caption{N-DA, $u_x'$, $y^\star=19.84$}
        \label{Fig:probe6_X}
    \end{subfigure}
    \caption{Power Spectra of the velocity field obtained with a Morlet wavelet transform. Results are shown for DA-LES2 (\textcolor{green}{$-$}), baseline LES (\textcolor{cyan}{$-$}) and the DNS (\textcolor{violet}{$-$}). For the latter, additional results obtained using a fast Fourier transform are shown in grey.}
    \label{Fig:psd}
\end{figure}

Finally, a spectral analysis of the velocity field $\mathbf{u}$ is performed in Fig. \ref{Fig:psd}. This flow variable has been sampled in time at four probes located at $y^\star \in [1.45,46.74]$. Power spectrums have been obtained using a Morlet Transform \cite{Torrence1998_bams} for the baseline LES, the run DA-LES2, and the DNS. The spectra are plotted over the dimensionless wave number, $\kappa^+ = \kappa\nu/u_\tau$ with $\kappa = 2\pi f/u_c$. $f$ is here the set of frequencies used for the Morlet transform.
On the first line, data for the streamwise component $u_x$ are shown at locations where observation is available and data from that sensor is used in the DA analysis phase (indicated as U-DA in the legend). Comparing the baseline simulation and the DA run, one can see that the accuracy of the spectra has been improved for every $y^\star$ investigated. The best result is observed in the proximity of the wall as shown in Fig. \ref{Fig:probe1_X}. For this location, the energy's amplitude is improved by approximately one order of magnitude. 
The comparison of the spectra from the DNS and the run DA-LES2 also indicates an offset of the wavenumber for which the spectral density starts to decrease fast. This offset, which is around one octave, is very close to the ratio of the mesh resolution in the streamwise direction (see Tab.\ref{Tab:grid_ resolutions}). For the baseline LES, this drop in energy begins at lower wavenumbers. This observation can be justified by the discrepancy in the prediction of $u_\tau$ (which is used to obtain $\kappa^+$) as well as by the Smagorinsky closure, which provides an unwanted dissipative effect at the large scales.
Results on the second line of Figs. \ref{Fig:probe5_Y} and \ref{Fig:probe5_Z} are obtained at a location where a DNS sensor is available and used for DA analysis, but the information assimilated (streamwise velocity) is not the one here investigated. More precisely, the power spectra for the spanwise and vertical components are shown. One can see that, similarly to what was observed for the spectra of the streamwise velocity, a global improvement is obtained for the DA-LES2 run. This result confirms the global beneficial DA effect over the complete flow field, and not just for the variables for which observation is available. The analysis is completed by the results in  Fig. \ref{Fig:probe6_X}, where the spectrum of the streamwise velocity sampled at a location not used in the DA analysis (N-DA) is shown. Again, one can see that the spectrum shows an improvement similar to what was observed at sensors actively used in the DA procedure, indicating that the optimization of the SGS closure is globally beneficial in particular to reduce the dissipation of the resolved energy at large scales.

In summary, on-the-fly DA using instantaneous measurements is able to improve the accuracy of LES via calibration of the SGS closure. An interesting point is that present results are similar to findings by Mons et al. \cite{Mons2021_prf}, which were however obtained via observation of the physical quantities used to evaluate the performance of the LES simulations. In this case, the optimization process is more complex, because of the instantaneous nature of the observation as well as for its sparsity in space and time. Thus, the present findings open perspectives of real-time optimization of scale-resolving CFD using tools based on the EnKF, once the computational architectures are strong enough to do so. However, similarly to what was observed by Mons et al. \cite{Mons2021_prf}, the parametric optimization can mitigate but not eliminate the discrepancy between Smagorinsky LES and DNS, due to the intrinsic limitations of the structural form of the SGS model. While this problem is difficult to challenge, one can arguably consider that on-the-fly DA has a higher potential to determine in real-time SGS model structural forms and correction for a specific case than offline EnKF approaches. Lastly, both strategies used in this analysis indicate that the best accuracy is obtained for very low values of the model constant $C_k$. Despite the run DA-LES2 provides a more sophisticated space distribution of this parameter, values are low enough to consider that the dynamic effect of the SGS closure becomes globally and locally minor, as shown by the profiles obtained for DA-LES1. These results are consistent with recent works presenting extensive comparisons between explicit and implicit SGS closures \cite{VicenteCruz2023_jot}.

\section{Synchronization of the flow field}
\label{Sec::synchro}

The synchronization capabilities of the DA algorithm are now investigated. With synchronization, we indicate the capability of the DA algorithm to progressively reduce the discrepancy between the instantaneous model solution and the observation, both in proximity and far from the sensors. If successful, the only state corrections applied by the analysis phase are due to the accumulation of error in forecast step(s), due to the lack of accuracy of the model. 
Even though synchronization is not necessary for the analysis of statistical moments, such as the ones investigated in Sec. \ref{Sec:results}, it has crucial importance for the analysis of instantaneous features of unstationary flows. In fact, in a digital twin system, efficient synchronization enables the model to identify extreme events and thus prevent critical occurrences for the physical counterpart. 
Tools based on the EnKF can naturally act on the synchronization of the instantaneous flow. Thanks to the flexibility of the quantity observed and the local correlation captured between the physical variables, their efficiency in this task is supposedly higher than classical Nudging. However, during the DA calculations, the variability of the ensemble tends to diminish relatively fast, potentially precluding an efficient synchronization. To avoid this issue, the hyperparameter known as inflation must be properly optimized. 

In this section, a number of DA runs are performed to study the effects of inflation over the rate of synchronization. In this case, the attention is focused on the very first analysis phases, and results are investigated over two advective times $t_A$. The DA analyses are now performed every two time steps i.e. $0.04t_A$, which corresponds to a total of $50$ DA state updated over the time window of investigation. Such a high frequency in updates has been imposed to ensure that errors due to the sparsity in time of the data are neglected \cite{Meldi2018_ftc}. The state estimation is obtained via the flow prediction of $40$ members, which are initialized using a different velocity field but share the same value of $C_k$ for the SGS model obtained in the DA-LES2 procedure. The $40$ velocity fields used as prior states have been generated running a single simulation with the same optimized SGS model obtained Sec. \ref{Sec:results} and sampling complete flow fields every $10t_A$. 

The inflation is here applied only to the \textit{state estimation} via the stochastic approach described in Sec. \ref{Sec:inflation}. The inflation applied to the parametric SGS description is here set to zero in order to exclude effects due to different behaviors of the LES closure. In addition, the covariance matrix $\mathcal{R}$ is also the same for each DA run and it is set to $R = \sigma_m^2 I$ with $\sigma_m = 5\%$. More details about these two last hyperparameters are provided in appendix \ref{Appendix::synchronization}, where parametric inflation is shown to have negligible effect for the purpose of this analysis. 

The effectiveness of the synchronization is evaluated using the following information: 

\begin{itemize}
    \item The velocity field obtained by the ensemble members is sampled in correspondence of three sensors, which are selected among the $10800$ sensors previously used in the reference DNS. Details about the sensors are given in Tab. \ref{Tab:sync_probes}. One can see that two of the probes are used in the DA algorithm, while the last one is not directly used. Still, the data obtained for the latter can be used for comparison.
    \item A global estimation of the normalized root mean square deviation (indicated as $\Phi$) for the velocity field is performed considering data from the $408$ sensors used within the DA algorithm and for $408$ sensors that were not used in the EnKF. The definition of $\Phi$ is given below, for an instant $k$ : 
    \begin{equation}
        \Phi_k = \sqrt{\frac{\sum_{j=1}^{N_o}{(\langle s_{j,k} \rangle-\alpha_{j,k})^2}}{N_o}}/\alpha_k^{mean}
        \label{eq:NRMSD}
    \end{equation}
\end{itemize}
with $\langle s_k \rangle = \sum_{i=1}^{N_e}Hu_{i,k}$, $\alpha_k^{mean} = \sum_{j=1}^{N_o}\alpha_{j,k}/N_o$ and $N_o$ the number of observations \\

\begin{table}
    \centering
    \begin{tabular}{|c|c|c|c|c|c|}
        \toprule
        \textbf{Probe ID} & $\boldsymbol{x/h}$ & $\boldsymbol{y/h}$ & $\boldsymbol{z/h}$ & $\boldsymbol{y^\star}$ & \textbf{Used in DA analysis} \\ \midrule
        1 & 7.372 & 0.0027 & 2.349 & 1.45 & yes \\ \midrule
        2 & 3.967 & 1.9631 & 1.233 & 19.84 & yes \\ \midrule
        3 & 7.464 & 1.9501 & 1.748 & 26.85 & no \\ \midrule
    \end{tabular}
    \caption{Details about the sensors used to study synchronisation.}
    \label{Tab:sync_probes}
\end{table}

\begin{figure}
    \centering
    
    \begin{subfigure}[b]{.33\textwidth}
        \centering
        \includegraphics[width=\textwidth]{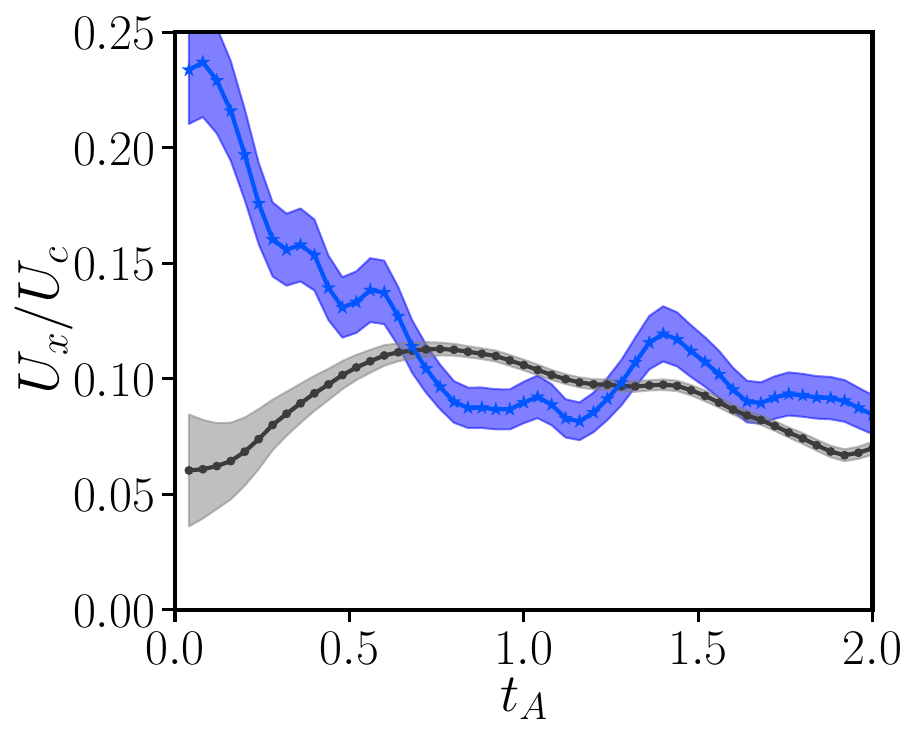}
    \end{subfigure}%
    \hfill 
    \begin{subfigure}[b]{.33\textwidth}
        \centering
        \includegraphics[width=\textwidth]{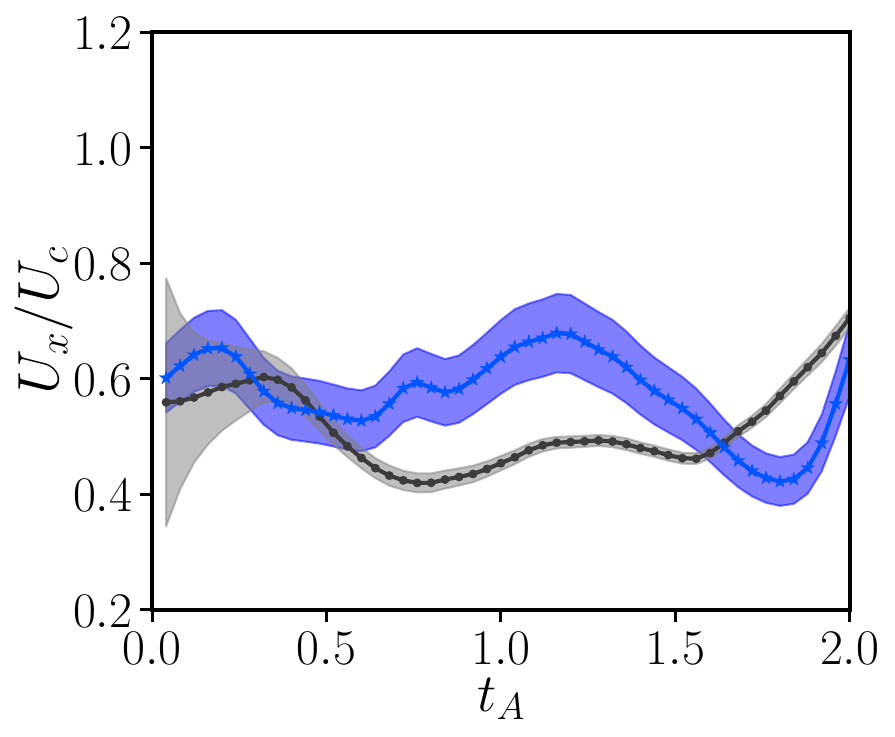}
    \end{subfigure}%
    \hfill 
    \begin{subfigure}[b]{.33\textwidth}
        \centering
        \includegraphics[width=\textwidth]{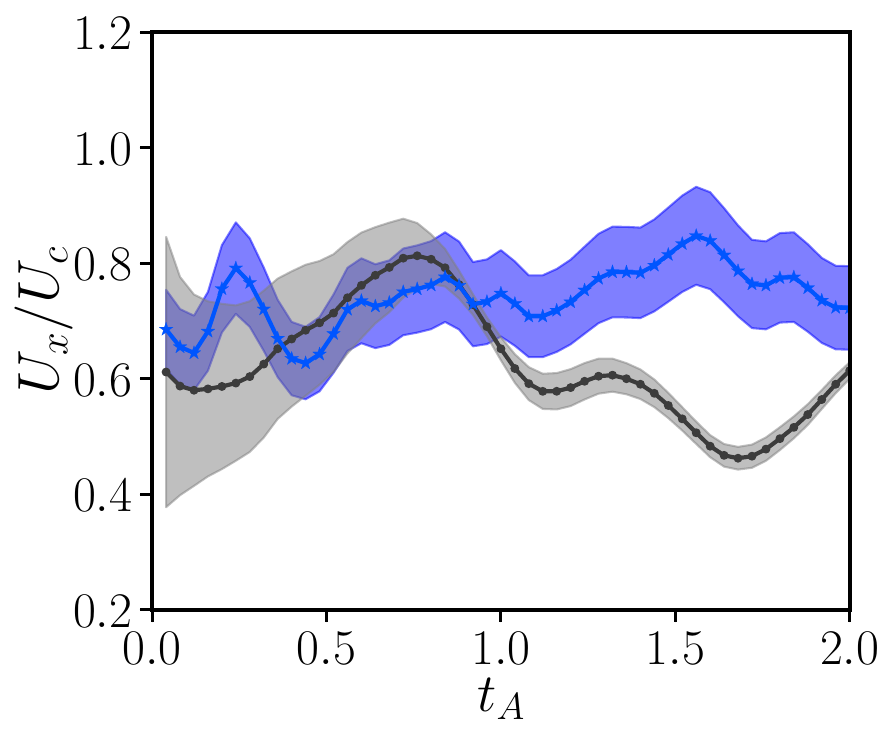}
    \end{subfigure}

    \begin{subfigure}[b]{.33\textwidth}
        \centering
        \includegraphics[width=\textwidth]{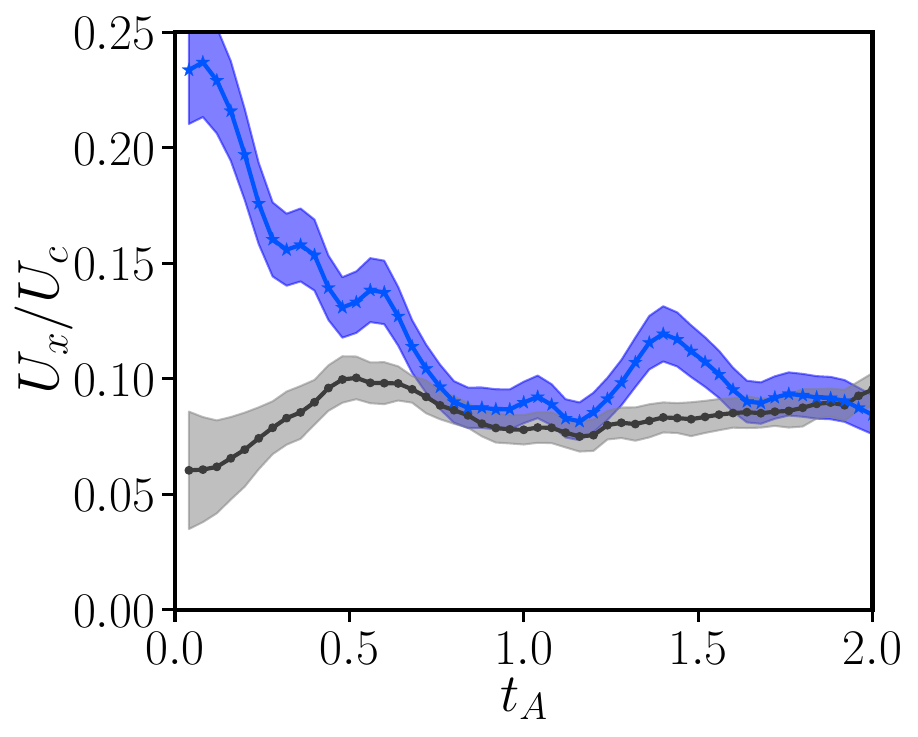}
    \end{subfigure}%
    \hfill 
    \begin{subfigure}[b]{.33\textwidth}
        \centering
        \includegraphics[width=\textwidth]{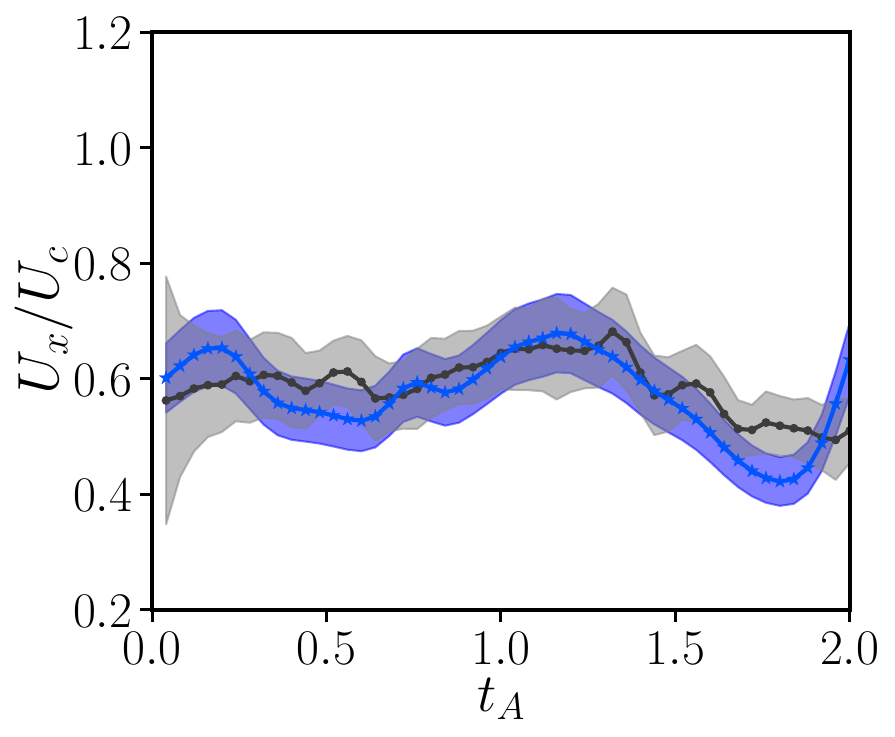}
    \end{subfigure}%
    \hfill 
    \begin{subfigure}[b]{.33\textwidth}
        \centering
        \includegraphics[width=\textwidth]{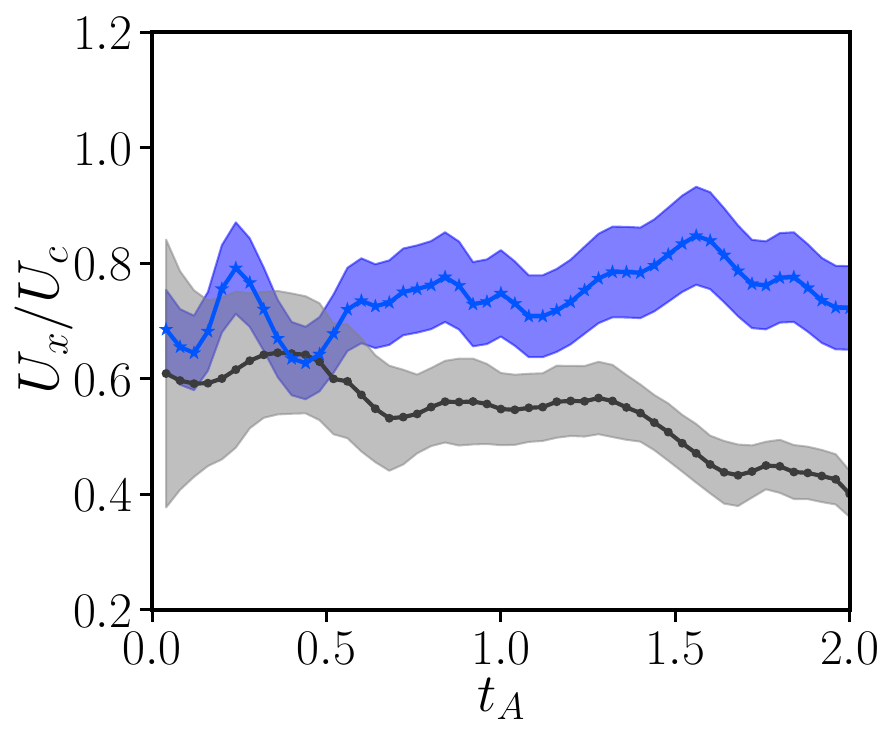}
    \end{subfigure}

    \begin{subfigure}[b]{.33\textwidth}
        \centering
        \includegraphics[width=\textwidth]{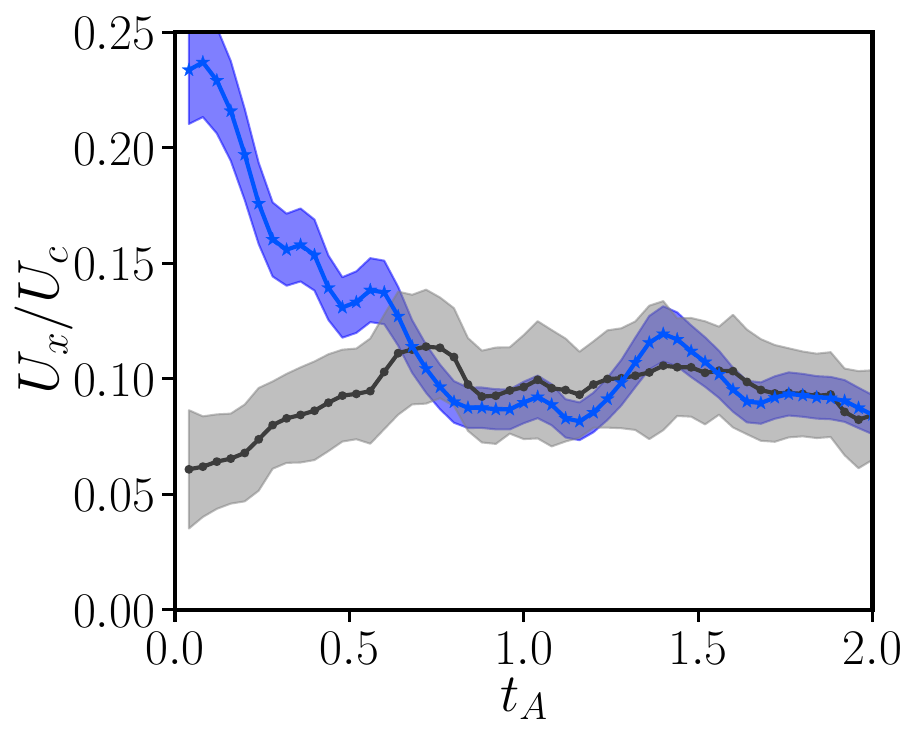}
    \end{subfigure}%
    \hfill 
    \begin{subfigure}[b]{.33\textwidth}
        \centering
        \includegraphics[width=\textwidth]{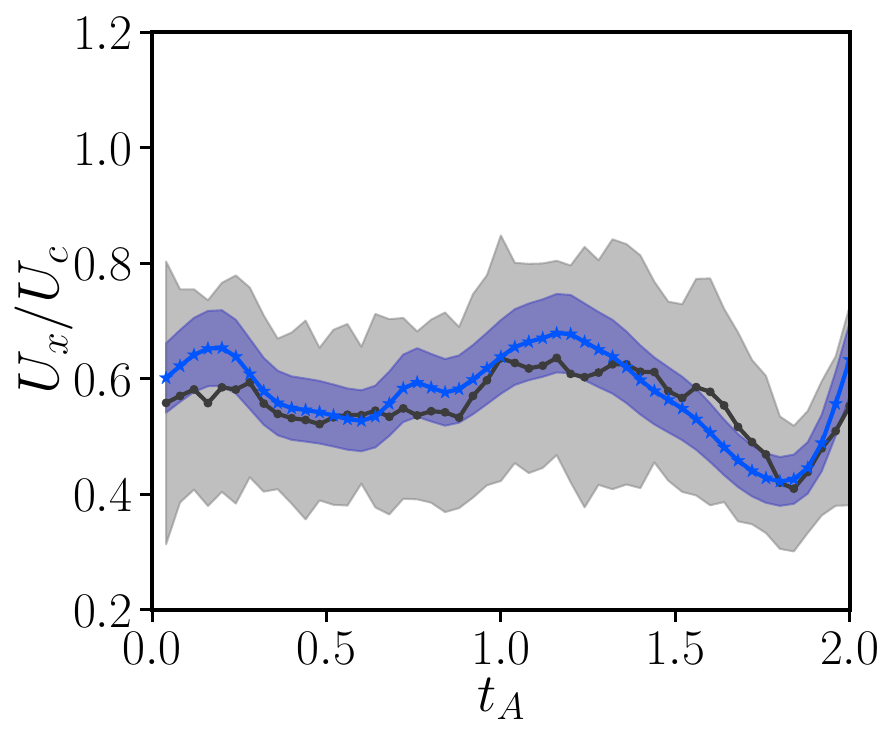}
    \end{subfigure}%
    \hfill 
    \begin{subfigure}[b]{.33\textwidth}
        \centering
        \includegraphics[width=\textwidth]{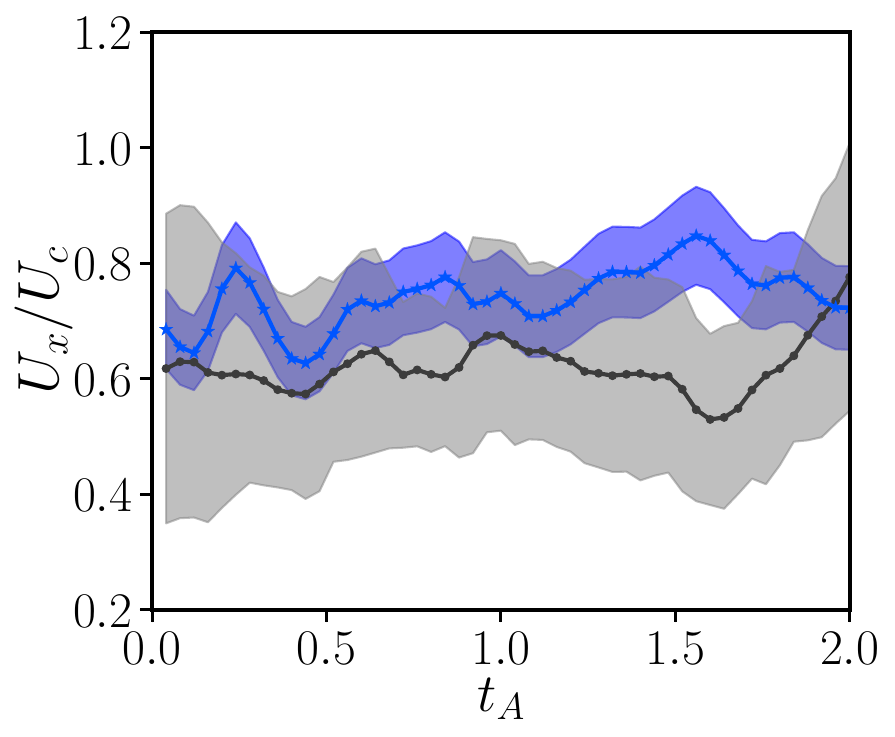}
    \end{subfigure}

    \begin{subfigure}[b]{.33\textwidth}
        \centering
        \includegraphics[width=\textwidth]{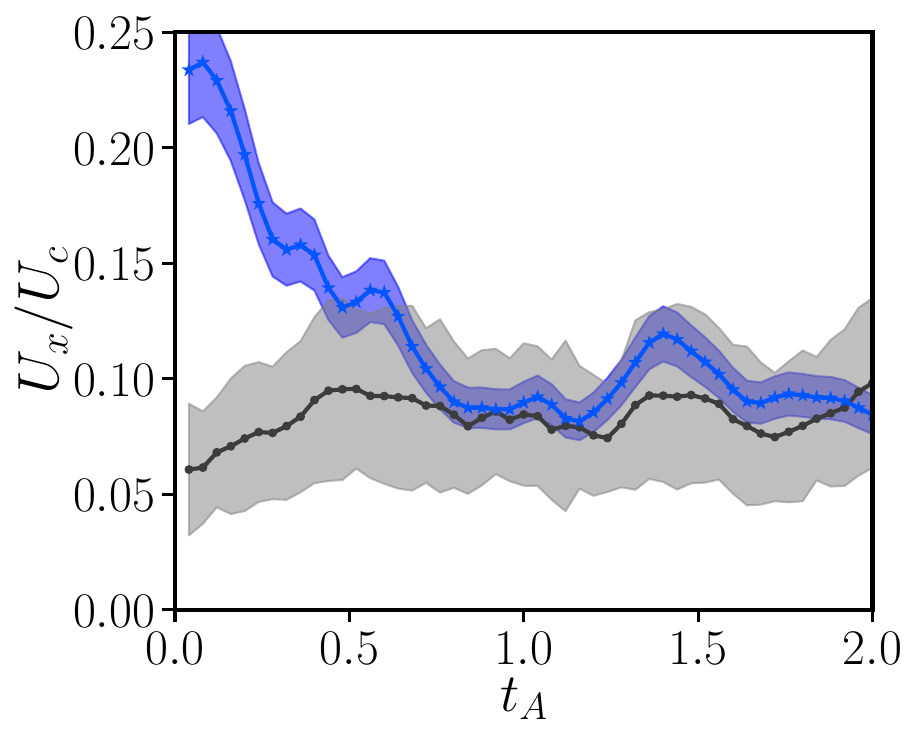}
    \end{subfigure}%
    \hfill 
    \begin{subfigure}[b]{.33\textwidth}
        \centering
        \includegraphics[width=\textwidth]{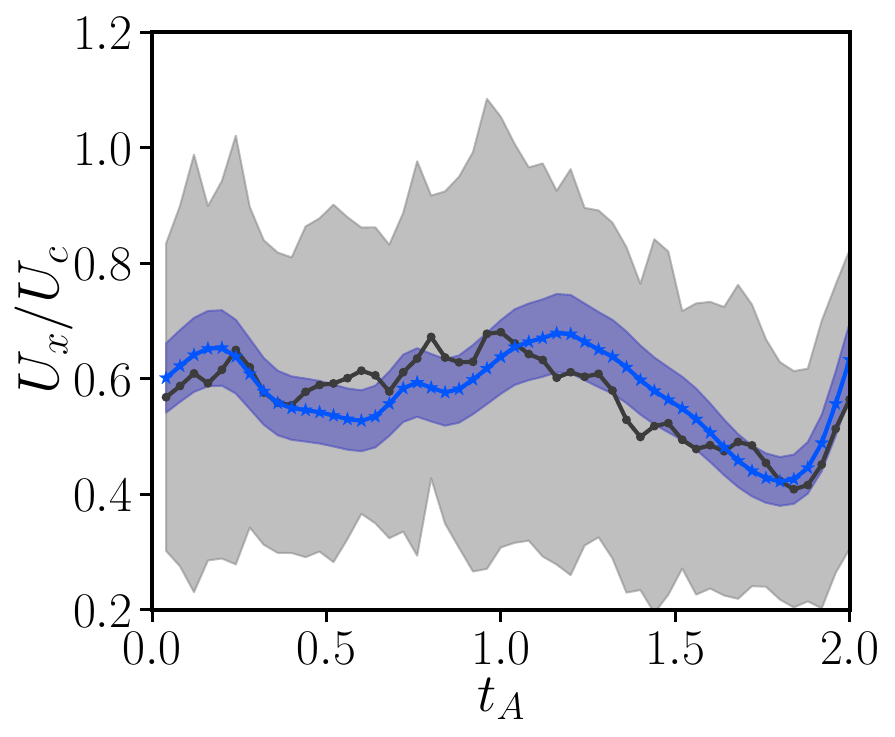}
    \end{subfigure}%
    \hfill 
    \begin{subfigure}[b]{.33\textwidth}
        \centering
        \includegraphics[width=\textwidth]{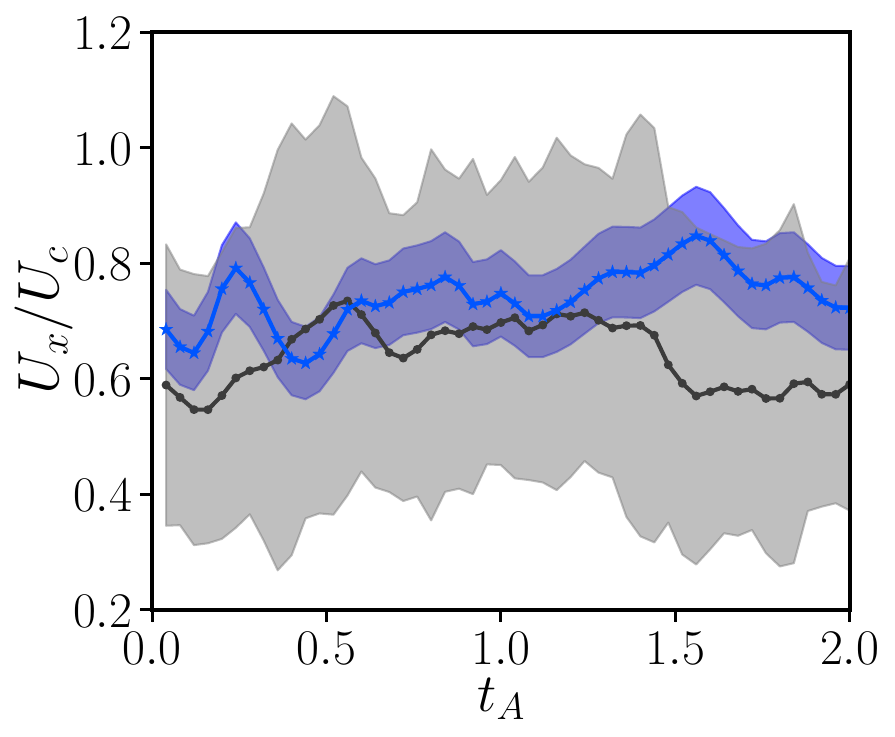}
    \end{subfigure}
    
    \caption{Synchronization of the velocity field. Data in blue represents the observation, while black lines indicate the model prediction. Results are shown for (left column) Probe 1, (center column) Probe 2, and  (right column) Probe 3. The magnitude of the state inflation is set to (first row) 0\%, (second row) 5\%, (third row) 15\% and (fourth row) 25\%.}
    \label{Fig:synchro}
\end{figure}

The evolution of the instantaneous streamwise velocity $u_x$ over the centerline average velocity $u_c$ is shown in Fig. \ref{Fig:synchro} for the three probes. The velocity sampled from the DNS, which is used as observation, is shown in blue. Data sampled in the same location from the ensemble members of the DA procedure is shown in black. In this case, the black line corresponds to an ensemble average. Shaded areas visually represent the confidence level/variability in the data. More precisely, the blue area is connected with the values included in the covariance matrix $R$ showing an area of thickness $2*\sigma_m$. On the other hand, the grey area represents the 95\% confidence interval for the model representation. This quantity is driven by the distribution of the prior states for $t_A = 0$, and it is progressively affected by the inflation applied to the physical state as more analysis phases are performed. 
The three probes have been selected to highlight different features of the flow field. Probes $1$ and $2$ correspond to sensors used in the DA procedures, but they are located at different distances from the wall ($y^\star = 1.45$ and $y^\star = 19.84$, respectively). On the other hand, the probe $3$ is located at $y^\star = 26.85$ and the corresponding sensor is not used in the DA analyses. One can see in the first line of Fig. \ref{Fig:synchro} that, if no \textit{state} inflation is used, the initial model variability due to the prior states collapses very rapidly with a drastic shrinking of the grey area. The grey and blue areas exhibit a very limited superposition, which prevents the model realizations from synchronizing with the observation. In the second, third, and fourth lines of Fig. \ref{Fig:synchro}, progressively more state inflation is used during the analysis phases. One can distinctively see an increase in the grey area associated with model variability, which does not decrease for larger simulation times. The analysis of the results for probes $1$ and $2$ clearly indicates that a threshold level of $15\%$ Gaussian state inflation appears to be enough to obtain a convincing synchronization of the velocity field in correspondence with the sensors. This threshold could potentially be even lower if more sophisticated algorithms for state inflation are used. Significant improvements with increasing inflation are observed as well for the probe $3$, even if the synchronization is not completely obtained. Therefore, these results confirm that the effect of the EnKF is not just local but, thanks to the scale interactions captured by the underlying LES model, a global improvement in the instantaneous flow prediction is obtained. One conclusion that can be drawn is that, once a significantly large superposition of the confidence areas is obtained for a sufficiently long time, a good synchronization is obtained. Similar behavior was also observed by Tandeo et al. \cite{Tandeo2020_mwr} but for a one-dimensional model.    

\begin{figure}
    \centering
    \begin{subfigure}[b]{.45\textwidth}
        \centering
        \includegraphics[width=1\textwidth]{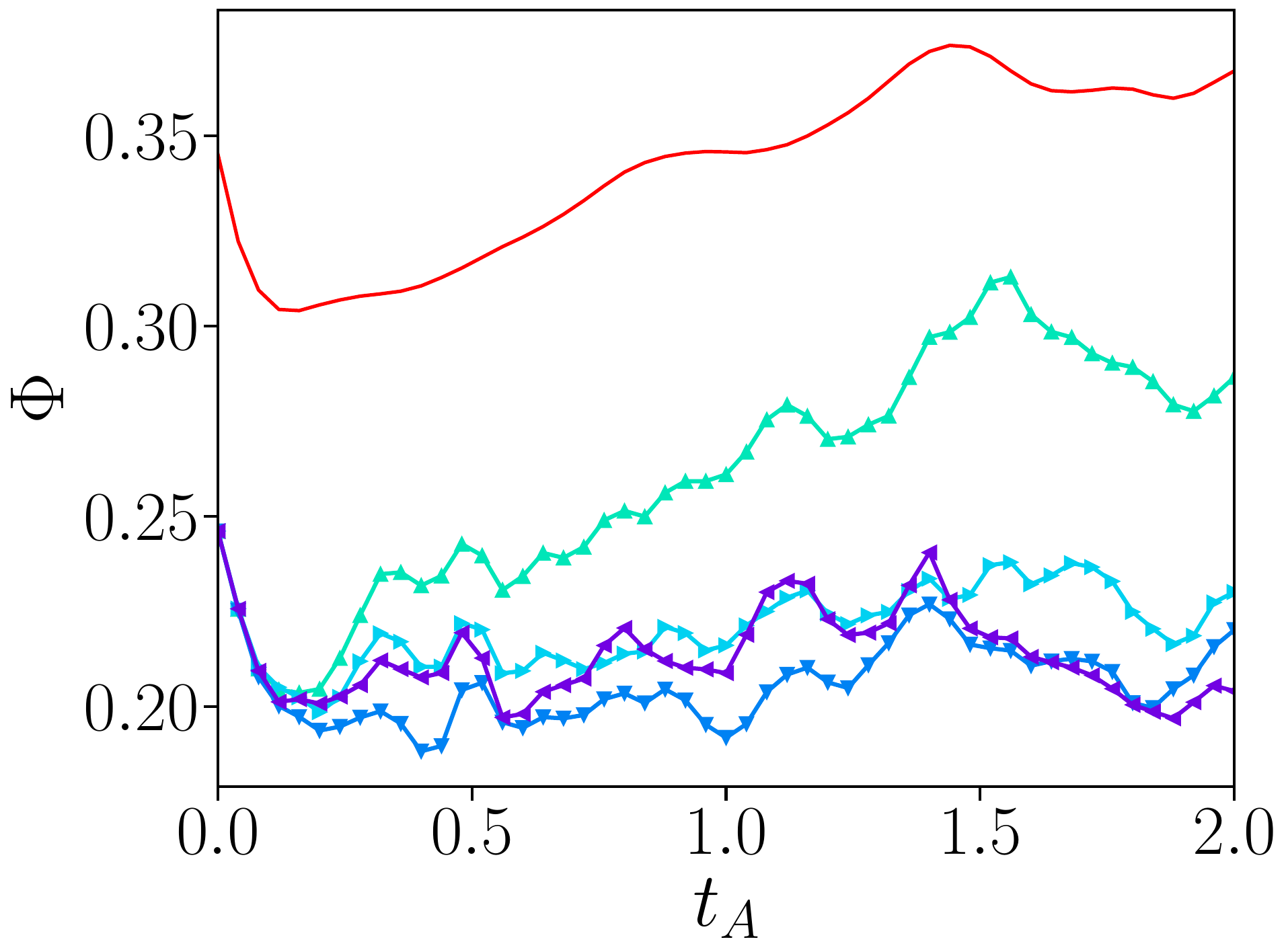}
        \caption{408 probes used for the DA}
    \end{subfigure}
    \hfill 
    \begin{subfigure}[b]{.45\textwidth}
        \centering
        \includegraphics[width=1\textwidth]{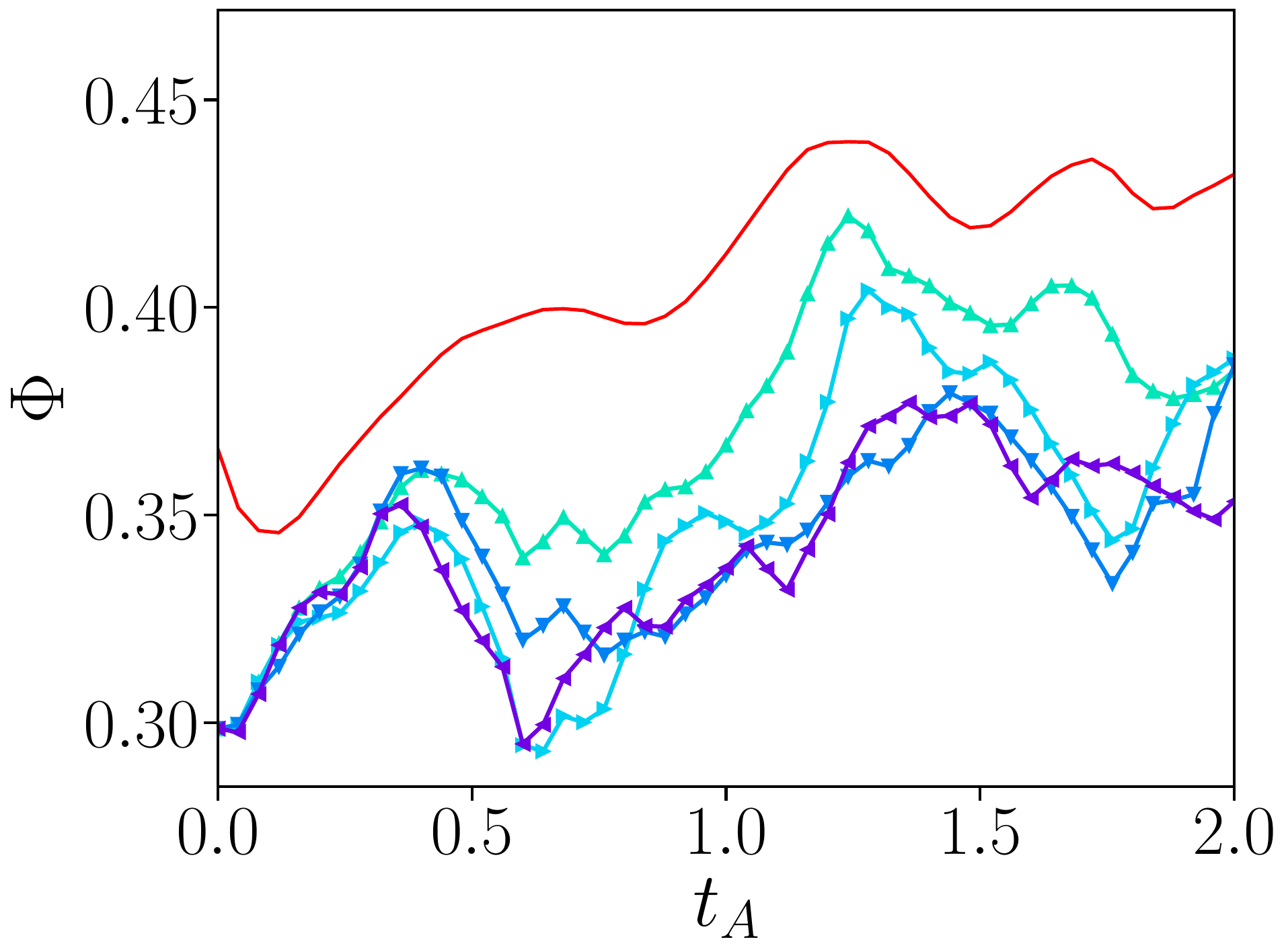}
        \caption{408 probes not used for the DA}
    \end{subfigure}
    \caption{Plots of $\Phi$ : From green to violet - 0\%, 5\%, 15\% and 25\% inflation. Red curve shows the average limit obtained comparing the values of the DA-LES simulations against 1000 observations times.}
    \label{Fig:error}
\end{figure}

The normalized root mean square deviation defined in equation \ref{eq:NRMSD} is now used to provide a global assessment of the capabilities of the DA algorithm to synchronize the LES model with the DNS available data. Results are shown in Fig. \ref{Fig:error} for the four cases previously analyzed i.e. $0\%$, $5\%$, $15\%$, and $25\%$ state inflation. The red line corresponds to a limit $\Phi_{lim}$ calculated comparing the values of the not inflated simulations against 1000 observations times. Therefore, for an infinite number of observations and locations, $\Phi_{lim}$ for Fig. \ref{Fig:error} (a) and (b) should be the same. It is here different due to the limited amount of probes and the heterogeneous set of coordinates of the probes. Results in Fig. \ref{Fig:error} (a) corresponds to the average discrepancy observed over the 408 locations where sensors are used for DA. The DA runs perform significantly better in the first stages, thanks to the variability initially provided with the choice of the prior states. However, results tend to degrade pretty rapidly for the DA run without state inflation. It could be expected to show very similar errors to $\Phi_{lim}$ after a sufficiently long time. On the other hand, the three DA experiments with non-zero state inflation behave very similarly. Their magnitude is significantly smaller than $\Phi_{lim}$ and it does not appear to deteriorate in the time window analyzed. Fig. \ref{Fig:error} (b) shows the results for the normalized root mean square deviation in correspondence of sensors where DNS data is available, but it is not used in the DA procedure. Results are qualitatively similar to what previously discussed for Fig. \ref{Fig:error} (a) even if, in this case, results for the DA runs are closer to $\Phi_{lim}$. This observation is due to the lack of correct representation of the correlation between variables, which is due to the limited number of ensemble members (sampling error). In this case, results seem to be more sensitive to the value of the state inflation, as very strong inflation seems to perform worse than moderate state inflation. One could expect in this case that the perturbations might be strong enough to introduce an unwanted noise effect on the flow prediction, degrading the global accuracy. Another potential issue with hyperparameters, which is not studied in the present work, is associated with the characteristic length used for covariance localization. If the selected length is large, spurious correlations may appear because of sampling errors. On the other hand, a short length may preclude an accurate representation of the correlation between the variables, working as a filter over the multi-scale non-local interactions observed in turbulent flows. 

In summary, the analysis of the global quantity $\Phi$ stresses how much DA can be important to provide a successful instantaneous state estimation, which could be even more important than an accurate parametric optimization for the prediction of rare extreme events and for the optimization of unstationary flows exposing a strong time evolution of its features. 

\section{Conclusion}
\label{Sec::conclusions}

An online DA strategy based on state-of-the-art techniques for the Ensemble Kalman Filter has been used to improve the predictive capabilities of Large-Eddy Simulation. The attention of the work is mostly devoted to the correct representation of instantaneous features, which can be essential to predict and anticipate extreme events affecting industrial applications. To perform the analysis, an on-the-fly coupling has been performed via the platform CONES, combining LES solver runs using OpenFOAM and localized instantaneous high-fidelity information obtained from a DNS. First, the DA runs used instantaneous values of the velocity field to optimize the parametric behavior of the Smagorinsky model used for subgrid closure. Two strategies have been proposed to obtain an optimized value of the model constant $C_k$. Despite the difference in complexity, both strategies provide a similar result, which is a significant reduction of the intensity of $C_k$ and of the SGS model. These conclusions support recent discussion in the LES community about the usage of explicit and implicit SGS modeling \cite{VicenteCruz2023_jot}. This optimization reduces the discrepancy of the statistical moments of the flow field with DNS data, but it does not eliminate it, as observed by Mons et al. \cite{Mons2021_prf}. The reason behind this observation is associated with the structural limitations of the Smagorinsky model, whose intrinsically dissipative nature is not able to fully take into account the effects of the filtered scales and their interactions with the resolved flow field. 
The DA model has then been used to analyze the efficiency in flow reconstruction and synchronization with the high-fidelity sparse data available. It was shown that DA is able to significantly improve the correlation between model results and observation, but the efficiency in such synchronization is governed by the state inflation applied. This hyperparameter is an essential key feature of the DA algorithm which deserves more specific studies in the future. Similarly, the effects of physical and covariance localization, which were excluded in the present analysis, will be extensively investigated in future research for online DA strategies. 

\begin{acknowledgments}
Our research activities are supported by the funding of the French Agence Nationale de la Recherche (ANR) through project PRC 2020 ALEKCIA. 
\end{acknowledgments}

\appendix
\section{Usage of multiple physical information for each sensor in the DA procedure}
\label{Appendix::DA-LESA}
\begin{figure}
    \centering
    \begin{subfigure}{.45\textwidth}
        \centering 
        \includegraphics[width=1\textwidth]{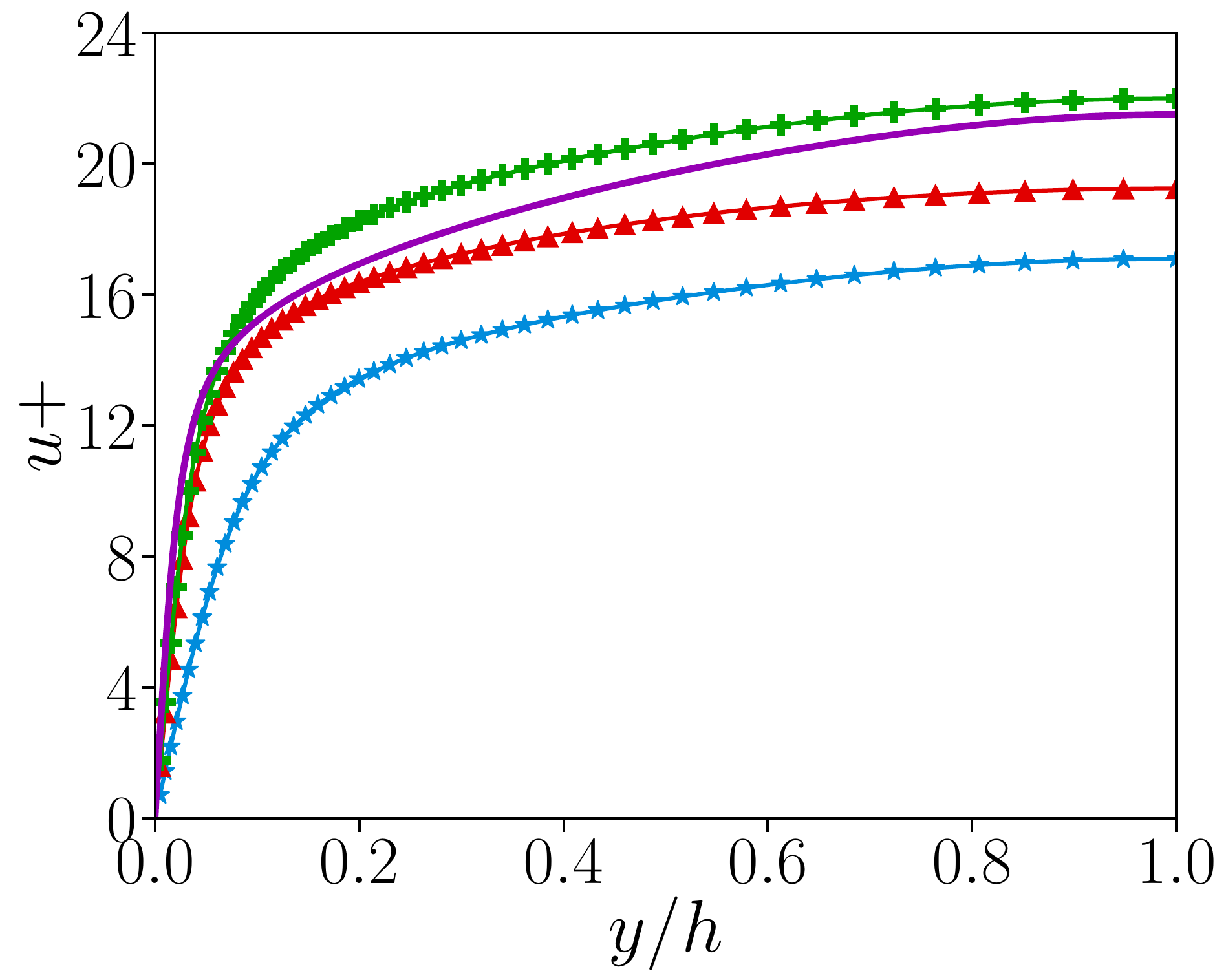}
        \caption{}
        \label{Fig:UPlusOverYh_comp}
    \end{subfigure}
    \begin{subfigure}{.45\textwidth}
        \centering 
        \includegraphics[width=1\textwidth]{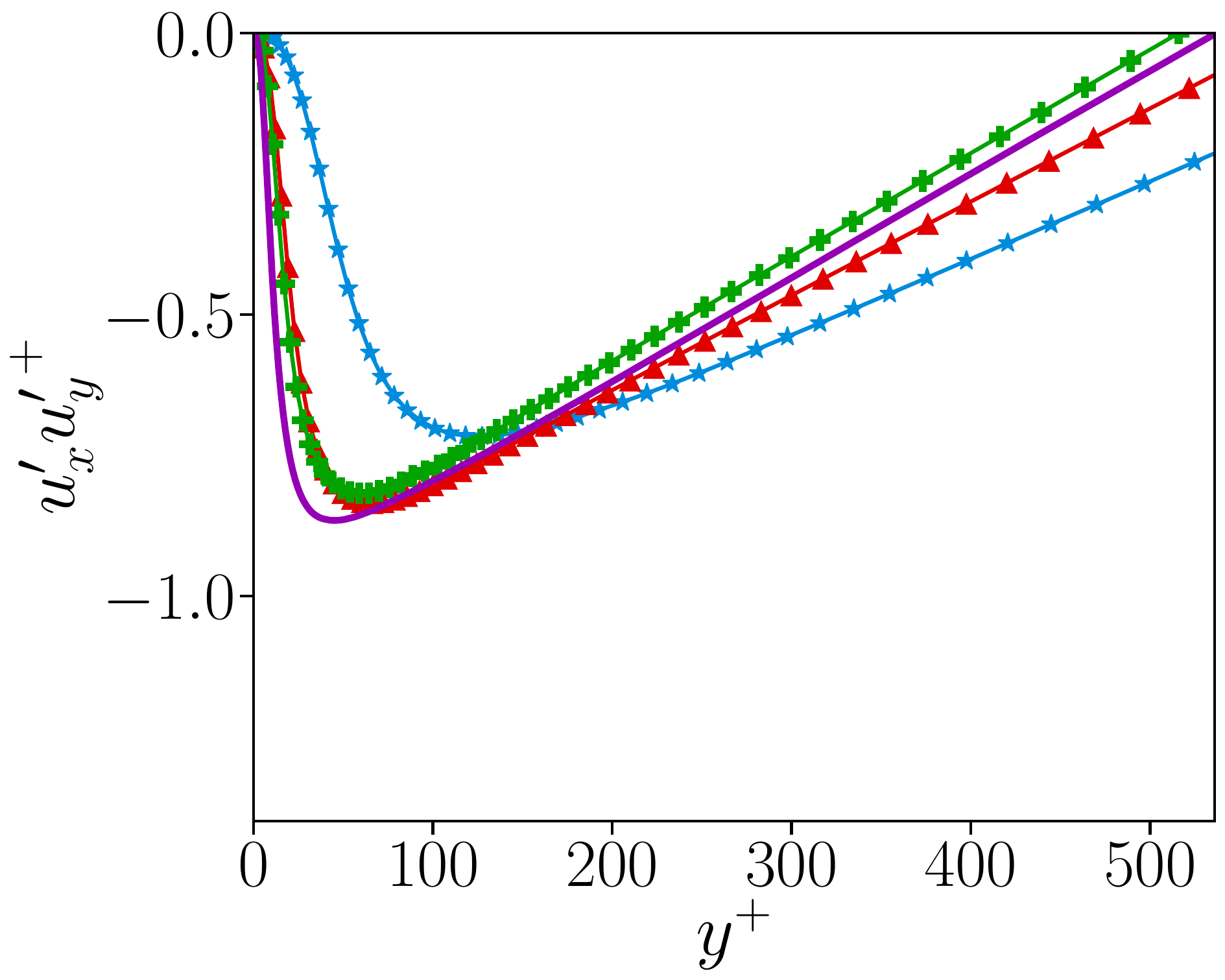}
        \caption{}
        \label{Fig:uvPlusOverYPlus_comp}
    \end{subfigure}
\caption{Half channel velocity profiles for the DA runs DA-LES1 (\textcolor{red}{$\blacktriangleright$}), DA-LESA (\textcolor{green}{$+$}), the baseline Smagorinsky LES (\textcolor{cyan}{$*$}) and the DNS (\textcolor{violet}{$\times$})}
\label{Fig:mean_profiles_comp}
\end{figure}

In order to test the sensitivity of the DA algorithm to multiple physical information available at one sensor, an additional DA run has been performed. This test, referred to as DA-LESA, is almost identical to DA-LES1. The only difference is that, for each sensor, the three components of the velocity field are here provided. We remind that for the runs DA-LES1 and DA-LES2 only the streamwise component of the velocity field was used as observation in the analysis phase. Therefore for DA-LESA, the observation matrix is composed of $408 \times 3=1224$ values at each analysis phase. The covariance matrix of the measurement error is expressed as $R = \sigma_m^2I$, where $\sigma_m$ quantifies the uncertainty of the measurements. In this case, $\sigma_m$ is the same for every sensor and it is calculated accounting for a $5\%$ uncertainty over the maximum velocity observed in the DNS to mimic the accuracy of experimental measurements. Therefore, $\sigma_m \approx 0.045$ in this case. This choice implies that the confidence in the DNS results is lower approaching the wall. This decision is beneficial to obtain a robust behavior of the EnKF because large discrepancies between DNS and LES can be observed very close to the wall.
The results of the optimization are similar to those of the other DA runs, indicating that the EnKF procedure is robust. The optimized value of the model constant is $C_k \approx 0.025$ which is $3.7$ times smaller than the baseline LES and corresponds to $C_S \approx 0.06$. During the DA run, values exhibit oscillations in the range $C_k \in [0.020, \, 0.030]$. The DA-LESA also shows a good improvement in the prediction of the friction velocity $u_\tau= 0.052$ with an over-prediction of the friction velocity of $8.3\%$, compared with the $28\%$ of the baseline LES. The normalized mean velocity $u^+$ over $y^+$ and the normalized resolved shear stress of DA-LES1 and DA-LESA are shown in Fig.\ref{Fig:mean_profiles_comp}. Other statistical moments of the velocity field are not shown here for the sake of conciseness, as they provide similar information. Differences between the DA runs for the prediction of the statistical moments are noticeable and mainly associated with the different prediction of the friction velocity, which is less accurate for DA-LESA. One possible reason is associated with the level of confidence in the observation, which was set at the same level for the three components of the velocity field. In the near wall region, the streamwise component is around one order of magnitude larger than the other two components, and uncertainties propagated in the observation vector act as a random noise for $u_y$ and $u_z$. The problem of determining an optimized hyperparametric description of the confidence level of observation, which degraded the global accuracy of the DA run in this case, deserves future investigation when such a quantity is not directly quantifiable.

\section{Computational resources required to perform the DA run}
\label{Appendix::cost}
The computational resources required to perform the DA runs are now discussed. Tab. \ref{Tab:gain_comp} shows information about preliminary tests performed varying a number of key parameters such as the number of mesh elements used for the LES model $N$, the amount of sensors/observations $N_o$ and the size of the ensemble $N_e$.  
In particular, the values investigated for $N$ ($350\, 000$ and $154\, 000$) correspond to the numbers of mesh elements for the complete and clipped physical domain used in the present work. Comparing the completion time between lines 1 and 2 of Tab. \ref{Tab:gain_comp}, one can see that the reduction of the degrees of freedom of the model is beneficial in terms of computational cost, dividing by $2.25$ the completion time. However, the most important parameter is the number of observations. The comparison of lines 2, 4, and 5 shows a dramatic reduction of the computational resources required with fewer sensors. This point stresses out the importance of the quality of observations used in the DA rather than quantity, as previously shown in \cite{Villanueva2023_arxiv}. At last, one can see that the comparison of results in lines 2 and 3, where a different number of ensemble members $N_e$ is used, has a lower impact on the computational cost when compared with the previous parameters of investigation.  

\begin{table}
\centering
\begin{tabular}{@{}cccccc@{}}
\toprule
$N$         & $N_o$    & $N_e$  & \begin{tabular}[c]{@{}c@{}}Operations\\ Complexity\end{tabular} & \begin{tabular}[c]{@{}c@{}}Completion\\ time (s)\end{tabular} & \begin{tabular}[c]{@{}c@{}}Operations/s\end{tabular} \\ \midrule
$350,000\times3$ & $1224$ & $40$ & $\mathcal{O}(1.6\times10^{12})$ & $161$  & $10^{10}$      \\
$154,000\times3$ & $1244$ & $40$ & $\mathcal{O}(7.2\times10^{11})$ & $71.5$ & $10^{10}$         \\
$154,000\times3$ & $1224$ & $10$ & $\mathcal{O}(7\times10^{11})$ & $54$   & $1.3\times10^{10}$       \\
$154,000\times3$ & $408$  & $40$ & $\mathcal{O}(8.4\times10^{10})$ & $8.9$  & $9.5\times10^{09}$      \\
$154,000\times3$ & $228$  & $40$ & $\mathcal{O}(2.8\times10^{10})$ & $3.4$  & $8.4\times10^{09}$      \\ \bottomrule
\end{tabular}%
\caption{Summary of the test performed to evaluate the computational costs required by the EnKF}
\label{Tab:gain_comp}
\end{table}

\section{Supplementary details about synchronization}
\label{Appendix::synchronization}

\begin{figure}
    \centering
    \begin{subfigure}{.45\textwidth}
        \centering 
        \includegraphics[width=1\textwidth]{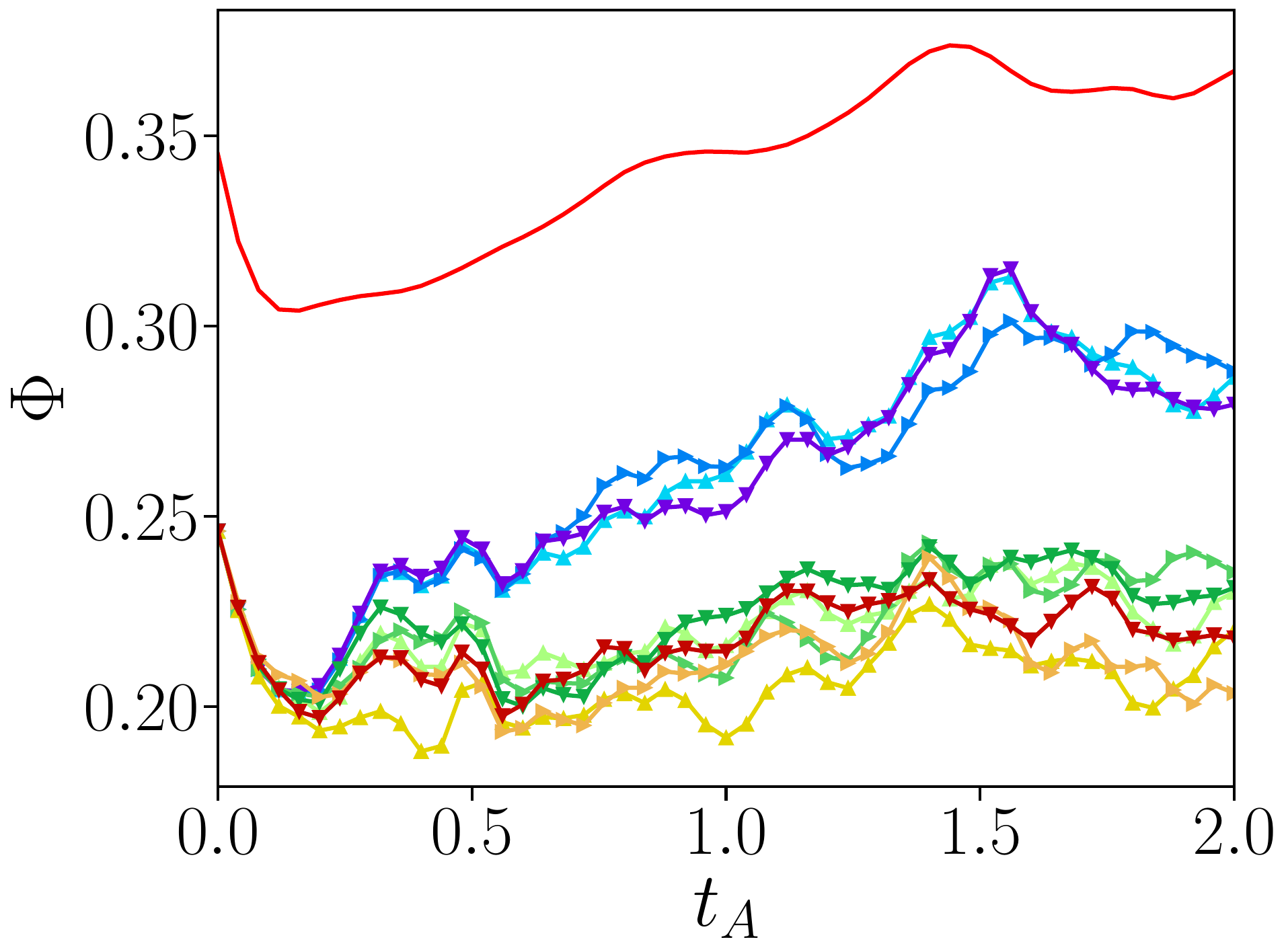}
        \caption{}
        \label{Fig:NRMSD_A1}
    \end{subfigure}
    \begin{subfigure}{.45\textwidth}
        \centering 
        \includegraphics[width=1\textwidth]{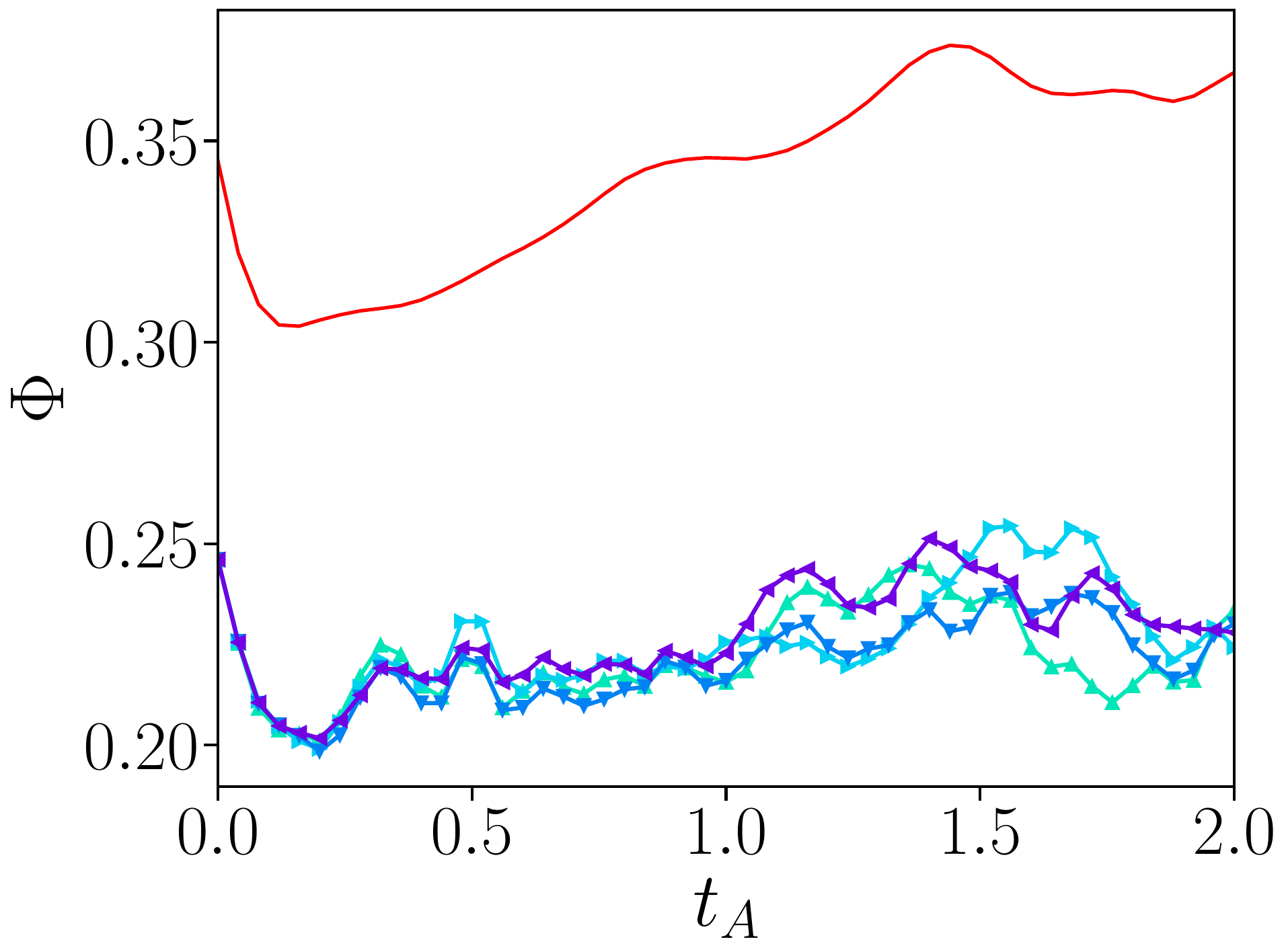}
        \caption{}
        \label{Fig:NRMSD_A2}
    \end{subfigure}
\caption{$\Phi$ of the 408 probes used in the DA analysis, using different levels of parameter inflation (left), from light to dark color: 0\%, 2\%, and 5\% inflation, and observation confidence (right) - from green to violet: 0.5\%, 1\%, 5\%, and 10\% confidence}
\label{Fig:Pi_conf_effect}
\end{figure}

The sensitivity of synchronization to inflation in the parametric description of the model and in the variance of the observation is here discussed. Fig. \ref{Fig:NRMSD_A1} shows the normalized root mean square deviation with variation of the inflation on the parameters and state inflation. Three levels of parameter inflation are used from light to dark color: 0\%, 2\%, and 5\%. State inflation is set to 0\% in blue, 5\% in green, and 15\% in orange colors. Inflation for the model parameters appears to have a negligible effect on the synchronization obtained via DA when compared with state inflation. Fig. \ref{Fig:NRMSD_A2} shows four levels of the prescribed variance for the observation, for 5\% inflation of the state. Again, synchronization does not seem to be affected by the level of confidence in the observations here tested, which is in the range of recommendations for robust application of the EnKF. Very low or very high confidence in the observation can lead to poor synchronization as well as inaccurate parametric optimization, as shown by Tandeo et al. \cite{Tandeo2020_mwr}.

\newpage
\bibliography{bibliography}

\end{document}